BULGARIAN ACADEMY OF SCIENCES

INSTITUTE OF SOLID STATE PHYSICS
"ACAD. GEORGI NADJAKOV"

# Dimo N. Astadjov

# *High-End State-of-Art Copper Bromide Vapor Lasers: Optical, Electric and Thermal Properties*

*Research and development of copper bromide vapor lasers in Laboratory of metal vapor lasers*

*Latest 20+ years*

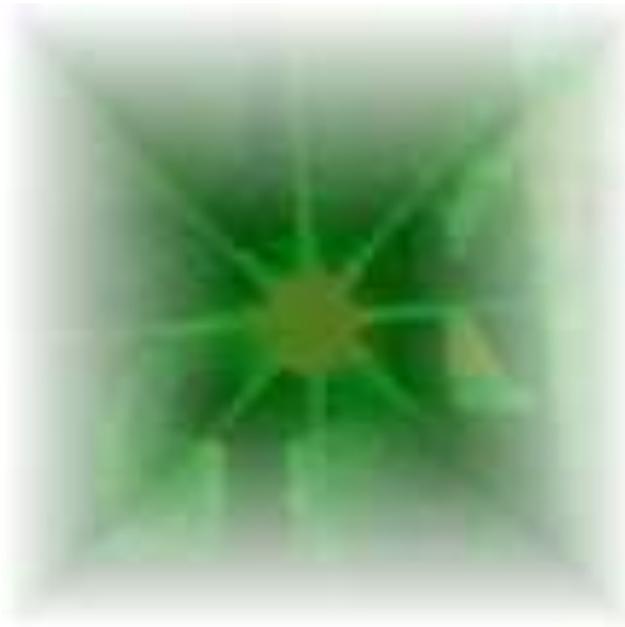

Sofia - Bulgaria • 2017

*To my parents, Kleopatra & Nikola,*

*This book is devoted.*

Copper bromide vapor laser is the most successful sealed-off version of a plenty of copper vapor lasers launched, studied and produced for the last half of century. Being first reported in 1975, it became the leading and long-lasting R&D&T project followed by the Laboratory of metal vapor lasers in Institute of solid state physics of the Bulgarian Academy of Sciences ever since. This manuscript gathers major research & development results of the last studies unfolding unexplored and unknown optical, electric and thermal properties of CuBr laser held in the Laboratory of metal vapor lasers during foregoing 20+ years. Some of these new aftermaths are record-high power, efficiency and lifetime of CuBr laser; comprehensive studies of energy dissipation and dynamics of the electric field of CuBr laser; achievement of high-brightness diffraction-limited throughout-pulse emission of master-oscillator power-amplifier (MOPA) CuBr laser and high focusability of CuBr laser emission low-affected by beam profile intensity variations.

Dimo Nikolov Astadjov, 2017

# CONTENTS

# *Preface*

The copper bromide vapor laser was born in the early 1970s when laser emission on atomic copper self-terminating transitions in copper bromide vapor was simultaneously reported by three research groups – a Bulgarian (Sabotinov *et al.,* 1975) and two Russian (Shuhtin *et al.,* 1975 and Akirtava *et al.,* 1975). In fact the real year was 1974 when first laser oscillation in CuBr vapor was obtained by Bulgarian physicists headed by Sabotinov but it was not uncovered for intellectual property reasons (Sabotinov, 1996). Ever since this became the leading area of research in the Department (Laboratory) of metal vapor lasers and important part of development and technology in Institute of solid state physics for many years. And this original-Bulgarian invent gave enormous yields not only in Bulgaria yet around the world.

Nobody imagined the long and prolific life of this new laser. It appeared as a version of fast developing copper vapor laser being an element of the boom in research and development of metal vapor lasers all over the world. Unlike some of the obvious advantages of CuBr laser technology as much lower temperature of operation and simplified construction design, other aspects of its technology and use in practice were reasons of discomfort and anxiety for good future. The major problem came from the halogen component of CuBr molecule – member of the family of most aggressive elements in the periodical table. Bromine and all its products corrode materials in contact and lead to failure of lasers. So the once very promising perspective of CuBr laser was going to be doomed. Short life expectancy is not a good advertiser for an emerging commercial product.

So the chance to survive in a competition with other lasers laid in the resolving of the problem with bromine and its products during the lifetime cycle of CuBr laser. First years following the invention were a time of intensive research of the processes that take place in CuBr laser and cause aging of construction components, and in the end the laser failure. Systematic and detailed research in the Laboratory of metal vapor lasers (ISSP, BAS) gave initial results in the middle of 1980s. The studies showed that laser deterioration can be mitigated if not fully eliminated. The lifetime of CuBr laser was dramatically increased from tens hours of unstable and low-power operation to thousands of hours of stable and failsafe work (Sabotinov, 1991). And that was accompanied by high-power high-efficiency performance never achieved before due to the strong effect of hydrogen additives discovered in our lab by Astadjov *et al.* (1985). This breakthrough in the copper vapor technology had a striking impact on the world community of metal vapor laser researchers (Sabotinov *et al.,* 1986). It actually brought about a worldwide renaissance and bright comeback of halogens as factors catalyzing and improving the overall performance of copper (and other metal) vapor lasers. The revival spawned families of new technological types as HyBrID (hydrogen bromide in discharge) lasers and kinetically-enhanced lasers (Little, 1999). The vitality of copper vapor laser is so high that it is still present firmly in the extremely competitive world laser market.

In this manuscript, I will give the last studies that unfold optical, electric and thermal properties of CuBr laser held in the Laboratory of metal vapor lasers during foregoing 20 years. Before that in a short survey, I will go through a timeline of milestone events happening in CuBr laser life paving its glorious 40+ - year history.

## *1.  Milestones in the history of copper bromide vapor laser*

The very first problem we came across and had to deal with was the short lifetime and unstable operation of CuBr laser coming from the presence of a halogen element − the bromine. For a 10-years' time, we put our efforts to make clear how to get through that severe problem. Is it possible or not? Most world research teams gave up very soon but that was not our case. We concentrate a lot of vigor and time to make it true that the problem with bromine can be overcome. Moreover, we found that it can be harnessed to give fruits that nobody could even imagine at that time. Two construction types of laser tubes were elaborately investigated in our lab – whole-heated tube (Vuchkov, 1984) and cold-ends tube (Astadjov, 1989). While the former let us measure the bromine concentration and find the links with laser operation, the latter helped us to put the bromine under control. The outcome was that lifetime of CuBr laser tubes was extended to thousands of hours and made the laser ready for the market.

During the years to come making the most of the strong effect of hydrogen additives mentioned above (Astadjov *et al.*, 1985), we obtained record-high values of 1.4 W/cm$^3$ of average power per volume (Astadjov *et al.*, 1994), and average laser power of 120W and efficiency of 3% from stand-alone CuBr laser tubes (Astadjov *et al.*, 1997). Besides we continued our research immersing deeply in physics of phenomena occurring in CuBr laser tubes. Consequently, the distribution of energy dissipated along the tube (*i.e.* electrodes and the lasing zone) was under detailed investigation by Astadjov & Sabotinov (1997). Later Astadjov & Sabotinov (1999) examined the electric field dynamics by applying an original technique of surface voltage scanning.

Quite a lot time we devoted to optical properties of CuBr laser emission. To begin with, in 2000, we reported of a high-spatial intensity 10 W-CuBr laser with hydrogen additives working with an unstable resonator (Stoilov *et al.*, 2000). Then the time came for investigation of cascaded laser tubes forming master-oscillator power-amplifier (MOPA) systems. Diffraction-limited throughout-pulse emission was attained via generalized diffraction filtering resonator (GDFR) and a CuBr MOPA laser of very high brightness was demonstrated by Astadjov *et al.* (2005). By means of an innovative reversal shear interferometer (our modification of the Michelson interferometer), we carried on with measurements of optical coherence for the first time in our lab. The GDFR CuBr MOPA laser showed greatly improved spatial coherence degree of up to 0.65 (Astadjov *et al.*, 2006). Also, we first reported the measurement of spatial coherence of low-cost 532nm green lasers in (Astadjov & Prakash, 2013b).

Astadjov *et al.* (2007) published a study on the beam propagation factor (known as $M^2$) of CuBr MOPA laser emission compliant with ISO 11146. Statistical parameters of 2D intensity profile of the near and far fields of MOPA laser radiation were measured by a beam analyzing technique as functions of timing delay between master oscillator (MO) and power amplifier (PA). For the first time, the influence on CuBr laser beam focusability (as $M^2$) of the gas buffer (causing the radiation profile to change from annular to top-hat and Gaussian-like) and light polarization was under investigation. For annular radiation $M^2$ range was from 13-14 (small delays) to 5-6 (large delays) and for filled-center radiation $M^2$ was 6-7 (small delays) and at the end of PA, gain curve was as much as 4. With polarized light, $M^2$ dropped to 3 at the end of PA gain curve. The brightness of laser emission with hydrogen increased 3-5 times. And with the linearly-polarized MO beam injected into PA, the MOPA laser brightness augmented further. It turned to be at least 40% brighter than that in the case of PA seeded with partial or non-polarized MO beams.

The years after, we worked on energy focusability of circular annular beams as in theory so in the experiment (Astadjov, 2009, 2010; Astadjov & Nakhe, 2010; Prakash *et al.*, 2013; Astadjov & Prakash, 2013a). We made a computer model of transformation of an annular beam with a certain intensity distribution (2D near-field input) into a beam with new intensity distribution (calculated 2D far-field output). The far-field intensity distribution of beam (the same as that of the focal plane of a lens) was mathematically completed by 2D fast Fourier transforms (FFT). This way we were able to estimate the influence of digression of beams from normal (as of Gaussian or top-hat profile) shapes on the energy distribution of the output beam in the focus. The simulations (Astadjov, 2009, 2010) of annular beams through 2D FFT revealed that the far-field central peak energy fraction is prevailing in the focal (far-field) energy distribution virtually regardless the (near-field) shape of the annular source beam. Beam annularity (the ratio of low-intensity core radius to high-intensity annulus radius) and beam core intensity were not critical parameters with the exception of annularity values higher than 0.65 and the annulus being almost black (core intensity is less than 10% of the ring intensity). So they had a negligible effect on the focal energy distribution. In practice, that means a less radiation-affected area in the vicinity of output beam central peak (part of the far-field energy distribution). Taking into account that the central peak energy is confined within a smaller spot area, the net impact of side energy spread diminishes furthermore. The simulations were in good agreement with experiments employing laser source of an alterable degree of spatial coherence (Prakash *et al.,* 2013; Astadjov & Prakash, 2013a).

## 2. *Power, efficiency and lifetime of copper bromide vapor laser*

### 2.1. *Small-bore CuBr laser with 1.4 W/cm$^3$ average output power*

One of the interesting questions in laser physics, especially from the point of view of high-power laser development, is how much laser power can be extracted per unit volume of the active medium of a laser. This question concerns all kinds of lasers, including the continuously pulsed copper vapor laser. Sabotinov *et al.* (1990) reported that a conventional CuBr laser with an active zone of 50-cm x 2-cm diameter delivered a specific average power (SAP) of 0.13 W/cm$^3$. Somewhat higher was the SAP (0.2 W/cm$^3$) obtained by Livingstone *et al.* (1992) from an active zone of 30-cm x 1.3-cm diameter in a laser operated with CuBr produced from elemental copper and Ne—HBr flow. However, this is much lower than the SAP reported by Vorobev *et al.* (1991): for a copper vapor laser with an active zone of 30-cm x 0.45-cm diameter they asserted that the SAP was 1.3 W/cm$^3$ while the maximum laser power measured was 3.1 W.

In order to demonstrate the possibilities of conventional CuBr vapor laser to produce high SAP and to compare it with its traditional competitor, the elemental copper laser, we had to undertake an experimental investigation (Astadjov *et al.,* 1994) of a small-bore CuBr laser with the same dimensions of the active zone as the tube of the small-bore copper vapor laser reported by the latter researchers - Vorobev *et al.* (1991).

*Experimental apparatus*

The construct of the laser tube of the small-bore CuBr laser that was studied is shown in Figure 2.1.1. Five quartz diaphragms were mounted in a fused silica tube of 15-mm inside diameter. The thin (1.5-mm) diaphragms were of 4.5-mm aperture, which defined the active zone. The distance between the diaphragms located close to each electrode was 300 mm, which was assumed for the length of the active zone. Thus, the active volume of the tube by construction was 4.77 cm$^3$, as given in (Vorobev *et al.,* 1991). The ends of the laser tube were set at the Brewster angle for λ540 nm, and quartz windows were stuck to them. The electrodes were of porous copper and had special side arms to prevent their contamination with bromine. CuBr was placed in four side reservoirs, which were heated externally. The laser tube was wrapped by a thin layer of fibrous insulation. The laser cavity consisted of two flat mirrors: a 99.8% reflectivity mirror and a plane-parallel quartz output coupler.

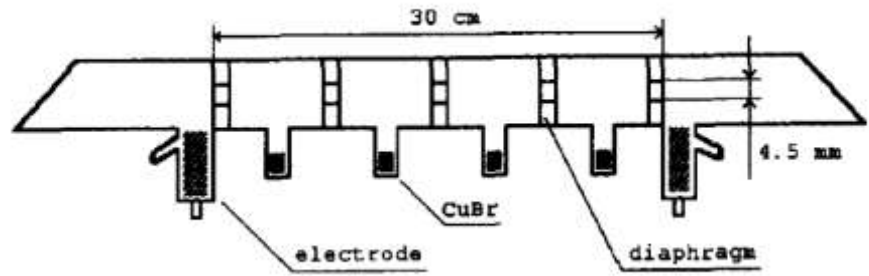

**Figure 2.1.1 Construct of the small-bore CuBr laser**

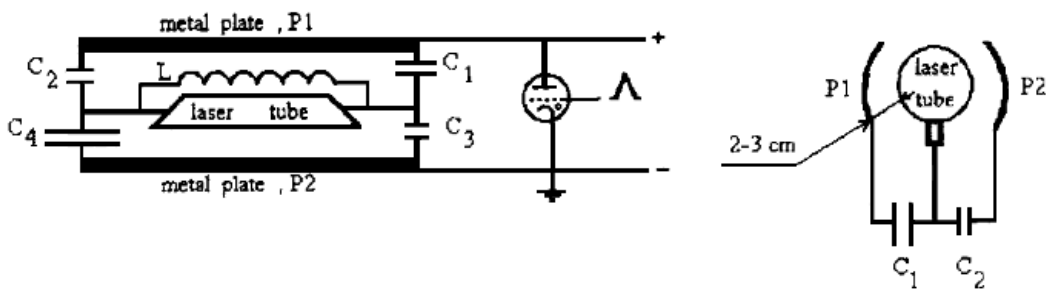

**Figure 2.1.2 Schematic diagram of the excitation circuit for the small-bore CuBr laser and its arrangement**

The laser tube was energized by an HV power supply (Figure 2.1.2) using a hydrogen thyratron TGI1 1000/25 switch. The laser tube was IC-excited (Vuchkov *et al.*, 1991) because the conventional circuitry with a peaking capacitor produced lower output laser powers. The input power was evaluated from the voltage and current of the high-voltage rectifier. The capacitors ($C_1$, $C_2$, $C_3$, and $C_4$) were of the low-inductance KVI-3 type; their capacitance and the bypass inductance, L were experimentally optimized. The metal plates P1 and P2 were made of duralumin sheet and were placed as close as possible (2 - 3 cm) to the laser tube but not so close as to cause electrical puncture of the silica tube.

The average laser power was measured with a Scientech 373 power meter. The current and voltage applied to the tube were displayed on a Tektronix model 2455A oscilloscope.

*Results and discussion*

The optimization of the small-bore CuBr laser performance required a multiparametric investigation. Below, we present the main experimental findings concerning that investigation. As Vuchkov *et al.* (1991) revealed IC-excitation permits the tailoring of the shapes of the tube voltage and current by using different capacitor sets, and thus the optimization of the laser performance. The capacitors varied as follows: $C_1$ from 330 to 600 pF, $C_2 = C_3$ from 100 to 1000 pF and $C_4$ from 1000 to 3300 pF (all values are nominal). The optimum capacitor set with regard to high and stable laser power operation was the set $C_1 = 470$ pF, $C_2 = C_3 = 100$ pF, and $C_4 = 1000$ pF. The majority of experimentation was carried out with that set of capacitors.

The vapor pressure of CuBr was varied from 0.1 to 0.4 Torr by alteration of the temperature of the CuBr reservoir heaters from 450 to 500° C. This means that the density of CuBr molecules over the reservoir surfaces was no higher than $5.10^{15}$ cm$^3$. The temperature of the laser tube was about 50 degrees higher than that of the CuBr reservoirs.

As a gas buffer, we used neon at pressures between 10 Torr and 760 Torr. Hydrogen of 0.3-0.4 Torr was also added each time to the neon gas fill. The dependence of the average laser power on neon pressure for the optimum capacitor set is plotted in Figure 2.1.3. For each experimental point, the pulse recurrence frequency, CuBr vapor pressure, and the input power were optimized. As can be seen, the optimum pressure was 20-30 Torr, although stable laser oscillation was obtained up to 760 Torr. The interpretation of the data needs more experimentation and simulation, which should be done in future.

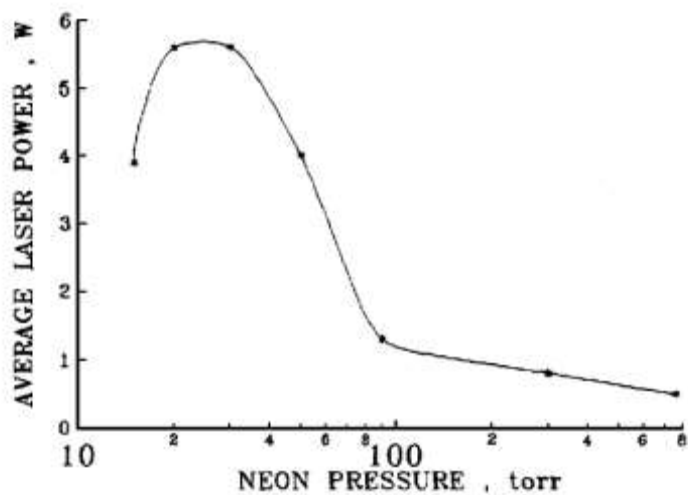

**Figure 2.1.3 Dependence of average laser power on neon buffer pressure**

The recurrence frequency of electric pulses had a strong effect on the average laser power. There are at least five locally optimal frequencies (in the range 22 to 52 kHz that was examined), which tended to shift upwards as neon pressure was reduced. The dependence of average laser power on recurrence frequency for 50-ton neon and the optimum capacitor set is plotted in Figure 2.1.4. As can be seen, the laser power had a tendency to go up with an increase in recurrence frequency. The extrema observed may be attributed to the tube design, especially to the discharge confining diaphragms in the active zone presumably favoring the appearance of standing acoustic waves. The maximum average laser power was attained at recurrence frequencies high than 50 kHz. We failed, however, to run the laser at recurrence frequencies above 52 kHz (where higher laser powers are expected) due to limitations in the thyratron trigger circuitry.

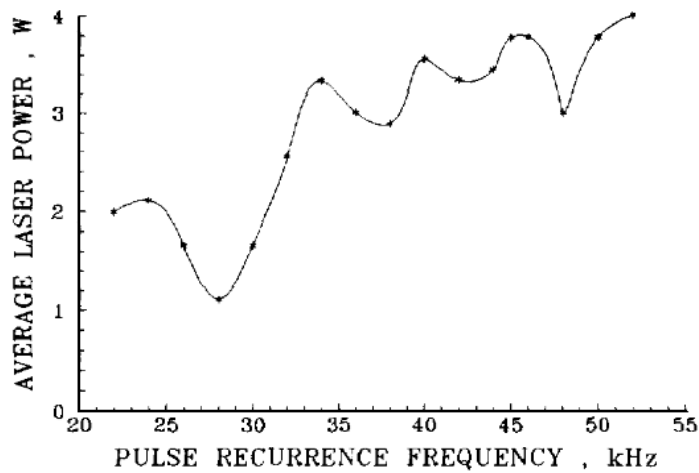

**Figure 2.1.4 Dependence of average laser power on pulse recurrence frequency**

The input electric power, measured at the high-voltage rectifier as the product of the average current and voltage, varied from 0.8 to 1.1 kW. Assuming that 30-40 percent of it is lost elsewhere in the circuit, the electric power deposited in the active zone would have amounted to around 0.8 kW or not more than 170 W per cubic centimeter.

Typical waveforms of tube current, electrode voltages (shown inverted) and laser emission are presented in Figure 2.1.5. The duration of a laser oscillation at 510.6 and 578.2 nm was 20-30 ns. The tube voltage is the difference between the electrode voltages.

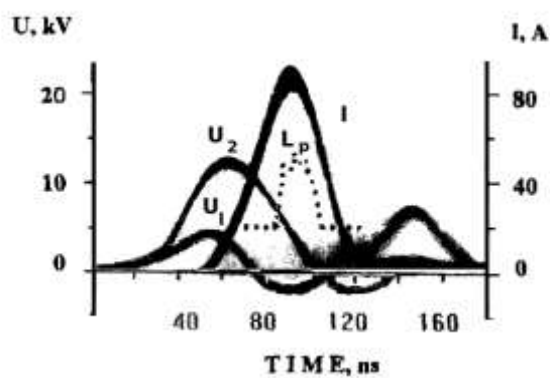

**Figure 2.1.5 Typical small-bore laser electric discharge waveforms: I - tube current; $U_1$ and $U_2$ (latter shown inverted) - electrode voltage (from ground) from the side of the capacitor couples (C2 & C4) and (C1 & C3), respectively. $L_P$ indicates laser pulse (dotted line).**

The highest laser power of 6.7 W was obtained at 20 Torr neon (as main gas, and with the 0.3 Torr $H_2$ as a mandatory additive to it), with a capacitor set of $C_1$ = 330 pF, $C_2$ = $C_3$ = 100 pF and $C_4$ = 1000 pF (all values are rated) and a bypass inductor L = 0.3 mH. The pulse recurrence frequency was 52 kHz, and the input electric power was 1 kW. Thus, the active volume as defined by tube dimensions was 4.77 cm$^3$, and it yielded 1.4 W/cm$^3$, which is the highest specific average laser power yet reported for a multi-kilohertz copper vapor laser. The laser efficiency *i.e.* laser power/input power ratio was more than 0.6 percent. Just to compare our results with those by Vorobev *et al.* (1991) we illuminated the laser tube through its windows using a low-divergence laser beam, and thus, we measured the real aperture of the active zone. We found it to be around 0.75 of the constructive aperture. Attenuation of the laser beam caused by scatterings at the laser windows was about 10 percent. If the "real" (as calculated by Vorobev *et al.* 1991) specific average power is needed, taken together that leads to a correction factor of 1.5. Thus, according to these researchers, the specific average power of our small-bore CuBr laser reached 2 W/cm$^3$.

## *Conclusion*

The small-bore CuBr laser investigation has demonstrated that longitudinally excited CuBr lasers can successfully produce output powers of the order of 5-10 W with very high densities of power extracted from the active volume. Further improvements are expected with an increase in pulse recurrence frequency.

## 2.2. *Influence on operating characteristics of scaling CuBr lasers in active length*

The copper halide laser is a variety of copper vapor laser which is an efficient source of high-power pulsed green/yellow radiation. Typical output parameters for a copper halide laser are 20 W average output power and an efficiency of 1.5% at pulse recurrence frequency (PRF) of 16 kHz (Sabotinov, 1996). Copper halide laser has a number of advantages over conventional copper vapor laser as a very fast start-up (due to low thermal mass and lower temperature of operation), higher wall-plug efficiency and not the least of which is completely sealed-off laser operation for a long time. The pseudo-Gaussian beam intensity profile could be better suited for many applications than the top-hat profile of the elemental laser.

The halide laser differs from elemental devices in that the copper lasant atoms are introduced to the discharge zone in the former as copper halide molecules rather than as an atomic vapor. Molecular dissociation in the discharge liberates free copper atoms which are excited in the same manner as in the elemental type of laser. The volatile nature of the copper halide molecules allows tube temperatures of around 500°C to be used rather than the 1600°C necessary to generate sufficient copper vapor. A lower operating temperature means that the tube structure of the halide laser is simpler than that of an elemental laser and amenable to sealed-off operation with the reduction in internal tube thermal insulation.

Elemental copper vapor laser oscillators have been demonstrated with average output powers of hundreds of watts (Chang *et al.*, 1994). On the other hand, to date, the highest average output power to have been reported for a sealed-off copper bromide laser is 112 W (Elaev *et al.*, 1989), but this is an isolated result. (Note that copper hybrid lasers which exhibit characteristics of both halide and elemental systems have been operated up to output powers of 201 W by Jones *et al.* (1994). However, they are flowing gas systems and utilize a different metal atom seeding technique to that used in copper halide lasers.) The maximum power reported for conventional sealed-off copper bromide lasers elsewhere has until recently been around 20 W, where laser performance has been monitored over hundreds of hours routine operation with the same gas fill (Sabotinov, 1996). Our intention is to take this well-proven technology at the 20 W level and examine means of scaling to higher average output powers. Here we discuss issues relating to the scaling of the average output power of sealed-off copper bromide lasers (Astadjov *et al.,* 1997a). By increasing the active length from 50 cm to 120 cm, we have increased the maximum average output power of 40 mm bore devices from 24 W to 58 W.

*Experimental details*

The laser tubes were of all fused silica construction and are shown schematically in Figure 2.2.1. The tubes were similar in design to that Astadjov *et al.* (1988) reported. In the present tubes, a major difference was the absence of annular silica diaphragms to keep the discharge away from the tube wall. The discharge in the region between the electrodes was bounded solely by the fused silica tube wall and was not confined closer to the tube axis by diaphragms as in other our studies. The electrodes were located in side-arms, which were filled with pressed copper granules that connected to the external electrical circuit via tungsten-quartz feed-through at the base of the side-arm. Surrounding each electrode was a fused silica annulus (see Figure 2.2.1) which acted as a trap for copper bromide generated by the reaction of excess gaseous bromine with the copper granules in the electrode side-arm. In this manner, the free bromine vapor pressure was controlled, making the long-life sealed-off operation of CuBr laser tube possible. Copper bromide powder of 99.9% purity was placed in each of the side-arm reservoirs which were located between the electrodes.

We investigated two 40mm-bore laser tubes of active lengths of 50 cm (Tube 1) and 120 cm (Tube 2). The separation between CuBr reservoirs was the same in each laser, being 15 cm. Tube 1 had three such reservoirs while Tube 2 had seven.

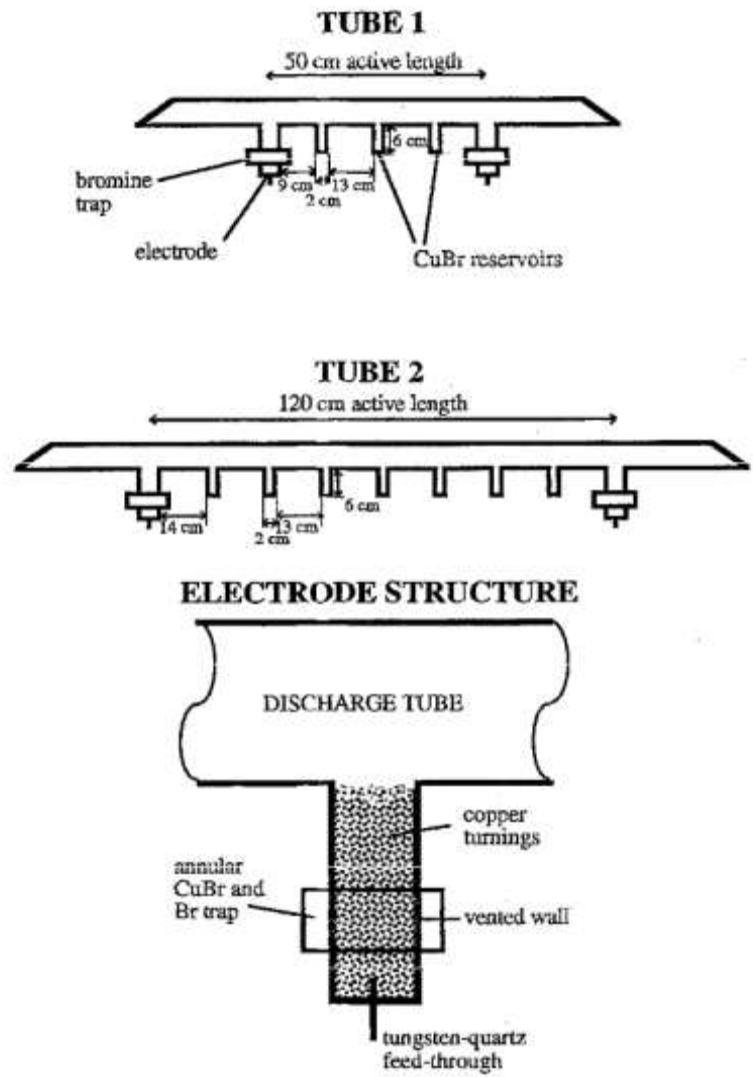

**Figure 2.2.1 Schematic diagrams of the two 40mm-bore laser tubes, together with a close up of the electrode construction**

We characterized each laser in turn. Before a tube was operated, it was filled with a neon-hydrogen gas mixture. The addition of hydrogen acts to double the output power and efficiency of copper bromide lasers as Astadjov *et al.* (1985) proved quite long ago. A pulsed discharge was excited in the tube using an interacting circuit (IC) (Figure 2.2.2), which is a variant of the more usual storage-peaking capacitor arrangement (Vuchkov *et al.,* 1991).

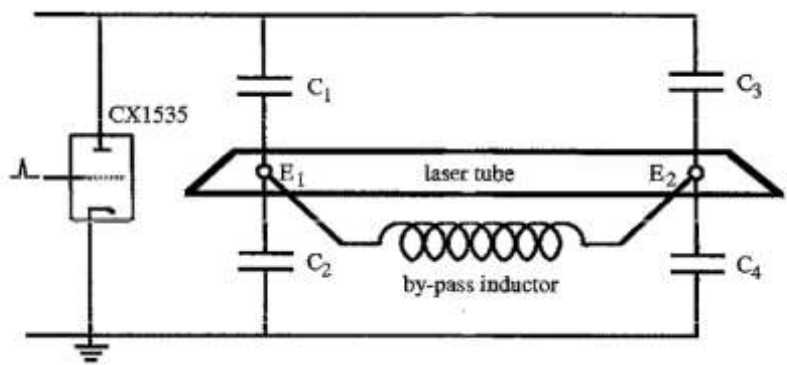

**Figure 2.2.2 Schematic diagram of the capacitor placements in the interacting circuit used to excite the laser tubes**

This circuit, in which the capacitor arrangement simulates the actions of a storage capacitor during charge up, and a peaking capacitor after the laser tube breaks down with the redistribution of charge on the capacitors, allows for more efficient excitation of copper bromide lasers. The optimum tube wall temperature was ~ 500 °C and CuBr reservoirs were heated to ~ 450°C to generate a vapor pressure of copper donor molecules in the main discharge tube. By switching off the reservoir heater elements some minutes before the discharge is terminated, we made CuBr disappear from main discharge so the laser tube's operation and characteristics are quite repeatable.

The tube was fitted with Brewster-angled windows, which were sealed onto the tube with vacuum wax. The optical cavity consisted of a flat multilayer dielectric-coated high reflector and a flat, parallel-surfaced quartz disc.

*Laser characteristics*

Both laser tubes were characterized with respect to the parameters of gas pressure, PRF, charging voltage, and circuit capacitances (matching and discharge current). The optimum values of capacitance for both tubes were $C_1$=1320pF, $C_2$=235pF, $C_3$=235pF and $C_4$=3360pF, corresponding to an equivalent storage capacitance of 1.0nF.

The gas mixture was 20 Torr of neon and 0.6 Torr of hydrogen (as usual, added to the cold tubes prior to operation) and it corresponded to maximum output power. The dependence of average output power on hydrogen additive partial pressure was investigated for Tube 2. There is a trend for the optimum partial pressure of added hydrogen to increase as the discharge cross section becomes larger. This observation has been confirmed in our most recent experiments.

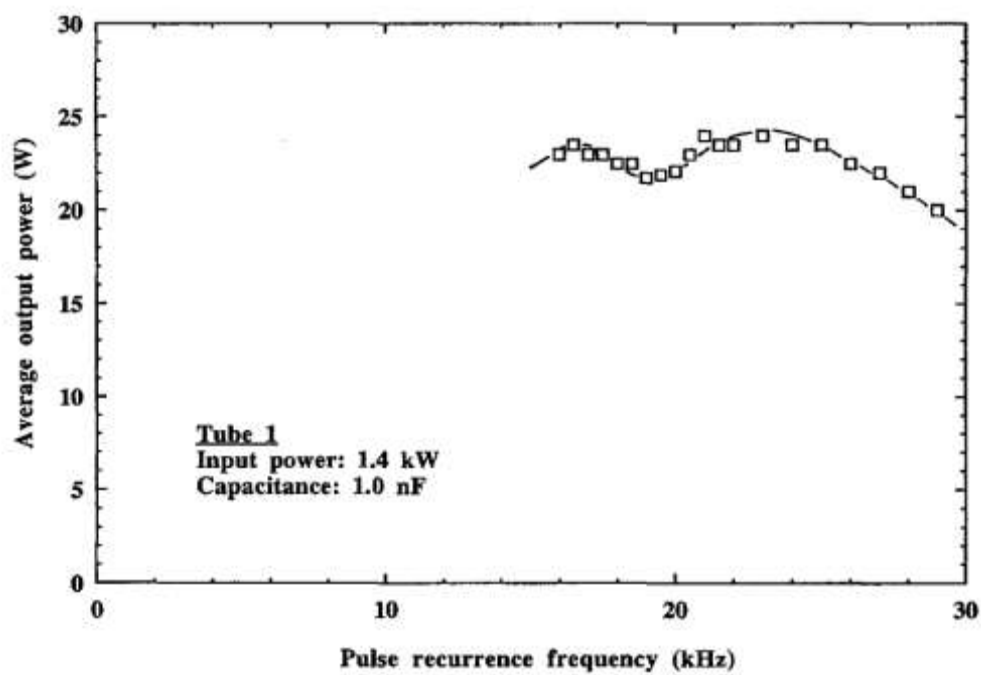

**Figure 2.2.3 Dependence of average laser output power on PRF for Tube 1; input power of 1.4 kW**

The dependence of average output power on PRF for each laser tube was determined with a constant input power of 1.4 kW. As the PRF was increased, the charging voltage (voltage on the thyratron anode, $V_a$) was reduced: for Tube 1 - $V_a$ was 12.2 kV at 16 kHz and 9.1 kV at 29 kHz; for Tube 2 - $V_a$ was 13.6 kV at 15 kHz and 10.6 kV at 25 kHz. For each tube, the PRF dependence of output power displayed two maxima. In the case of Tube 1 (Figure 2.2.3), laser power was maximum around 16.5 kHz PRF ($V_a = 13$ kV) and 23 kHz ($V_a = 11$ kV). Laser Tube 2 displayed maxima near 17.5 kHz ($V_a = 12.8$ kV) and 21.5 kHz ($V_a = 11.3$ kV), as shown in Figure 2.2.4. The positions of the maxima did not depend on input power or capacitance. (We define input power as IC-stored energy multiplied by PRF. We did not measure tube voltage during the experiments.)

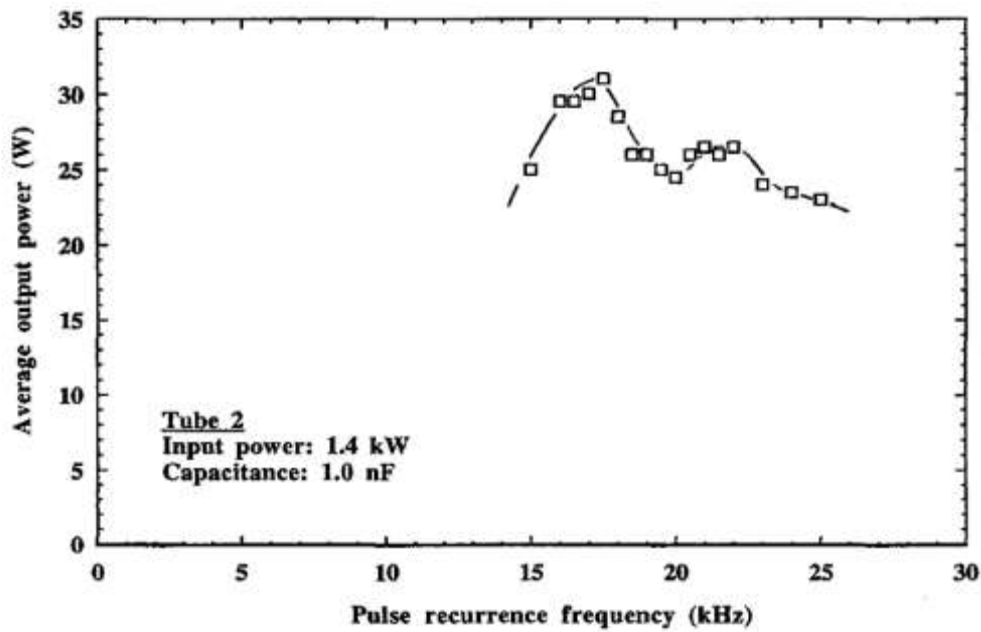

**Figure 2.2.4 Dependence of average laser output power on PRF for Tube 2; input power of 1.4 kW**

For an input power of 1.2 kW to Tube 1 (2.4 kWm$^{-1}$), the maximum output power measured was 24W at 23 kHz, with an efficiency of 2%. For Tube 2, under the almost similar power per unit length loading of 2.25kWm$^{-1}$ (the input power of 2.7 kW was the power supply limit) the greatest output power recorded was 58 W at 17.5 kHz, with an efficiency of 2.15% (Figure 2.2.5). We can infer that CuBr laser efficiency tends to increase with active length, given the same input power per unit length. Or if the input power per unit volume decreases (as in this case – from 1.9Wcm$^{-3}$ to 1.8Wcm$^{-3}$), we could certainly make a similar inference.

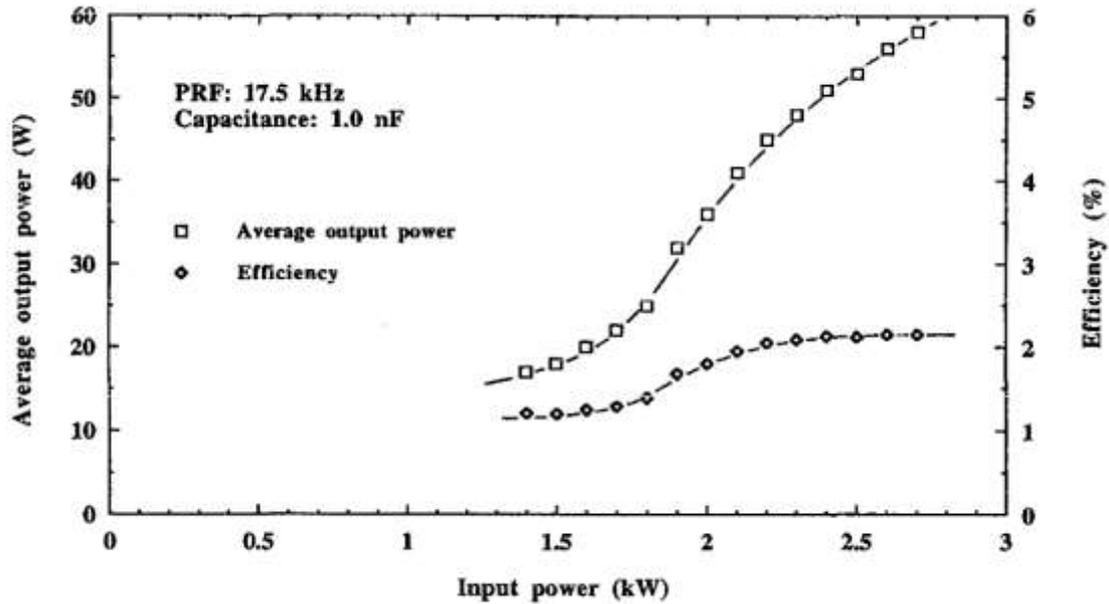

**Figure 2.2.5 Dependence of average laser output power of Tube 2 on input power; PRF of 17.5 kHz**

Figure 2.2.5 displays the dependence of output power of Tube 2 on input power for a fixed PRF of 17.5 kHz. It can be seen that the maximum power of 58 W reached in these investigations could have been exceeded if a higher input power had been available, although the laser efficiency had saturated at 2.15%.

*Discussion*

The complicated dependences of output power on PRF are not observed in laser tubes which do not have side-arm reservoirs, even though diaphragms may be present. The laser tube reported by Sabotinov *et al.* (1995) was of similar construction to that in by Astadjov et al. (1994) but the latter had side-arm reservoirs whereas the former did not. The laser with side-arm reservoirs displayed a rather complex dependence of output power on PRF while the laser without side-arms exhibited a smooth dependence of output power on PRF from 20 to 70 kHz as observed by Sabotinov & Little (1993). We consider that the maxima in output power exhibited at various frequencies are related to the establishment of acoustic resonances in the laser tube. We have examined the various resonances that can be excited in the Tubes 1 and 2, with differences in frequency between adjacent harmonics of 6.5 to 4.5 kHz (the frequencies of modulation of laser power in Tubes 1 and 2 are fairly similar: 6.5 ± 2 kHz for Tube 1 (Figure 2.2.3), and 4.5 ± 2 kHz in the case of Tube 2 (Figure 2.2.4)). Taking a gas temperature of 1300°C as a radial average in the active zone , a temperature within the reservoirs of 490°C (both from Astadjov *et al.,* 1988a), and assuming that the speed of sound in the laser tube is governed by the neon gas, we have found that the only acoustic resonances which can sensibly account for the observed PRF dependence correspond to harmonics of longitudinal acoustic waves set up within the CuBr reservoirs, taking each to be a pipe with one open and one closed end. Only odd harmonics are present, and for an effective reservoir depth of 6 cm, we expect a fundamental at 3 kHz, and the fifth, seventh and ninth harmonics to occur at 15 kHz, 21 kHz, and 27 kHz, respectively. These harmonics correspond, within the margin of uncertainty (initially, the reservoirs are filled with CuBr to the extent that differs little from each other, but these differences may enhance, especially, after hours of laser operation), to the minima in laser power in the PRF curves of Figures 2.2.3 and 2.2.4. The attenuation of laser oscillation when the PRF is set to a resonant frequency will modulate what would, in the absence of acoustic resonances, have been a smooth dependence of output power on PRF with a broad maximum around 20 kHz. The influence of various acoustic oscillations on laser generation has been observed and discussed before by Astadjov *et al.* (1994), Ilyushko *et al.* (1985) and Kravchenko *et al.* (1982). In the present work, we expect that the atomic and molecular number densities above each reservoir in Tubes 1 and 2 will be altered away from the ideal densities when resonance occurs, and thereby the effectiveness of laser generation will be reduced. It may also be that under acoustic resonance conditions, where there is an anti-node above each reservoir, the entry of CuBr vapor into the laser tube is impeded, reducing laser output power. Note that since the depth of each CuBr reservoir was approximately the same in Tubes 1 and 2, the minima (or maxima) in laser power occurred at similar PRFs, despite the different active lengths. Astadjov *et al.* (1994) also attributed the power modulation to acoustic standing waves set up as a result of tube geometry. The complex behavior of laser power with varying PRF points out the importance of designing laser tubes with geometries chosen so as to avoid acoustic resonances coinciding with the optimum PRF for laser generation.

Some of the increase in efficiency with scaling of active length can be attributed to a reduction in power dissipated in the electrodes. We measured the temperature high-voltage electrode (cathode $E_1$) and the low-voltage electrode (anode $E_2$) using thermocouples placed in contact with the electrode side-arms, for Tubes 2.2.1 and 2.2.2 with three different input powers. The results are summarized in Table 2.2.1.

**Table 2.2.1 Dependence of electrode temperature on input power**

| Electrode | Electrode Temperature (°C) | | | | | |
|---|---|---|---|---|---|---|
| | $E_1$ | $E_2$ | $E_1$ | $E_2$ | $E_1$ | $E_2$ |
| Tube 1 | 230 | 210 | 310 | 240 | | |
| Tube 2 | 200 | 190 | 220 | 200 | 300 | 270 |
| Input power (kW) | 1.2 | | 1.4 | | 2 | |

Clearly, losses in the electrodes decrease with increasing active length. This is consistent with the general trend we observe for the efficiency of laser generation to increase with active length. The measurements of electrode temperature are also consistent with observations we have made in previous experiments with laser tubes of different length. In particular, for the small-bore copper hybrid lasers described by Sabotinov *et al.* (1995) when reducing the active length from 30 cm to 15 cm to 2.5 cm, we noticed that electrode heating greatly increased - in the case of the 15 cm active length laser tube, the copper electrode glowed red hot under ordinary operating conditions, while the copper electrode of the 2.5 cm tube melted after a short period of operation (Sabotinov & Little, 1993). A systematic study of the dependence of electrode heating on active length in copper bromide lasers is currently underway in order to establish the important mechanisms. Understanding these observations is of practical use in developing copper bromide lasers with long electrode lifetime.

Another practical observation we have made is that in order for maximum laser power to be attained whilst varying PRF during characterization, it is necessary to readjust the CuBr reservoir temperature at each PRF. In general, as PRF is raised under constant input power conditions, so the reservoir temperature must be increased. This is illustrated in Table 2; the temperatures given in the table seem to be higher than the actual temperatures of the CuBr reservoir, and should be treated as relative.

**Table 2.2.2 Copper bromide reservoir temperature needed to maximize laser output power at three PRFs for laser generation in Tube 2**

| PRF (kHz) | Reservoir temperature (°C) | Maximum laser output power (W) |
|---|---|---|
| 16.5 | 480 | 23.5 |
| 21.5 | 530 | 24 |
| 26 | 590 | 23 |

The behavior may be explained as follows. As the PRF is increased, for constant input power per unit length, the time available for interpulse thermal relaxation of the gas in the laser active volume reduces. Consequently, the gas number density, including that of the copper atoms, reduces on axis and increases near the tube wall. In order to maintain the number of copper atoms on the axis of the laser tube as PRF is raised, a higher vapor pressure still of CuBr must be set near the tube wall, or equivalently, within each reservoir. For that reason, the temperature of the reservoirs needs to be increased in order to maintain conditions on the axis of the laser tube, where excitation is most intense, optimal for laser oscillation. This depletion mechanism was also invoked by Liu *et al.* (1977) to explain why the reservoir temperature of continuously pulsed lasers operate at higher temperatures than in double-pulse excited lasers, where radial temperature gradients do not affect laser kinetics.

*Conclusions*

We have studied laser oscillation in two copper bromide lasers of tube bore diameter of 40 mm. In contrast to the traditional design, these tubes have no diaphragms to confine discharge axially (and by this way, laser radiation as a consequence too). The only difference in their constructions was that Tube 1 was of an active length of 50 cm, whilst Tube 2 was 2.4 times longer active zone - 120 cm. Here we outline major findings.

Under nearly same specific input power loading, Tube 1 (2.4kW/m, equivalent to 1.9Wcm$^{-3}$) gave an output power of 24 W at an efficiency of 2.0% and Tube 2 (2.25kW/m, equivalent to 1.8Wcm$^{-3}$) produced 58 W at 2.15% efficiency. So for similar power loading, the output power increases linearly with the active zone length. Moreover, with great accuracy (<1%), efficiency increases with the decrease of specific input power loading *i.e.* in a reciprocal manner (see values given above) into the scaled laser tubes. The increase in laser efficiency with average power scaling by length can be attributed in part to reduced losses at the electrodes with the lengthening of the active zone. This is evident from the easing of operation of electrodes (their temperature gets low) when the active length of this type of laser is increased.

The dependence of output power was found to have a complex dependence on PRF, very likely as a result of the influence of acoustic resonances in the CuBr reservoirs which alter the number density of CuBr above the reservoir and afterward, in the active region.

We found (if it is not an unidentified artifact of our measurement apparatus) that the CuBr reservoir temperature has a trend to go up extremely fast as PRF is raised. If this really takes place, that should surge the consumption of CuBr and exhaust it for a short time.

Finally, we found that the optimum pressure of hydrogen additive (as part of Ne-$H_2$ mixture) in the laser tube is increased when the effective diameter ('aperture') of the pulsed glow discharge is raised from 20 mm to 40 mm.

These practical findings are all important for the future development of this type of laser when higher average output powers are sought. Our studies will continue to elucidate the mechanisms that are responsible for these observations.

## 2.3. Lifetime of 30W/40W CuBr lasers

In 1999-2000, we tested completely sealed-off CuBr lasers of 30W/40W average powers for long-term operation. 50mm-bore laser tubes were of 30mm-diameter ('aperture') active zone determined by axial silica diaphragms. The 30W CuBr laser was of 100cm electrode separation and the 40W CuBr laser had the electrode separated by 140cm.

### Design and fabrication of the 30W/40W CuBr laser tube

The construct of the 30W/40W CuBr laser tube is given in Figure 2.3.1. The CuBr laser tube is made of fused silica (Heraeus Company). The full length of laser tube is 156cm and the inside diameter is 50mm (thickness of 2.6±0.5mm). Porous copper electrodes are placed symmetrically and their separation is 100cm/140 cm for the 30W/40W laser, respectively. The copper, they are made of, is purified in vacuum by electron beam evaporation. The electrodes are surrounded by a trenchlike arrangement which prevents from forming copper dendrites on top of electrodes. Six baffling rings (aka diaphragms) of 30mm inside diameter are fused equidistantly between electrodes. Five side arm reservoirs (containing totally 80 g of CuBr) are placed between the rings along the length of discharge zone. Out of the discharge zone, one more side arm is also attached from the side of the grounded electrode. This is the container of an internal hydrogen generator (IHG).

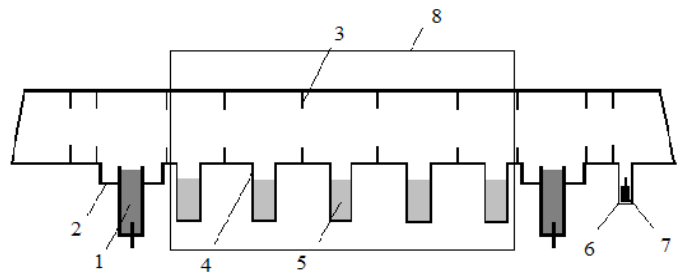

**Figure 2.3.1 Construct of 30W/40W- CuBr laser tubes: electrode, 1 surrounded with trenchlike arrangement, 2; silica ring, 3; reservoir, 4 with CuBr, 5; side arm, 6 with an internal hydrogen generator, 7 and a stainless steel enclosure, 8**

The discharge zone and CuBr reservoirs are in a stainless steel enclosure. It is used as a hot discharge zone curb as well as a return current path. The steel enclosure is connected to the grounded electrode and to the 0.5nF- capacitor of a Blumlein circuit formed by a 0.5nF- capacitor and a 1.1nF- capacitor. On the bottom of the enclosure, there are long heaters just under the CuBr reservoirs. A temperature controller maintained the temperature inside the enclosure ≥500°C. The hydrogen generator is also operated by a temperature controller. To diminish the IHG contamination, the part of the tube just above its side arm is heated too. IDG ensures conditions for long-term CuBr laser performance: hydrogen pressure ≥ 0.2 Torr (Astadjov *et al.,* 1985) and bromine pressure < 1 Torr (Vuchkov, 1984). The test began with a hydrogen-enriched (hydrogen pressure = 0.5 Torr) Ne-$H_2$ mixture of 19Torr. We knew that depletion processes would lower hydrogen pressure with the progress of the test. Via the IDG controller, we varied temperature between 400-500°C to maintain optimum conditions for maximum laser power throughout the test.

The power supply (with a TGI1000/25 hydrogen thyratron, Russia) provided maximum 13 kV to the thyratron anode for long-term operation and the maximum possible repetition frequency was 18.3±0.1kHz which gave 2.4 kW of average input power. The laser tube was steadily run at 12.6kV at 18.3±0.1kHz maintaining 2.25kW. Every 5-6 hours, for a short time of 10-15 minutes the laser was checked at 13kV and average laser power was recorded.

The evolution of average laser power of 30W/40W CuBr lasers with test time is plotted in Figure 2.3.2. The 30W CuBr laser started at 33W; after 300h, it degraded to 30W but then kept up to the 1000h. Similar was with the 40W CuBr laser: gradually decreasing from 44W to 40W for 500h and then producing 40W till the end of test time.

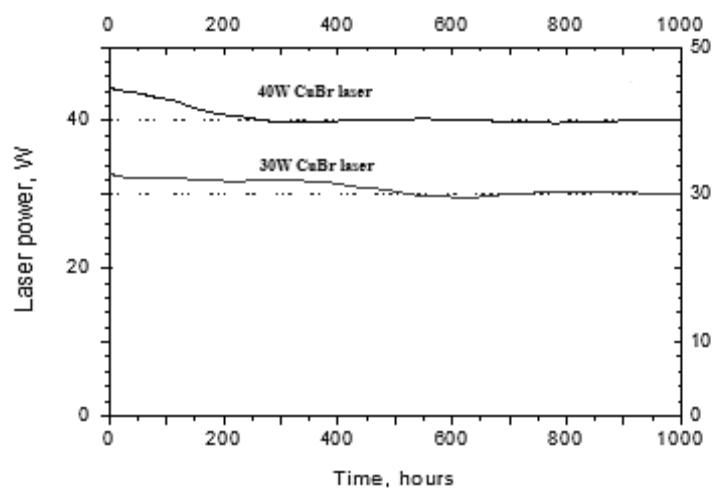

**Figure 2.3.2 Average laser power of 30W/40W CuBr lasers during long-term operation test; Average input power is 2.4 kW**

*Conclusions*

The highest laser power obtained from the 30W CuBr laser tube was 38W (efficiency of 1.6%) at 13kV (2.4 kW input power). When, for a short time, the laser was run at 2.5 kW energized by a power supply with an EEV CX 1535 thyratron (the best UK-made) we got 53W (efficiency of 2.1%) recorded at 14kV and 20 kHz. These record-high laser characteristics uncover the importance of quality of power supply used to excite CuBr lasers.

The long-term operation test outlines typical behavior of laser output of completely sealed-off 30W/40W CuBr lasers. Laser power slowly degrades (~10%) in the beginning (first 250-500 hours) and then it stays actually constant (within some fluctuations) till the end of the 1000h test. We conclude that completely sealed-off 30W/40W CuBr laser performance is steady for 1000 hours and higher power, efficiency, stability, *etc.* can be expected with the improvement of the electric power supply employed.

## 2.4. *Large-bore 120W CuBr laser of 3% efficiency*

Copper bromide lasers are the most powerful and efficient of the copper halide lasers, able to match the average output powers of conventional copper vapor lasers (CVLs) up to powers of ~100 W, but with efficiencies twice those of CVLs. They typically produce output powers of tens of watts at efficiencies of ~2.5% on the 578.2nm and 510.6nm transitions of atomic copper. Besides the report by Elaev *et al.* (1989) of an average output power of 112 W from a copper bromide laser, the development of copper bromide lasers since the mid-eighties has concentrated on the engineering of long-life, reliable, sealed-off devices. Nowadays, sealed-off Ne-$H_2$-CuBr lasers of small (20mm aperture x 50cm active length) and medium (30mm aperture x 100cm/140 cm active lengths) size exist with lifetimes of 1000 h at 10W-, and at 30W/40W output level, respectively.

The development of copper bromide lasers has turned to the scaling of discharge tubes in bore and length to increase the available output power. Most recently, Astadjov *et al.* (1997) have demonstrated an increase in the average output powers from 24 to 58 W, with scaling in active length from 50 to 120 cm of 40mm-bore CuBr lasers. The increase in active length was also accompanied by an increase in efficiency from 2% to 2.15%, due to a reduction in losses at the electrodes.

Our research has now led to the development of a copper bromide laser which has produced a record average output power of 120 W at an efficiency of 2.4%. A record efficiency of 3% was achieved at the 100-W output level (with the actual power of 90W). Previously, this combination of high output power and high efficiency had only been demonstrated from another variant of the copper laser, the copper hybrid laser by Jones *et al.* (1994). Below we describe the operating regime of the high-power copper bromide laser in terms of the dependence of output power and efficiency on discharge parameters and present some characteristics the electrical excitation of this large-bore CuBr laser (Sabotinov *et al.,* 1996 and Astadjov *et al.,* 1997b).

### *Experimental details*
The laser discharge tube is shown schematically in Figure 2.4.1. The discharge was contained within a quartz tube of a 60-mm bore, between electrodes which were separated by 200 cm—these dimensions defined the active volume of the laser. No diaphragms were necessary in order to stabilize the discharge, as has been found for smaller devices in the past (Sabotinov, 1996). The total length of the laser tube, including the extended window regions, was 323 cm. Nine heated side arm reservoirs of CuBr, distributed equally along the length of the tube, were used to seed the discharge zone with CuBr vapor. The temperature of the reservoirs was typically 490 °C while the discharge channel was held at the slightly elevated temperature of 550 °C so that the reservoir temperature-controlled the vapor pressure of CuBr in the main tube. Each electrode was a quartz side arm with a tungsten-quartz feed-through in its base, and which was filled with copper filings, to which the discharge attached. Quartz discs at the base of the electrode acted as cooling fins. A CuBr/bromine trap, similar to that described by Astadjov *et al.* (1997), was also attached to each electrode. The laser cavity was formed by a flat dielectric-coated high reflector, and uncoated quartz flat acted as an output coupler.

The excitation circuit employed was as that described by Vuchkov et al. (1994) and dubbed interacting circuit. The equivalent storage capacitance for optimal operation varied between 1.0 and 1.3nF. Charging voltages of up to 21 kV were employed and the thyratron (an EEV Ltd type CX1535) switched average powers of up to 5 kW in the circuit at pulse recurrence frequencies (PRF's) of up to 21.5 kHz.

The laser power was measured through a beam chopper with a Scientech model 364 power meter. Current and voltage waveforms were observed on a Tektronix model TDS420 oscilloscope (150-MHz bandwidth).

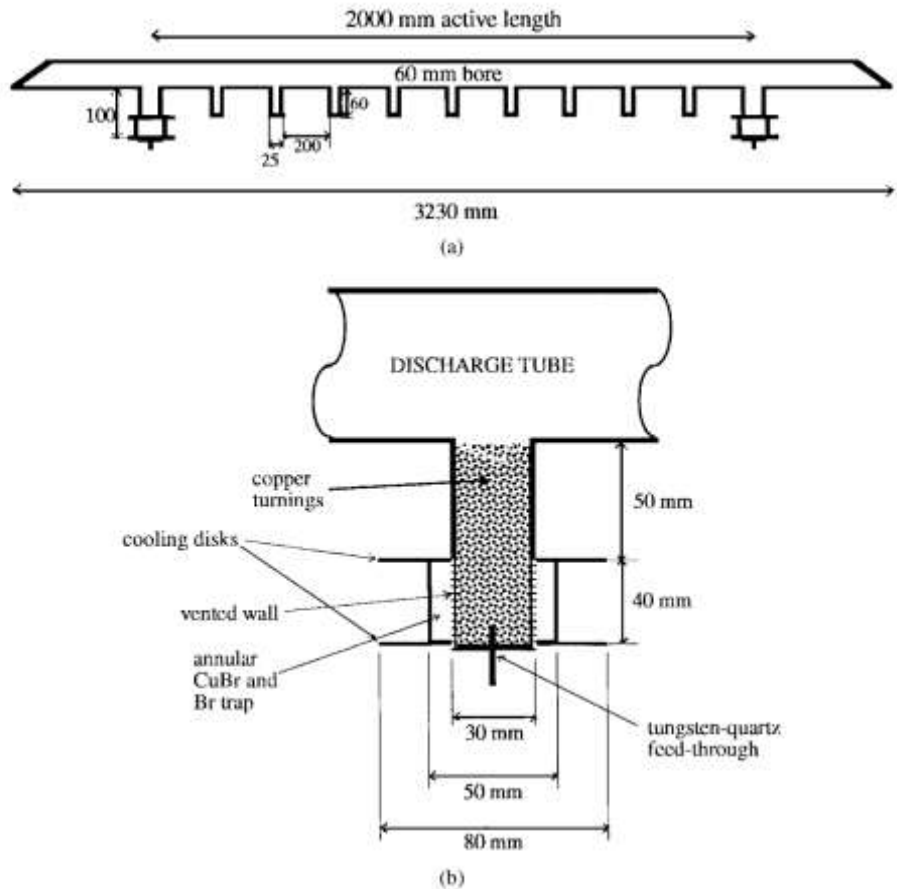

**Figure 2.4.1 Schematic plan of the copper bromide laser tube and electrodes**

## *Results and discussion*

The input power was varied from 2 to 5 kW, the PRF from 14 to 21.5 kHz, and the Ne buffer gas pressure from 15 to 70 Torr. Maximum output power and efficiency were obtained when the hydrogen additive pressure was 0.6 Torr. The dependence of average output power and efficiency on input power (charging voltage) is shown in Figure 2.4.2 for capacitors of 1.0 and 1.3 nF, respectively, for a PRF of 17.5 kHz, and a Ne pressure of 20 Torr. It can be seen that in each case the output power is a monotonically increasing function of input power up to the present limit (determined by the thyratron) of the power supply. In the case of 1.0-nF capacitance, the efficiency saturates at 3% for output powers of up to 100 W. For 1.3 nF capacitance, the efficiency is a decreasing function for input powers of 3 kW and above. However, the larger capacitance allows a higher output power of 120 W to be achieved, when the efficiency was 2.4%.

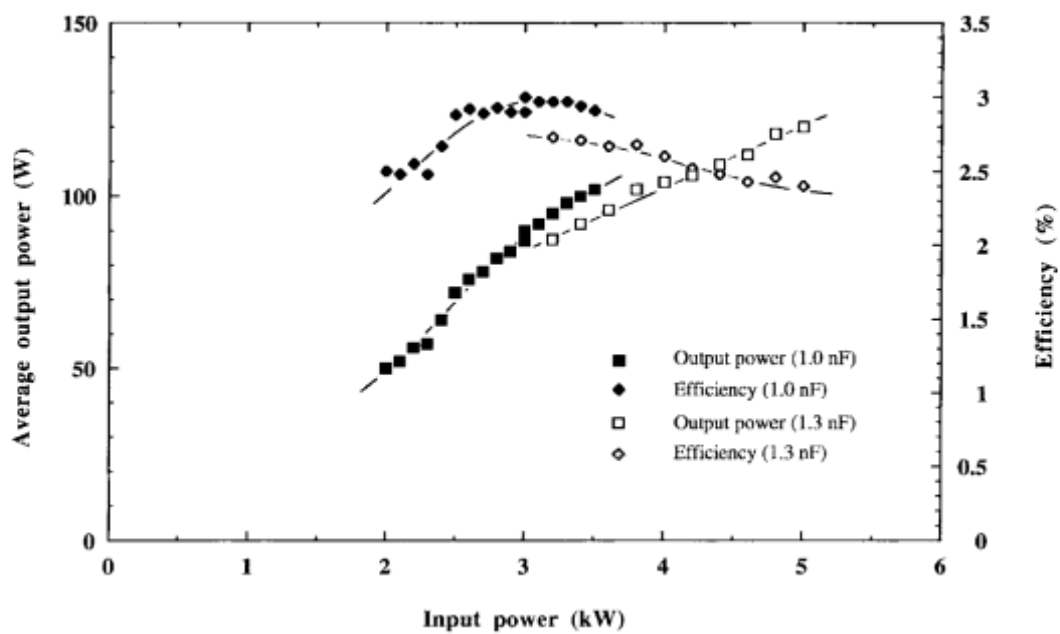

**Figure 2.4.2 Average output power and efficiency as functions of input power, with a Ne pressure of 20 Torr, an $H_2$ pressure of 0.6 Torr, a PRF of 17.5 kHz, and with equivalent storage capacitance of 1.0 nF (filled symbols) and 1.3 nF (open symbols)**

The output power of 120 W is the highest reported to date for conventional copper bromide lasers. The efficiency of 3% is also a record value for copper bromide lasers. Such high efficiencies at the 100-W power level make routine operation with one thyratron feasible. Copper vapor lasers that produce 100-W output powers are typical ~1% efficient and require three thyratrons operating in parallel to excite the tube - such a configuration is not favorable for industrial use.

The dependence of average output power and efficiency on PRF is shown in Figure 2.4.3 for a Ne pressure of 20 Torr, an $H_2$ pressure of 0.6 Torr, a PRF of 17.5 kHz, and a capacitance of 1.0 nF. The input power was held constant by adjusting the charging voltage, which is also shown in the figure. Output power and efficiency peak at PRF's between 15 and 18 kHz. In our earlier experiments with a 40-mm bore copper bromide laser (Astadjov *et al.,* 1997), acoustic resonances, corresponding to standing waves in the side-reservoirs, caused the output power at resonance to be reduced, which led to a minor but characteristic modulation of laser output power with PRF. The harmonics, which depend on the lengths of the reservoirs, will occur at approximately the same PRF's in the present laser as in (Astadjov *et al.*, 1997) because the lengths of the side reservoirs are the same. We, therefore, expect some modulation (reduction in laser output power) around 15 and 21 kHz for the same reasons, namely that the resonances alter the CuBr vapor pressure above each reservoir away from optimum, and therefore in the discharge channel. It is certainly true that for PRF's below 15 kHz, the output power falls quickly, and at 21 kHz the output power has fallen significantly. We, therefore, cannot rule out the influence of acoustic resonances on the output power-PRF behavior presented in Figure 2.4.3.

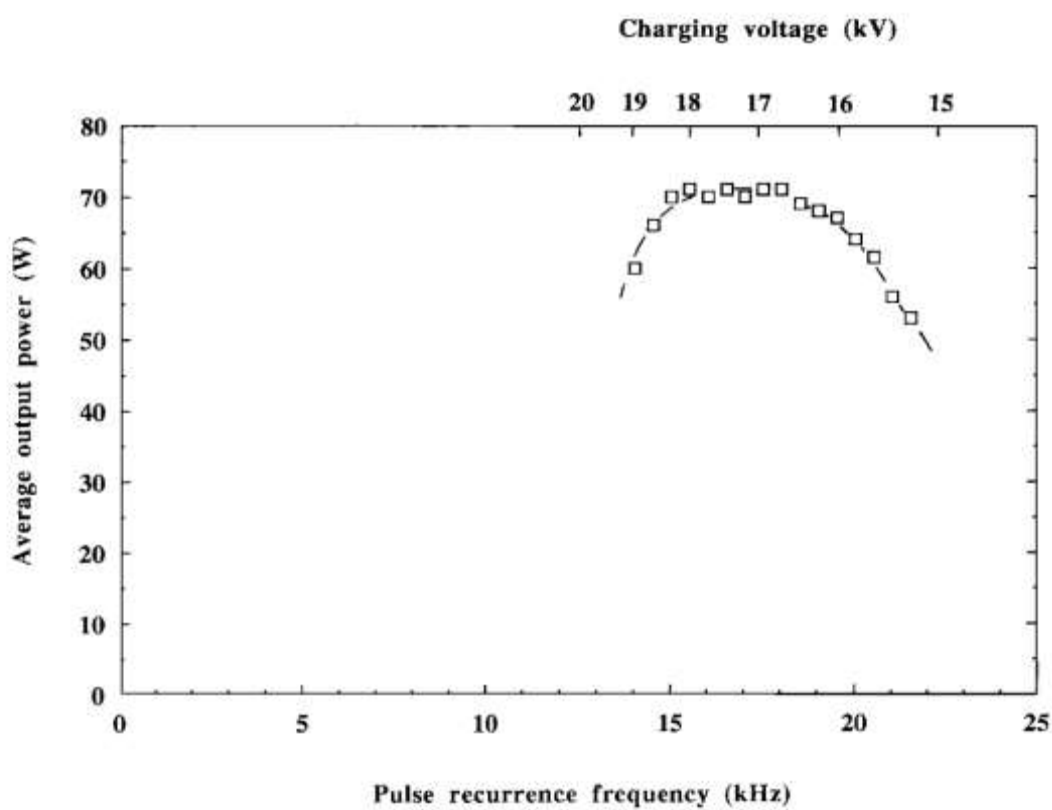

**Figure 2.4.3 Average output power versus pulse recurrence frequency, with a Ne pressure of 20 Torr, an $H_2$ pressure of 0.6 Torr, a PRF of 17.5 kHz, and a capacitance of 1.0 nF**

One of the significant advantages of applying copper bromide technology to generation on atomic copper transitions, as opposed to conventional CVL technology, is that the thermal mass in the former is small by virtue of the simple quartz tube construction and low operating temperature. As a result, warm-up times are significantly smaller for the copper bromide laser, as evidenced in Figure 2.4.4. Whereas a CVL of 100-W average output power takes 90 min or more to reach operating temperature, As seen in Figure 2.4.4 in large-bore CuBr laser, it takes <½ hour to full power (after turning on the laser tube from cold) and just ~11 min lasing to begin. Warm-up times and times for reaching maximum output power could be shortened significantly further by additional forced heating of the CuBr reservoirs. Short warm-up times are important to reduce downtime in commercial applications.

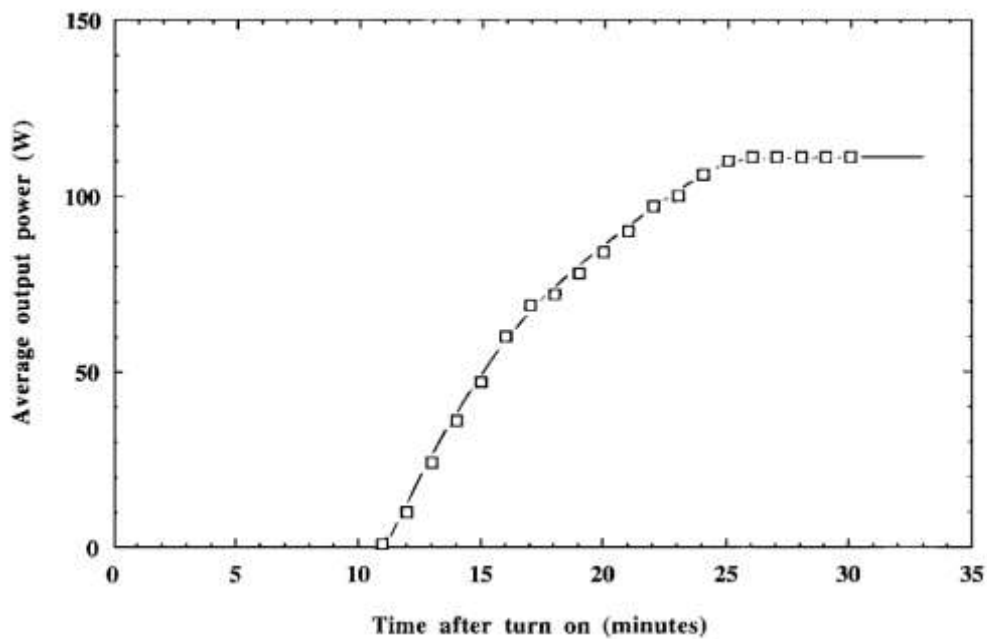

**Figure 2.4.4 Average laser power as a function of time from a cold start-up; input power - 4.6 kW, Ne - 20 Torr, H$_2$ - 0.6 Torr, PRF =17.5 kHz, and capacitance - 1.3nF**

In Figure 2.4.5 are oscillograms of the tube voltage, discharge current, and laser intensity waveforms, together with the instantaneous input power which was calculated without removing the inductive component of the voltage. The discharge conditions corresponded to those for 5-kW average input power and 120-W average output power in Figure 2.4.2, with the equivalent storage capacitance of 1.3 nF.

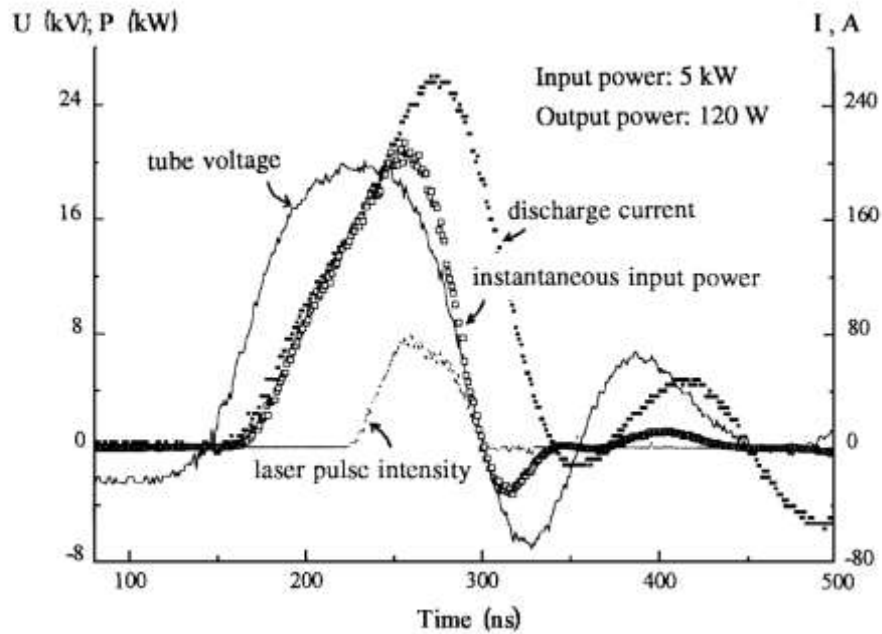

**Figure 2.4.5 Oscillograms of the tube voltage, discharge current, and laser intensity waveforms; instantaneous input power = tube voltage x discharge current**

## *Conclusion*

A Ne-$H_2$-CuBr laser has been shown to generate average output powers of up to 120 W at an efficiency of 2.4% and an output power of 100 W at 3.0% efficiency. The sealed-off mode of operation, fast warm-up time, and long laser pulse make this laser more suitable than an equivalent output power copper vapor laser for use in an industrial environment for materials-processing applications. This laser is now undergoing lifetime tests, which we hope to report on in the future.

# 3. Energy dissipation and dynamics of the electric field of CuBr laser

## 3.1. Energy dissipation in the electrodes of CuBr laser-like discharges

Copper vapor lasers (CVLs), being a source of hundreds of watts of laser power in the visible range, have been developed in various variants, namely CVL utilizing elemental copper or copper halides, or the so-called HYBrID CVL operating on the basis of in situ production of CuBr vapor after the reaction of elemental copper with flowing hydrogen bromide. Electric power of the order of ten kilowatts is normally deposited in the laser tube. This power goes not only to the discharge plasma but also to the electrodes. Information on the magnitude of and factors involved in energy dissipation in the electrodes has not yet been reported to our knowledge. Papers dealing with this matter in the metal vapor lasers are practically absent except, perhaps for the review by Chang *et al.* (1996). However, the great energy losses in electrodes can be one of the causes of efficiency in CVLs as low as a few percent. The purpose of this paper is also to focus attention on the significance of this limiting factor. Recently Astadjov *et al.* (1997) described an experimental observation of lowered electrode heating with laser scaling up. Here we report our study under conditions for a typical CuBr laser but the results are also indicative for other CVL variants (Astadjov & Sabotinov, 1996, 1997).

*Experimental set-up and technique*
Electrode power losses have been measured as functions of the gas mixture, power input and electrode separation in a small-bore silica tube of 17 mm inside diameter. The discharge tube (Figure 3.1.1) was fitted with seven equidistant electrodes, allowing six different lengths of electrode separation (minimum = 14 cm, maximum = 84 cm) to be studied. The electrode arrangement was similar to that in most of our work on CuBr lasers reported so far. The electrodes (copper tablets) were placed in cylindrical silica side arms perpendicular to the tube axis. Each electrode was 35 mm long and contained 20 g of copper tablets. The surface of the electrodes was flush with the inner tube surface and was of a circular shape of 15 mm diameter. The active area of the electrodes was normally the whole open (to the discharge) surface area of the electrodes.

Neon and hydrogen were used as gases. CuBr vapor was introduced into the discharge from side-arm heaters. The pressure of the gas species was in the range for optimum operation of CuBr lasers: Ne 15–20 Torr, $H_2$ 0.3– 0.5 Torr and CuBr 0.2–0.4 Torr.

A storage capacitor of 1.24 nF was discharged through a thyratron and the capacitor voltage was applied to an electrode pair (formed by the end electrode and one of the other electrodes). The pulse recurrence frequency was 16.9 kHz. The thyratron anode voltage determined the average rectifier power. In most of the cases, it was ~7 kV, which meant a rectifier power of ~500 W. Some experiments were carried out at thyratron anode voltage of ~4 kV corresponding to rectifier power of ~160 W.

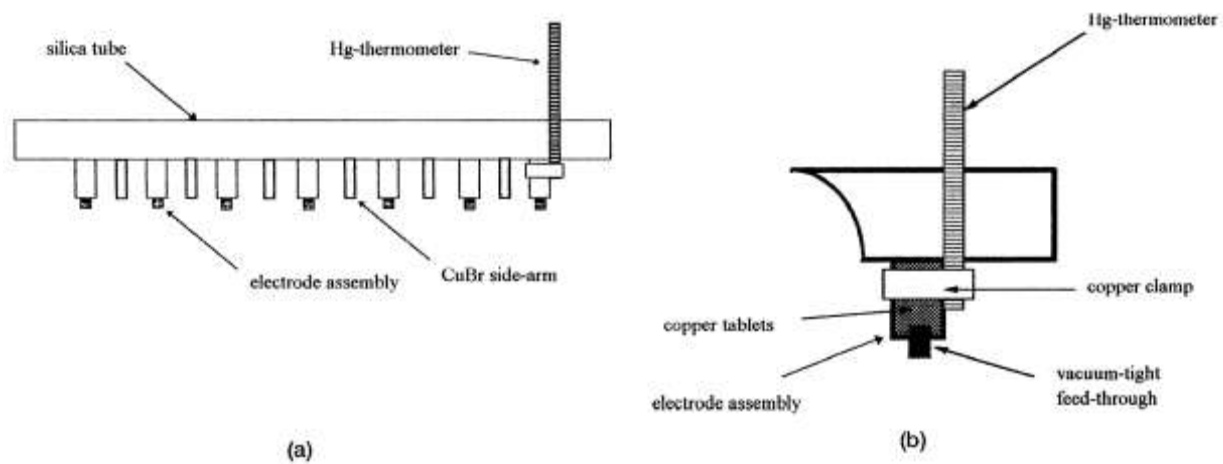

**Figure 3.1.1 (a) Schematic diagrams of the tube with seven equidistant electrodes; (b) Enlarged view of the end electrode**

The waveform, amplitude, and duration of current pulses depended on the test parameters (gas composition, electrode separation and power input). The waveform was aperiodic with different degrees of damping. The pulse duration varied from 200 ns to 1 μs and the pulse amplitude changed from 120 A to 20 A. Periodically a peaking capacitor of 470 pF was connected parallel to the electrode pair tested. This change diminished the circuit inductance but did not affect the general trend of electrode power losses. A sample of a typical current pulse is given in Figure 3.1.2.

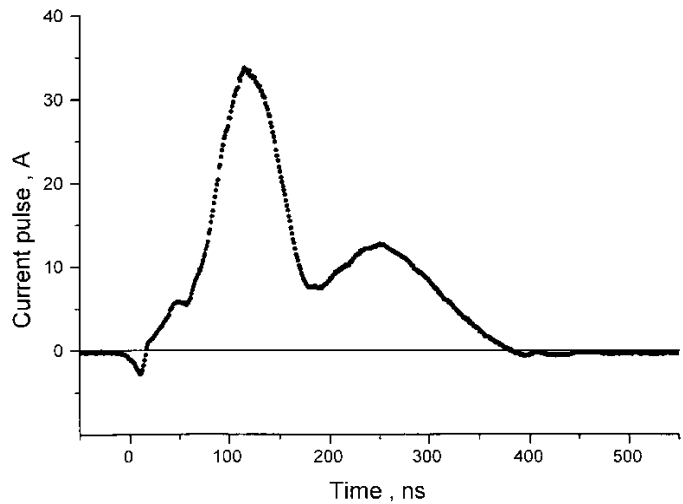

**Figure 3.1.2 A sample of a current pulse waveform**

The temperature of the end electrodes (as part of the electrode pair tested) was continuously measured by an Hg thermometer which was clamped to the outside wall of the silica side arm. We did not use an electronic thermometer because of electric shock hazard and the very high RF noise generated by the experimental equipment.

The Hg thermometer had been calibrated as 'power versus temperature' meter. The calibration was made by substituting in situ the electric discharge (as an energy source) with a miniature electric heater placed into the electrode just under the electrode surface. We evaluated that with electric discharge underrun, the evaporated/ejected copper carried away a power of <0.1 W. Meanwhile, the power deposited in the electrodes during calibration and later with the discharge operating, was of the order of tens of watts. Radiation and air convection were considered the same in both cases. The relative error of calibration was estimated within 10% and was mainly due to the experimental spread of calibration data. The individual errors of the devices (voltmeter, ammeter, and Hg thermometer) employed in calibration were not considered in the net error. The experimental error in separate measurements did not exceed 10%, leading to an overall net error below 20%.

From the measured temperature of the end electrode, we calculated its average power dissipation. As the polarity of this electrode was changed, we considered it as an anode or a cathode. Having the ratio of the average power lost in the electrode to the rectifier output power, we obtained the electrode power losses in percents, making possible a comparison between data obtained at different rectifier output powers. Thus, if not stated otherwise, the percentages below refer to power losses in the electrodes.

*Experimental results*

The energy losses in the electrodes were measured for each electrode separation at which a stable pulsed gas discharge was possible for a thyratron anode voltage of about 7 kV. This is the reason why all possible lengths of electrode separation (denoted by $i\Delta = i \times 14$ cm, i having discrete values from 1 to 6; so $1\Delta=14$ cm, $2\Delta=2\times14$ cm$=28$ cm, and so on) were not examined. The rectifier power output varied due to the negative overswing capacitor voltage and this made us keep it constant by fine adjustment. This was especially important when CuBr vapor was admitted and small changes would have been expected.

Regardless of the buffer gas, 18 Torr Ne (Figure 3.1.3) or 18 Torr Ne with 0.4 Torr $H_2$ (Figure 3.1.4), the power lost in the cathode is always higher than in the anode. Normally around 3/4 of electrode losses are in the cathode. With an increase in electrode separation, losses in the electrodes are reduced. The actual percentage is 22% and 8% for the cathode and 9% and 2% for the anode. As can be seen, the power losses in the electrodes go up as CuBr vapor (at sufficient pressure for laser oscillation, *i.e.* a few tenths of Torr) was introduced into the discharge.

In the case of 18 Torr Ne (Figure 3.1.3) the cathode losses increase by $\leq 1.6\%$, while the anode losses rise by $< 1\%$; the increases in total (*i.e.* the sum of cathode and anode losses) loss vary between 2% and 2.4%.

In the case of 18 Torr Ne and 0.4 Torr $H_2$ (Figure 3.1.4), the cathode losses increase $\leq 2.4\%$ and the anode losses $\leq 1.6\%$; the increase in total loss varies between 2.4% and 3.4%.

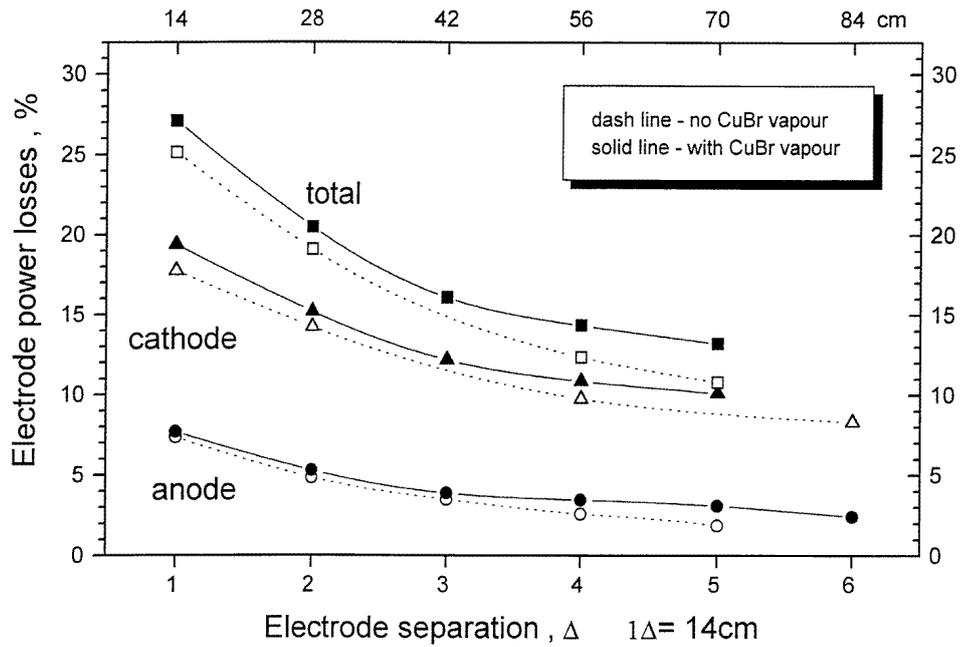

Figure 3.1.3 Electrode power losses with buffer gas of 18 Torr Ne versus electrode separation

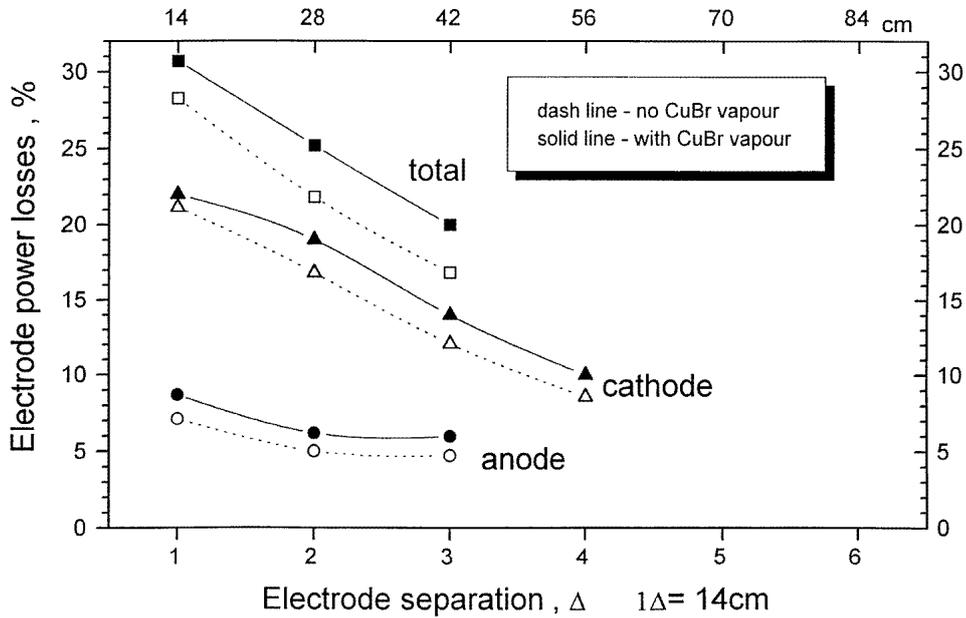

Figure 3.1.4 Electrode power losses with buffer gas of 18 Torr Ne and 0.4 Torr $H_2$ vs electrode separation

Figure 3.1.5 shows the electrode power losses versus electrode separation with neon as the buffer gas with/without hydrogen when the CuBr vapor pressure in the discharge is sufficient (a few tenths of Torr) to produce laser emission.

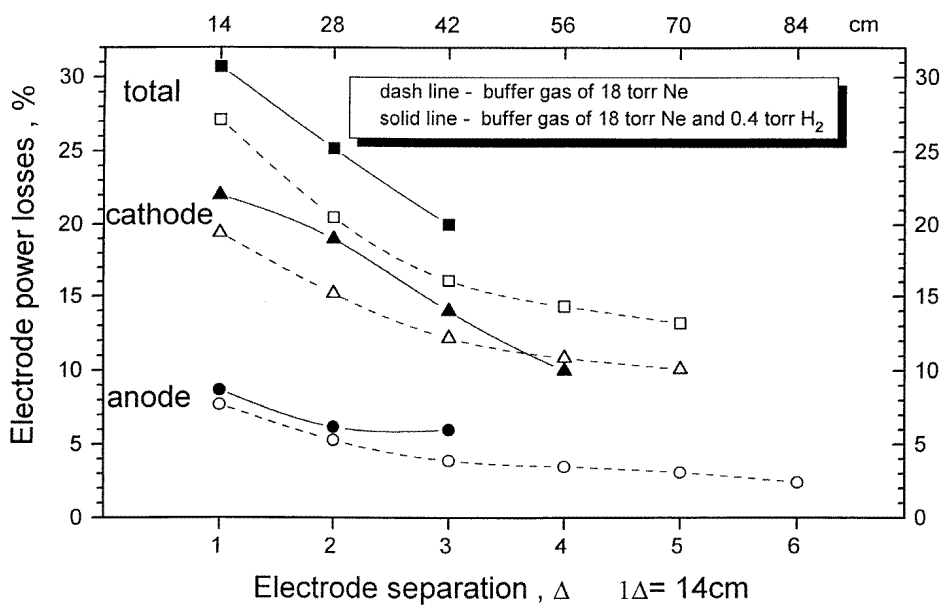

**Figure 3.1.5 Electrode power losses with CuBr vapor and different buffer gasses versus electrode separation**

The electrode power losses in the case of 18 Torr Ne and 0.4 Torr $H_2$ are generally higher than in the case of 18 Torr Ne buffer gas. The total power loss in the electrodes with the 0.4 Torr $H_2$ admixture comes down from 30.7% for $\Delta=1$ to 20% for $\Delta=3$, while the respective values with pure Ne buffer are 27.1% and 16.1%, falling to 13.2% for $\Delta=5$. Note that in the case of the 0.4 Torr $H_2$ admixtures the cathode and anode losses show different trends with the increase of electrode separation: cathode power loss drops faster while the anode loss tends to be nearly constant. This can be related to the obstructed gas discharge in these circumstances since the tube electrode voltage is not high enough for discharge breakdown. This obstructed form of the gas discharge is characterized by the transition of the current waveform from aperiodic to damp periodic. There is no obstructed gas discharge in the case of neon and such trends are not observed there. Consequently, the cathode power loss for $\Delta=4$ with pure neon is higher than that with the $H_2$ admixture.

The dependence of electrode power losses in the gas discharge of pure buffer gasses (*i.e.* before CuBr is loaded into the tube) is plotted in Figure 3.1.6.

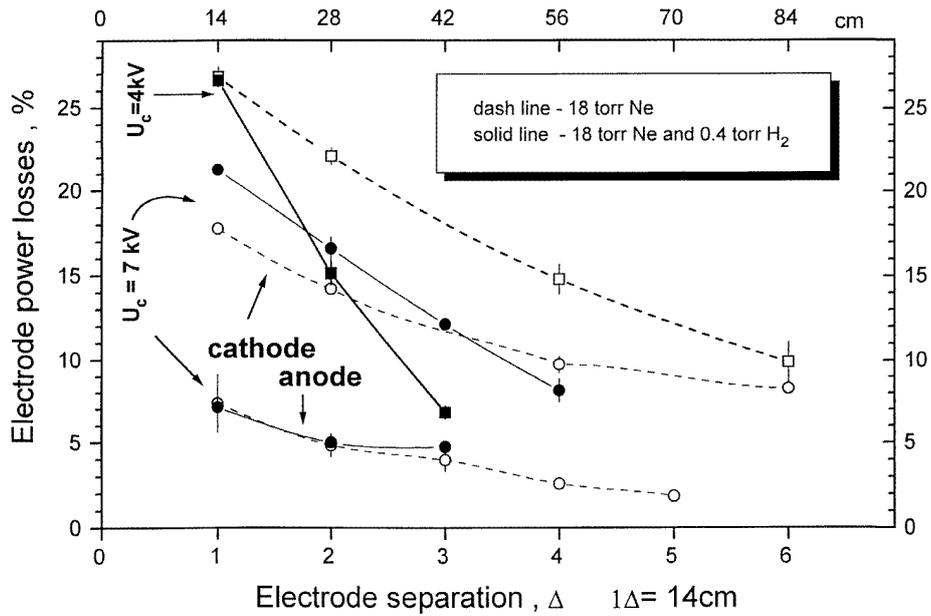

**Figure 3.1.6** Electrode power losses in pure gasses versus electrode separation for thyratron anode voltages of 4 kV and 7 kV

Clearly, it is quite the same as those with CuBr vapor (*cf.* Figure 3.1.5). The power lost in the cathode is less with pure neon at short distances but it is higher at Δ=4. As the thyratron anode voltage was lowered to 4 kV the gas discharge with the $H_2$ admixture became obstructed and the cathode power loss with pure neon was higher for any electrode separation.

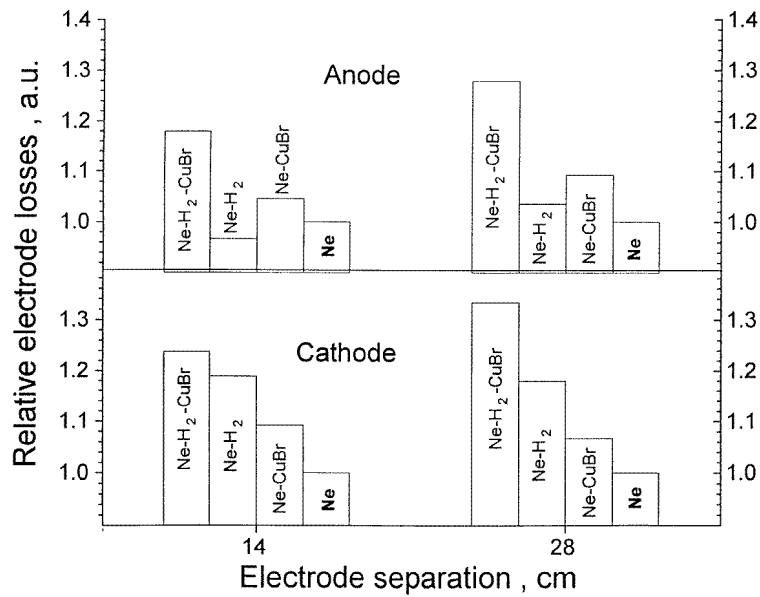

**Figure 3.1.7** Relative electrode power losses in different gaseous environments at short electrode separations of 14 cm and 28 cm. *NB*: At each point (*i.e.* separation and electrode) the electrode power losses in Ne are taken as unity

Figure 3.1.7 depicts anode and cathode losses at short electrode separations (14 cm and 28 cm) where the discharge breaks down into an aperiodic current shape (*cf.* Figure 3.1.2), very similar to that obtained in larger tubes. The losses are plotted relative to losses with 18 Torr Ne taken as unity for each distance and electrode. If 0.4 Torr $H_2$ is mixed with Ne at 18 Torr, the cathode losses increase by ~20%. The situation for anode losses is not regular. As CuBr vapor is also added to the Ne–$H_2$ mixture the power losses increase further for both electrodes. They become between 18% and 35% higher compared to losses with Ne. The CuBr vapor addition to Ne is not so dramatic. The energy losses in the electrodes are ≤10% higher than those in Ne. Remarkably, the cathode losses increase stepwise in the sequence Ne→Ne-CuBr→Ne-$H_2$→ Ne-$H_2$-CuBr. This is not so with the anode: power losses in Ne-$H_2$ are less than in Ne-CuBr, even less than that in Ne at the shortest separation of 14 cm.

Figure 3.1.8 shows the dependence of the energy per single charged particle on the electrode separation for either electrode in different gaseous environments.

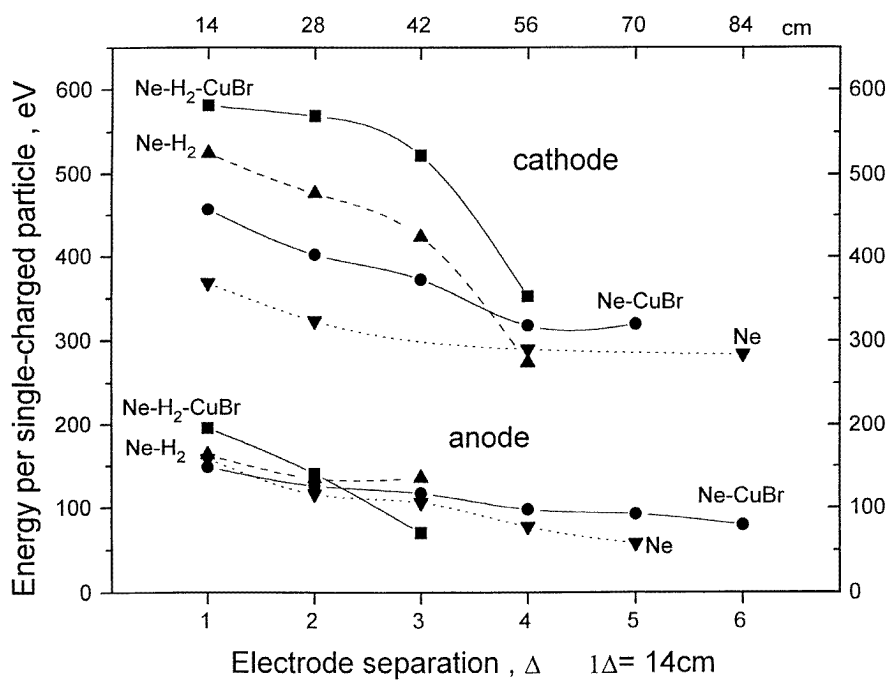

**Figure 3.1.8 Energy per single charged particle in different gaseous environments *vs* electrode separation.**

The energy per single charged particle, $E_e$ is easily derived as

$$E_e = eE_{el}/Q_i = eP_{el}/I_a$$

where $e$, $E_{el}$, $Q_i$, $P_{el}$ and $I_a$ are respectively the elementary charge, the pulse energy dissipated in the electrode, the quantity of electricity per pulse, the average electric power dissipated in the electrode and the average discharge current. The quantity $E_{el}/e$ is known as 'voltage equivalent of heat flux' and it is an important part in the energy balance relation for electrodes (Raizer, 1987).

The $E_e$ dependence on the electrode separation is not simple but some general comments can be made. At short electrode separation, $E_e$ changes in the same way as the relative losses given in Figure 6 with respect to the gaseous environment. The abrupt decline of $E_e$ in the $H_2$ admixture at electrode separations greater than 42 cm is due to obstructed discharge in these circumstances. For the cathode, $E_e$ varies between ~300 eV (in Ne) and ~600 eV (in the Ne–$H_2$–CuBr mixture); for the anode these values are two–three times lower.

We also made use of the multi-electrode tube to measure the evolution of the longitudinal potential of the electric field in 18 Torr pure Ne with a thyratron voltage of 4 kV (Figures 3.1.9 and 3.1.10).

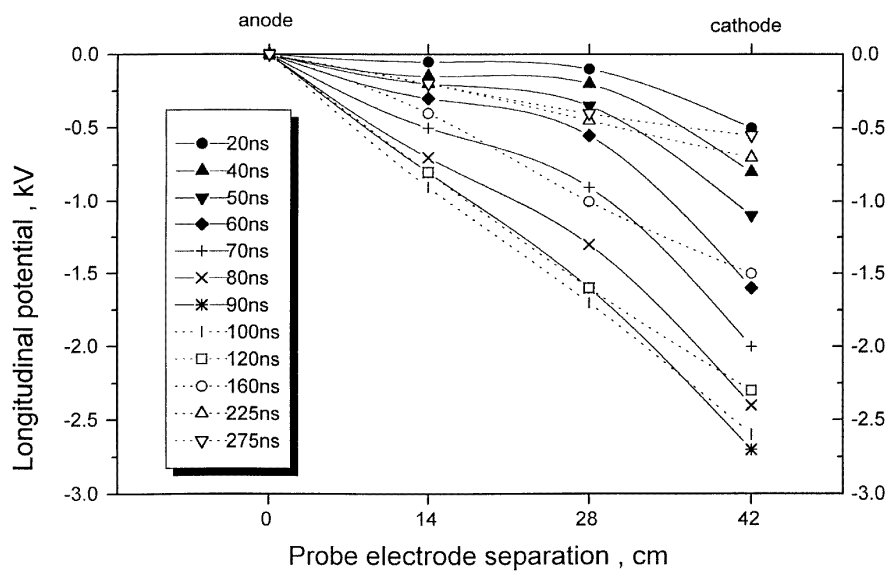

**Figure 3.1.9 Time evolution of the longitudinal potential distribution of electric field for a tube electrode separation of 42 cm. The times are from the potential set on cathode**

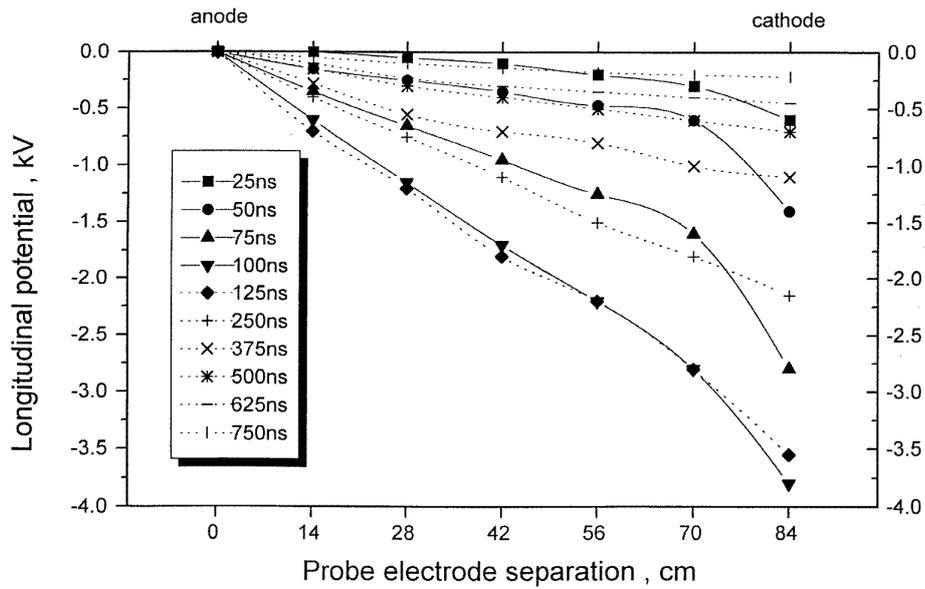

**Figure 3.1.10** Time evolution of the longitudinal potential distribution of electric field for a tube electrode separation of 84 cm. The times are from the potential set on cathode

As Knyazev (1971) and Kukharev (1984), we also noted the existence of a two-stage pulse development. The first stage is a glow discharge featured in the cathode and anode high-field regions and low-field positive column. This lasts ~70–80 ns and gradually transforms into a discharge with a uniform field distribution along the tube. The maximum field in this second stage is ~65 $Vcm^{-1}$ and ~40 $Vcm^{-1}$ for tube electrode separations of 42 cm and 84 cm, respectively. These maximum field values correspond to the moments in time just after the peaks of electric field potential on oscillograms recorded during the study.

*Discussion*

The observed behavior of energy losses in electrodes can be partly explained by the change in efficiency when energy is transferred from the storage capacitor to the gas discharge tube, *i.e.* the change in 'matching'. The power supply efficiency has already been reported to be dependent on the energy of the electric pulse switched and pressure of the buffer gas (Isaev & Lemmerman, 1987). Our experience is that 'matching' is actually improved <10% through addition of small amounts of hydrogen to the neon buffer gas of a CuBr laser at constant electric pulse energy switched by a thyratron. Meanwhile, the total power lost in the electrodes increase is ≤5%. So this increase, to some extent, may be caused by better

'matching' as an additive of 0.3(±0.1) Torr $H_2$ is present in neon gas when electric discharge is initiated.

Astadjov *et al.* (1986) pointed out some characteristics of pulsed excitation for a CuBr laser with an active zone of size 2 cm × 50 cm. We found that with a buffer gas of 15 Torr Ne, plasma impedance increases as CuBr vapor is added to the discharge. However, with a buffer gas of 15 Torr Ne and 0.3 Torr $H_2$, the situation is just the opposite, *i.e.* plasma impedance decreases with CuBr vapor addition. The outcome of the current study with our multi-electrode tube supports the results by Astadjov *et al.* (1986) However, the energy loading of electrodes increases in all cases as the composition of the gaseous medium becomes more compound. Therefore, the observed behavior of electrode power losses cannot be due to a simple redistribution of electric power between the discharge plasma and the electrodes proportionally to their impedance. Obviously, pre-electrode processes have to be considered too.

The role of electrodes in the initiation of gas discharges is very substantial. For cold cathodes, it is usually determined by the coefficient of secondary electron emission γ (Gama), which is a sum of the γ-coefficients for the processes of ion bombardment, bombardment by neutrals (excited or unexcited) and by photons. For anodes, a very important role is also played by bombardment by electrons. In gas discharges like that studied here, bombardment processes by heavy particles are predominant (Raizer, 1987); this is evident from visual sputtering of the electrodes, especially the cathodes. Their γ-coefficients are very sensitive to the cleanliness of the electrode surface. Usually oxide films or other films formed by active gasses in the discharge are present on the electrodes. After a variable clean-up time, these films can change γ by factors of 5–10 and lead, in many cases, to the contraction of a diffuse glow to active localized discharge spots on the cathode and anode (Loeb, 1959). As In that paper, in the same manner, Raizer (1987) also noted that generally all contaminated cathodes have poorer γ values except most oxidized cathodes. According to our observations, such bright spots are typical for non-porous electrodes made of a sheet or solid metal. These optically observed spots are often moving. Sometimes we can see them forming filamentary structures in the discharge volume or/and symmetric figures on the electrode surface. Similar phenomena are mentioned by Raizer (1987) and reported in XeCl lasers by Boettocher *et al.* (1992). All this reaffirms our speculations about the nature of our observations.

With a pyrometer, we measured the average temperature of the red-hot cathode surface - <1400 K. Accordingly Raizer (1987), if we assume only thermoelectronic emission from copper electrodes, the achievable thermocurrent is $10^{-7}$ A cm$^{-2}$. For our current densities of ~100 A cm$^{-2}$ there must be separate hot spots with temperatures >3000 K. Even if the porous structure of the copper electrodes is taken into account the hot-spot temperature must be ~2500 K. The existence of very hot arc-like spots on electrodes is a well-established fact. At these temperatures the evaporation of copper proceeds through high-velocity jet bursts of metal and for Cu this velocity is $\sim 1.5 \times 10^5$ cm s$^{-1}$ as reported by Knyazev (1971). So, the thin copper layers formed around the electrodes could also be a result of extreme electrode evaporation, not only sputtering. The rate of this electrode erosion for Cu is $\sim 10^{-4}$ g C$^{-1}$ (*cf.* Daalder, 1975 and Raizer, 1987) or, under our conditions, 1 g of Cu ejected over 10–20 h. This was actually observed in our tubes. Considering that copper electrodes cannot be free of all impurities, the most likely factor (*cf.* Loeb, 1959; Westberg, 1959 and Raizer, 1987) is the thermo-autoelectronic nature of pre-electrode processes, which include intense γ-processes. A similar explanation was given by Knyazev (1975). Because of the quite complex nature of pre-electrode processes we cannot at present elucidate the exact processes responsible. However, any variations in the gaseous environment of electrodes will obviously cause changes in their emission properties.

The differences in power loading of cathode and anode (Figure 3.1.7) are presumably due to bombardment by different charged particles. This can be attributed to the production of ions such as Cu$^+$, Br$^+$, Br$^-$, *etc.* which enhance the bombardment of respective electrodes. With different buffer gases, namely 18 Torr Ne and 0.4 Torr H$_2$ compared with pure 18 Torr Ne, the increase in power lost in the electrodes can be attributed to H$^+$, H$_2^+$, HBr$^+$, H$^-$, *etc.*, the increase in tube voltage and, then, possibly the velocity of ions bombarding the electrodes. The evolution of longitudinal potential of the electric field (Figures 3.1.9 and 3.1.10) shows definitely that, in copper vapor lasers (CVLs), the build-up of key processes happens in an inhomogeneous electric field. This should be taken into account for a better understanding of CVL kinetics.

*Summary and conclusions*

Energy dissipation in the electrodes of a CuBr laser pulsed discharge has been studied in gaseous environments (pure neon, neon–hydrogen mixture and neon–hydrogen–copper bromide vapor) relevant to CVLs as well as with discharge length scaling. The total energy losses in both electrodes range from ~10% to ~30% of the storage capacitor energy. Power losses in vacuum-arc copper cathodes of the order of 30% were, in the past, reported by Daalder (1977). Normally around 3/4 of electrode energy losses happen in the cathode. Total energy losses in electrodes can even be augmented further if losses in the thyratron circuit are high as well. It is likely that half of the energy delivered to the discharge tube is wasted in the electrodes instead of in the discharge volume for a CuBr laser-like discharge.

Not only relative power losses in electrodes drop with the increase of electrode separation. Similar is the trend for absolute power losses in electrodes. So in cathode as the major contributor to the losses, electric power deposition decreases of a factor of ~1.6 (from 80-100W at 14 cm to 50-60W at 84cm) in non-hydrogen mixture of 18 Torr neon and copper bromide vapor. The addition of 0.4 Torr $H_2$ to 18 Torr Ne results of a dramatic fall of power deposition with the increase of electrode separation, namely at 84cm (~10W) it amounts to 1/8 to 1/10 of that at 14cm (~100W). Hence, length scaling (getting electrodes more separated) of laser tubes leads to less power loading of electrodes and has a positive impact on the efficiency of lasers. Thus, one of the reasons for the low CVL efficiency, which has been overlooked for a long time, could be these excessive energy losses in electrodes.

The power lost in the electrodes increases significantly not only with decreases in electrode separation but also as the composition of the gaseous medium becomes more compound *i.e.* Ne→Ne-$H_2$→Ne-$H_2$-CuBr. The latter logically causes an increase in number and mass of particles bombarding the electrodes. The experimental results on electrode separation scaling cannot be explained by a simple redistribution of electric power (between the discharge plasma and the electrodes) proportionally to their impedance. Obviously, pre-electrode processes must also be considered. Further investigations are necessary to characterize the mechanisms responsible for the specific behavior of electrodes in CuBr laser-like discharges. Special electrodes designed for low-energy losses and efficient electron emission should be developed too.

The evolution of electric field longitudinal potential in copper vapor lasers shows beyond doubt that the majority of processes builds up in an inhomogeneous electric field. This and the processes in/around the electrodes should be considered in both experiment and modeling to promote a better understanding of CVL kinetics for the ultimate progress of CVLs.

## 3.2. Dynamics of the electric field in CuBr laser-like discharges

Copper vapor lasers of very high efficiencies of ~3%, producing an average laser power of >100W, have recently been demonstrated *e.g.* Astadjov *et al.* (1997). A crucial point in CVL development was overcome by Astadjov *et al.* (1985). They established and reported the great positive impact of small amounts of hydrogen added to the principal neon buffer gas on boosting CVL efficiency and power. The deliberate addition of $H_2$ to the ultra clean Cu–Ne environment of a CVL leads to a 2–4-fold improvement of CVL characteristics.

Elastic/inelastic processes, the formation of specific plasma species (such as negative ions), changes in electron energy distribution, *etc.* have been proposed to be responsible for the great jump in CVL characteristics. However, a complete explanation of the effect of hydrogen has still to be presented.

The current work deals with the changes in electric field (E-field) dynamics as hydrogen is added to a buffer gas of neon in CuBr laser-like discharges (Astadjov & Sabotinov, 1999). These gas discharges are excited through the electric pulsed power of many megawatts delivered on a submicrosecond time scale at a kilohertz pulse repetition rate. They are similar for all types of CVLs and there are reports concerning the matter (Knyazev, 1971; Kukharev, 1984; Satoh et al., 1997; *etc.*). The gas pressures studied are those typical for optimum operation of CVLs: hydrogen gas pressure is a few tenths of a Torr and neon gas pressure is ~20 Torr.

### *Experimental apparatus*

The formation and time evolution of E-field in a CuBr laser-like discharge as a function of neon–hydrogen gas mixture proportions have been studied in a 40mm-bore silica tube with two electrodes separated 52 cm from each other. The electrodes are hollow thin copper cylinders inserted into the tube. The electric gas discharge is produced by repetitive discharging of a 1.1 nF capacitor through the tube. The capacitor voltage is 10–11 kV which provides an average power of 1.1–1.2 kW switched at a pulse repetition frequency of 17 kHz. A peaking capacitor of 470 pF is connected between the electrodes too.

## Nonperturbation E-field measurement technique

E-field measurements are conducted by a *nonperturbation* technique employed in our laboratory for the first time (to our best knowledge). Here follows a description of this new technique.

The electric field potential (briefly, E-potential) is measured at points on the outside surface of the electric gas discharge tube which is under test. A high-voltage Tektronix P6015A probe (with reliable electrical ground and rise time <5 ns) gauges E-potential of each point and its waveform is saved thru digital oscillograms. The measurement points are on a line from the cathode to the anode and are nearly equidistant (Figure 3.2.1). The points adjacent to the electrodes are spaced more closely to allow better scanning of the electric field in the near-electrode domains. The number of points is 15 and their locations are fixed for all scanning series.

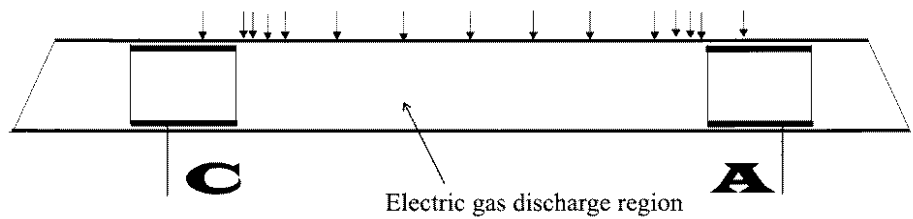

**Figure 3.2.1 A schematic diagram of the gas discharge tube used in the electric field surface scanning experiment in CuBr laser-like discharges. The arrows show the locations of the measurement points. C is cathode and A - anode**

The oscillograms of E-potential are processed by a computer to get a longitudinal distribution of E-potential for different moments of time during the electric gas discharge of submicrosecond length. At the beginning of electric pulses (of the full length of ≤350ns), the time scan increment is 10 ns going up to 30 ns at the end of pulses. So 15–20 curves of the longitudinal distribution of E-potential can be obtained.

## Experimental data processing and final results

A typical profile of the longitudinal distribution of E-field potential derived from E-potential oscillograms is given in Figure 3.2.2. We can distinguish three parts in that profile: two near-electrode domains and one discharge domain far from the electrodes.

E-field of the far-from-electrode (FFE) discharge is characterized by a linearly changing potential. That implies an actually constant E-field throughout the FFE domain. To obtain the value of the E-field in the FFE domain we perform linear fitting of the experimental time profiles. The derivative of the linear fit gives the E-field on the tube surface in the FFE domain for the respective moments of time.

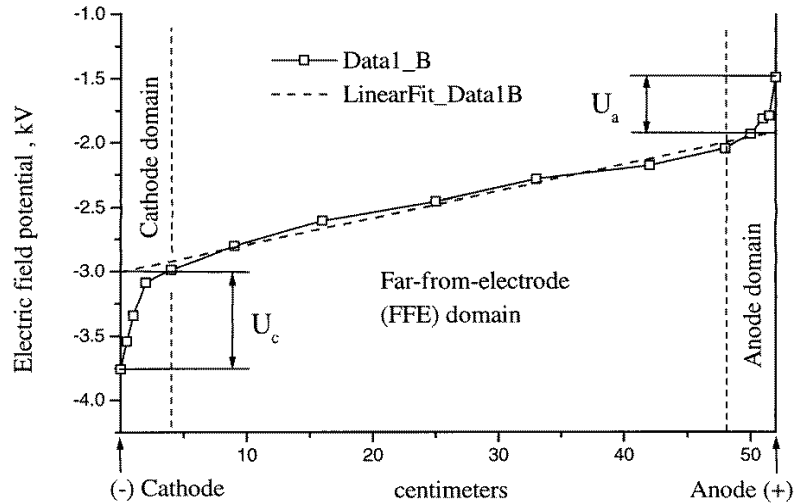

**Figure 3.2.2 Characteristics of a typical longitudinal distribution of E-potential on the tube surface and the processing of longitudinal distribution to derive near-electrode voltages ($U_c$ and $U_a$), and far-from-electrode E-field (FFE-field)**

There are also two near-electrode E-field domains: the near-cathode domain of voltage $U_c$ and the near-anode domain of voltage $U_a$. The near-cathode domain is more pronounced than the near-anode domain. E-field of the near-electrode domains is calculated from the longitudinal distribution of the E-potential as follows. For the near-cathode domain, we take the difference between the lowest measured value and the extrapolated potential of the FFE domain at the near-cathode domain. For the near-anode domain, we take the difference between the highest measured value and the extrapolated potential of the FFE domain at the near-anode domain. We illustrate that graphically in Figure 3.2.2.

The evolution of near-electrode voltages ($U_c$ and $U_a$), and FFE electric field for various gas mixtures of fixed neon pressure of 20 Torr and partial hydrogen pressures of 0.02 Torr, 0.1 Torr, and 0.37 Torr is given in Figure 3.2.3.

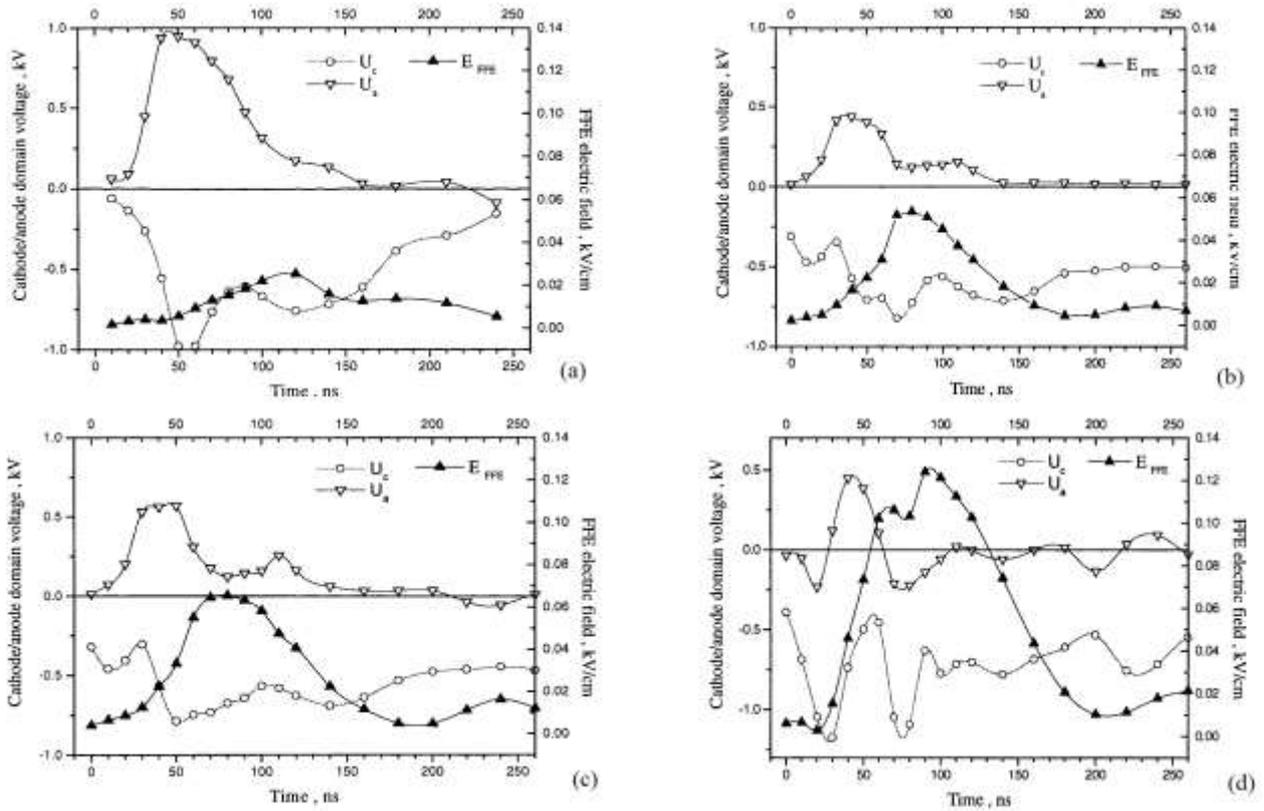

**Figure 3.2.3** Evolution of the near-electrode voltages ($U_c$ and $U_a$), and FFE electric field during the electric pulse for various gas mixtures: (a) 20 Torr Ne, (b) 20 Torr Ne plus 0.02 Torr $H_2$, (c) 20 Torr Ne plus 0.1 Torr $H_2$ and (d) 20 Torr Ne plus 0.37 Torr $H_2$

The evolution curves for various gas mixtures are presented in plots designated by letters. The highest pressure of the hydrogen additive corresponds to the optimum for CuBr laser performance. As seen the evolution of near-electrode voltages and FFE electric field is complex. We can generally characterize it as pulsed. The peaks of near-electrode voltages are 30–60 ns after the voltage onset. (The zero time for voltage onset is the time when the tube electrode voltage reaches about 300 V). The maximum FFE electric field is 80–100 ns after the voltage onset. Except for Ne gas (Figure 3.2.3 (a)) - it is 120 ns after the voltage onset.

The most apparent finding is the great jump of the FFE field as hydrogen is introduced and the partial pressure of hydrogen increases. With Ne, the maximum of FFE field is only ~20 $Vcm^{-1}$. As hydrogen additive pressure is getting gradually higher, FFE field becomes stronger and stronger. Reaching ~120 $Vcm^{-1}$ (with a hydrogen of 0.37 Torr) it is 6 times as much as that in neon. Besides, the peak of FFE field is 30–40 ns earlier than with neon gas only.

The maxima of the near-electrode voltages are 30–60 ns after the discharge onset. They are afore the peaks of the FFE field in all cases. For neon, the maxima of $U_a$ and $U_c$ are ~1 kV. As hydrogen is added they either drop down to ~0.5 kV for $U_a$ and 0.7–0.8 kV for $U_c$.

Now we shall describe the time evolution of discharge domains' voltage from another point of view. Let's put it on again. With Ne, the FFE voltage peak is ~1 kV whereas with 0.37 Torr $H_2$ additives it rises to ~6 kV. Meanwhile with hydrogen addition, the near-electrode fields become weaker (as we already pointed out above) and they are nearly half of the values in the case of neon. Thus, the addition of hydrogen results in a 6-fold increase of FFE voltage whereas the near-electrode fields fall by half.

For that purpose, we define the FFE coefficient, $k_{FFE}$

$$k_{FFE} = \text{FFE voltage} / (\text{FFE voltage} + U_a + U_c).$$

Which allows us to compare FFE voltage as a fraction of total surface voltage between electrodes in different cases (*e.g.* with and without hydrogen). The time evolution of FFE coefficient is plotted in Figure 4 for Ne of 20 Torr and for a mixture of 20 Torr Ne and 0.37 Torr $H_2$.

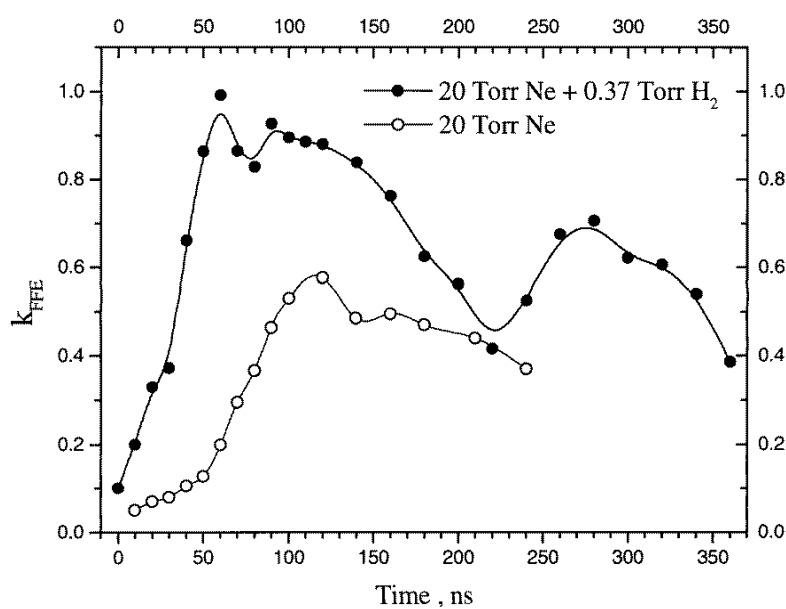

**Figure 3.2.4 The time dependences of FFE coefficient ($k_{FFE}$) during the electric pulse in the cases of 20 Torr Ne and 20 Torr Ne plus 0.37 Torr $H_2$**

Let's analyze the FFE coefficient in Figure 3.2.4. For neon case, its maximum is ~0.5. With 0.37 Torr $H_2$ additives, it is much higher ~0.9. In the first case the maximum of $k_{FFE}$ is ~120 ns after the voltage onset but in the second case, it appears ~60 ns earlier. The average FFE coefficient (averaged over the time of electric discharge pulse) is 0.32 for Ne case and with 0.37 Torr $H_2$ additives it doubles - $< k_{FFE} > = 0.64$. So in the case of Ne, a third of the electric power delivered to the discharge tube goes into the FFE domain. In the case of 0.37 Torr $H_2$ added, that fraction is twice as much - two-thirds of power is dissipated into the FFE domain. Thus, the electrical conditions in mixtures of Ne–$H_2$ turn into favoring the FFE domain where major laser generation processes take place.

More findings are shown in following graphs. We normalized the tube electrode voltage and the FFE electric field and plotted them in Figure 5 as functions of the partial pressure of hydrogen of a mixture with 20 Torr neon.

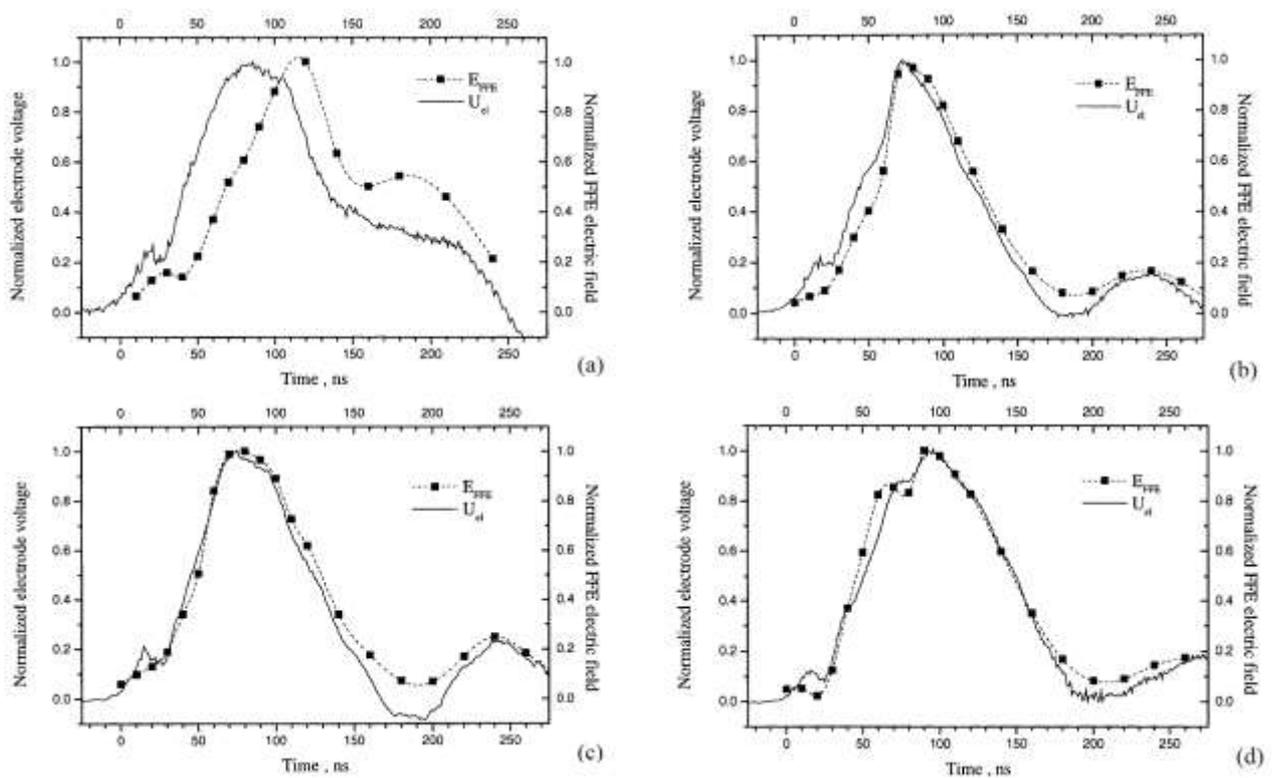

**Figure 3.2.5 The evolution of the normalized electrode voltage and the normalized FFE electric field during the electric pulse for various gas mixtures: (a) 20 Torr Ne; (b) 20 Torr Ne - 0.02 Torr $H_2$ ; (c) 20 Torr Ne - 0.1 Torr $H_2$ and (d) 20 Torr Ne - 0.37 Torr $H_2$**

The situation with various Ne–H$_2$ gas mixtures is presented in graphs designated by letters. As the additional partial pressure of H$_2$ increases, the coincidence of the two curves gets better. In the case of Ne (Figure 3.2.5(a)), there is a time delay (or dephasing) varying from 10 to 40 ns throughout the pulse. This dephasing gradually diminishes with increasing H$_2$ pressure. At H$_2$ pressures >0.1 Torr (Figures 3.2.5(c) and (d)) the dephasing virtually disappears.

*Discussion and conclusions*

A nonperturbation technique of E-field potential surface scanning is successfully employed for the first time. The major issue concerns the registered differences between the waveforms of the electric voltage applied to the electrodes and the waveforms of E-field on the tube surface. With the assumption that the tube surface E-field is an image of the E-field of the discharge plasma near the inside surface of the tube, we can draw some conclusions.

The first conclusion is that addition of a few tenths of Torr of H$_2$ to Ne gas (20 Torr) leads to redistribution of the longitudinal E-field (and thence of the electric power, *etc*.) between the near-electrode domains and the FFE domain. Although the role of near-electrode domains should be considered, the FFE domain is the main scene of laser generation, for at least two reasons. Because it is the region in which laser emission develops, the homogeneity of the FFE domain is of the highest importance. Also, the FFE domain is more scalable than are the spatially confined near-electrode domains. Our measurements of the longitudinal E-field show that proportionally much more electric energy goes into this FFE domain in comparison with the non-uniform near-electrode domains. Thus, the new distribution of the E-field (being in favor of the FFE domain) contributes much to the improvement of CuBr laser performance.

A dephasing of the tube electrode voltage and FFE electric field in neon was revealed. It was found that this dephasing disappears with the addition of a partial pressure of H$_2 \geq 0.1$ Torr. At present, the dephasing effect cannot be satisfactorily interpreted. However, we could assume that the synphasing effect when hydrogen is added to neon is a consequence of a redistribution of the *radial* E-field. The decrease in dephasing is probably due to changes in radial E-field distribution, which is probably getting more uniform (*i.e.* flatter) as hydrogen is added.

As an overall result, the addition of hydrogen leads to a *higher longitudinal* E-field and, probably, to a *more uniform radial* E-field in the major part for laser generation – the far-from-electrode domain of discharge volume. Since the CuBr laser discharge is similar to the discharges of all types of CVLs, the issues should apply to all CVLs.

The newly presented nonperturbation technique of E-field potential surface scanning is still under development in our laboratory, but obviously, it is simple, fast and accurate. As well as being a nonperturbation technique, it needs no intrusion of E-field probes into the construction of the device being tested. However, the interpretation of the data obtained by this technique is not yet complete. Probably, much more information about discharge-plasma parameters can be gathered through its sophisticated application.

# 4. Properties of the laser emission of copper bromide vapor laser

## 4.1. Influence of hydrogen on the optical properties of CuBr laser

In 2000, we reported our first study (Stoilov *et al.*, 2000) on the optical properties of CuBr laser. It was a medium-scale (2cm-bore x 50cm) low-power (10-20W) CuBr laser. Positive branch unstable resonator of two different magnifications was employed in the experiments. The first resonator comprised one concave mirror (curvature radius of 220 cm and diameter of 40 mm) and a convex mirror (curvature radius of 3 cm and diameter of 3 mm). The cavity length was 108.5 cm and the resonator magnification was 73. The second cavity was with the same convex mirror but the 40mm-diameter concave mirror was of 300cm curvature radius. The second cavity length was 148.5 cm and the resonator magnification was 100.

The measurements of the influence of hydrogen on the CuBr laser optical properties were performed with a laser where the electric discharge was run through different buffer gas mixtures of 17 Torr neon with hydrogen additives of three values of pressure - 0.2 Torr, 0.3 Torr, and 0.5 Torr. As it was found by Astadjov *et al.* (1988)a these values are the pre-optimum, optimum and post-optimum pressures of hydrogen additives for a stable-cavity low-power CuBr laser. The results were compared with the measurements with no hydrogen added. No separation of the green and yellow laser lines was done. The radial width of the laser beam was measured at the $1/e^2$ - level of the intensity profile maximum which is suitable for many technology applications. Here are the most important results of that study.

*Experimental results*

Figure 4.1.1 depicts the beam divergence evolution (θ) during the laser pulse. As can be expected the laser beam divergence goes down with the successive cavity round trips (RT) of light. The decrease is more pronounced in the presence of hydrogen, especially, of 0.2 Torr. In that case (0.2 Torr $H_2$), the beam divergence change from 2 RT to 3 RT is ~25 µrad downward. The divergence after 3 RT is ~ 90 µrad and at the end of the laser pulse, the beam divergence is ~ 65µrad. With 0.3 Torr $H_2$, the beam divergence changes by as much as ~10 µrad and the final beam divergence is ~80 µrad. In the case of no hydrogen, the change in beam divergence is even less, and since it is within experimental data spread, one could not take it for real.

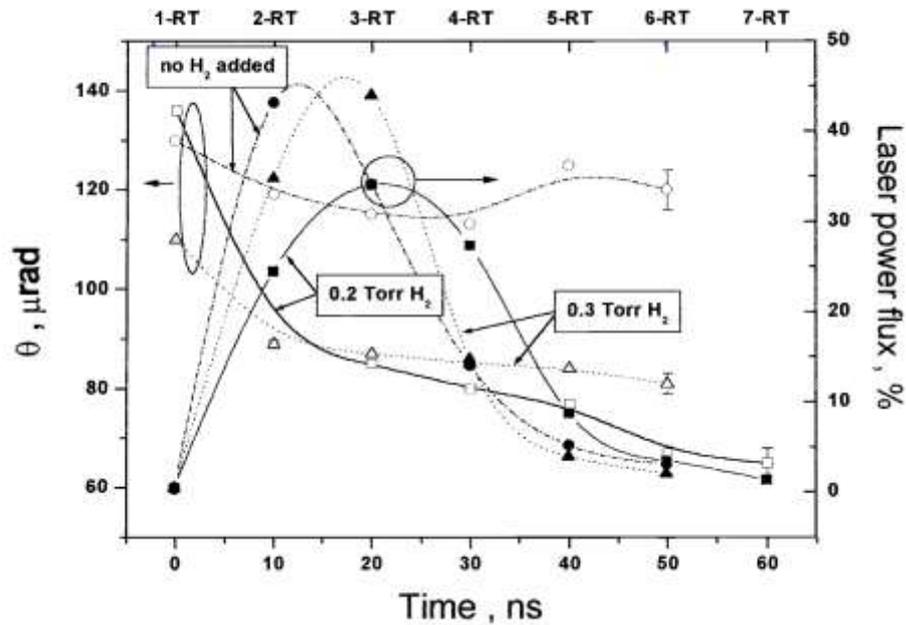

**Figure 4.1.1 Evolution of CuBr laser beam divergence (θ) and laser power flux (%) during the laser pulse successive cavity round trips (RT)**

In all cases, the major part of laser power is concentrated between the 2 RT and 4 RT time slices. With hydrogen most of the laser power is with a beam divergence of < 90 μrad whereas in the case of no hydrogen it is a beam divergence of ~120 μrad. In fact, all laser power flux without hydrogen is a 'bad' beam of ~120 μrad (*i.e.* ~2 x diffraction-limited). With hydrogen, the percentage of laser power flux of a 'good' divergence of ≤90 μrad (*i.e.* ≤1.5 x diffraction-limited) is very high. With 0.2 Torr $H_2$ it is 87% and with 0.3 Torr $H_2$, it is 82%. Thus, the addition of hydrogen not only makes the laser power to go up but also significantly decreases the beam divergence.

An important laser system property is the laser power density, $P_s$ obtained after focusing the output radiation. It is equal to

$$P_s = 1/\pi F^2 \times P_{out}/\theta^2.$$

Where $P_{out}$ is the average laser output power, θ is the divergence of laser output radiation and F is the focus length of output focusing optical system. It is clear that $P_s$ is proportional to the quantity $P_{out}/\theta^2$ which is an inherent property of the light emitted from a laser. This laser power density ($P_s$) represents the laser power spatial intensity[1] because $\theta^2$ is actually the solid angle in which laser output is concentrated. Laser power spatial intensity can be quantified in W.mrad$^{-2}$, for example. The inverse-square dependence of laser power spatial intensity on the divergence of laser output power means that higher laser power density can be more easily obtained by improving the beam divergence. Thus, a serious decrease in laser beam divergence should sharply reduce the demand for too high laser output.

---

[1] It can also be referred to as brightness. This term is used in our later work.

The foremost results of pulse-averaged (*i.e.* time-averaged) non-selective (for both laser lines) measurements are given in Tables 4.1.1 and 4.1.2. The cavity magnification tested is 100 and 73, and the electric input power is 1.3 kW and 1.1kW

**Table 4.1.1 Characteristics of CuBr laser emission with a positive branch cavity of magnifications 100 and 73; electric input power is 1.3 kW**

| Hydrogen pressure, Torr | Average power, W | | Pulse length, ns | | Beam divergence, μrad | | Laser spatial intensity, Wmrad$^{-2}$ | |
|---|---|---|---|---|---|---|---|---|
| - | 2.8 | 3.3 | 40.1 | 32.1 | 100 | 129 | 280 | 198 |
| 0.2 | 7.4 | 8.25 | 52.6 | 43.4 | 84 | 86 | 1049 | 1115 |
| 0.3 | 11.5 | 11.0 | 33.0 | 33.0 | 91 | 89 | 1389 | 1389 |
| 0.5 | 10.25 | 11.0 | 31.1 | 38.2 | 94 | 130 | 1160 | 651 |
|  |  |  |  |  |  |  |  |  |
| Cavity magnification | 100 | 73 | 100 | 73 | 100 | 73 | 100 | 73 |

**Table 4.1.2 Characteristics of CuBr laser emission with a positive branch cavity of magnifications 100 and 73; electric input power is 1.1 kW**

| Hydrogen pressure, Torr | Average power, W | | Pulse length, ns | | Beam divergence, μrad | | Laser spatial intensity, Wmrad$^{-2}$ | |
|---|---|---|---|---|---|---|---|---|
| - | 2.6 | 2.8 | 32.9 | 29.3 | 108 | 111 | 223 | 227 |
| 0.2 | 5.8 | 6.3 | 39.8 | 44.1 | 79 | 77 | 929 | 1063 |
| 0.3 | 10.0 | 9.25 | 32.0 | 37.1 | 93 | 110 | 1156 | 764 |
| 0.5 | 9.5 | 9.0 | 31.3 | 35.4 | 97 | 112 | 1010 | 717 |
|  |  |  |  |  |  |  |  |  |
| Cavity magnification | 100 | 73 | 100 | 73 | 100 | 73 | 100 | 73 |

Laser beam divergence is at minimum and laser pulse duration is at maximum for 0.2 Torr

H$_2$. The lowest beam divergence is 77 μrad for 44.1ns-long pulse (Table 4.1.2) while for the longest pulse of 52.6 ns it is 84 μrad (Table 4.1.1). The behavior of pulse length and beam divergence for the 100-fold magnification is more logical for laser emission build-up – longer pulse produces lower divergence. Same is not valid for the 73-fold magnification. So in our conditions, longer laser pulse does not always result in lower (*i.e.* better) beam divergence. Average laser power and its spatial intensity peak at 0.3 Torr H$_2$. With hydrogen, laser output spatial intensity is normally ≥ 1000 W.mrad$^{-2}$. The highest spatial intensity is 1389 W.mrad$^{-2}$ and it is for 100-fold magnification at 0.3 Torr H$_2$. With no hydrogen, laser power spatial intensity is 5-7 times lower (200-280 W.mrad$^{-2}$). Its highest value (280 W.mrad$^{-2}$) is again at a magnification of 100. A great benefit of harnessing unstable cavities is seen if compare laser power spatial intensity of a stable-cavity copper vapor laser normally being just a few W.mrad$^{-2}$. The percentage of the laser power having the divergence quoted in the tables ranges from 65% to 80%.

If compare the dependences of average laser power and that of laser spatial intensity on hydrogen pressure it is clear that the latter is not so crucial to hydrogen pressure variations around the optimum of 0.3 Torr (*i.e.* from 0.2 to 0.5 Torr). Thus, any long-term operation changes in hydrogen pressure (and/or average laser power as well) would have a minor impact on such an important beam parameter as laser spatial intensity.

*Conclusions*

The addition of hydrogen to our medium-scale (2cm Dia. x 50 cm) CuBr laser leads to beam divergence reduction and laser power lift-up, which finally yield 5-fold increase in laser power spatial intensity compared with the case of no hydrogen.

Experimentation with this CuBr laser fitted with a simple positive branch unstable resonator proved that CuBr laser radiation of practical (pulse-average) divergence of ~ 80 μrad is easy-feasible.

The better overall high-quality beam performance of CuBr laser is due to the improved radial uniformity of gain and the slow-down of laser pulse build-up. This is a result of a generally total decrease in kinetic process rates as hydrogen is present in the gain medium (Astadjov, 1996).

## *4.2. High-brightness master-oscillator power-amplifier (MOPA) CuBr laser with diffraction-limited throughout-pulse emission*

Copper vapor lasers (CVLs), their variants and the associated master-oscillator power-amplifier (MOPA) systems have quite long ago come to age which can be regarded as their maturity *(cf.* Little *et al.,* 1996 and Little, 1999). Nevertheless, there are still important issues worth paying attention that keep the stand of CVLs in the frontline of visible laser's research while gaining in power, beam quality and utility for many of the applications. One such issue is the characteristics evolving within a laser pulse namely beam divergence, spatial coherence, focal spot size, brightness *etc.*, from a CVL oscillator with commonly used stable and unstable resonators (Hargrove *et al.,* 1979; Omatsu *et al.,* 1992, 1993; Chang, 1994; Coutts, 1995; Prakash *et al.,* 2002). This is due to the high Fresnel number resonator optics used coupled with limited inversion time available. For an MOPA scheme based on such a master oscillator (MO), the output also suffers from same deficiencies of non-constancy of output beam parameters (Prakash *et al.,* 2002). More importantly, the MOPA beam characteristics become highly prone to circuit jitters inherently present in high repetition rate pulsed electronics (Little, 1999). This limits the application potential of such beams. However, a good exception is a CVL MO based on a filtering resonator (Bhatnagar *et al.,* 1989; Pini *et al.,* 1992; Dixit *et al.,* 1993; 2001, Dixit, 2001). It had been demonstrated that a CVL with generalized diffraction filtering resonator (GDFR) (Dixit *et al.,* 1993), an output beam with diffraction-limited divergence constant throughout the pulse (Prakash *et al.,* 2002, 2003) is produced. Dixit *et al.* (2001) and Prakash *et al.* (2003) showed that these constant spatial coherence MOPA features were independent of delay between the oscillator and amplifier, and GDFR CVL MOPA had also produced diffraction-limited high-power pulses with constant characteristics (divergence, spatial coherence, flux *etc.*). This is the perfect pulsed laser light source to be utilized.

For the first time, here (Astadjov *et al.,* 2005) we report the performance of a GDFR CuBr MOPA laser system. A CuBr laser (with added $H_2$) featured a low-temperature (~$500^0$ C) construct, high-repetition rate, compact & sealed off version of CVL with axially peaked gains and low thermally induced wavefront distortion (Astadjov *et al.,* 1988b). It is expected for comparable dimensions of CuBr laser and higher-temperature CVLs, the beam quality performance of the former will be much better. The present study focuses on time-resolved as well as time-averaged beam divergence and brightness characteristics of GDFR CuBr master oscillator and one-stage MOPA system.

Their performance is compared with typical unstable (UR) and plane-plane resonator (PPR) CuBr MO. The CuBr MOPA laser system used is of low-power variety since our attention was mainly concentrated on comparative beam quality features rather than high energy/power output. It is established that GDFR CuBr MOPA system considerably outperforms PPR/UR MOPA in terms of divergence and brightness. Constant divergence, as well as high brightness also almost constant throughout the pulse, was the special features of GDFR CuBr MOPA not available from other resonator geometries.

*Experimental setup*

The basic CuBr MOPA setup is given in Figure 4.2.1. The laser system comprised a master oscillator (MO) and a power amplifier (PA) having respectively bore diameter of 14 mm and 20 mm, and electrode separation of 60 cm and 55 cm. The lasing medium was formed from CuBr, Ne of 18 Torr and hydrogen of 0.3 Torr. Two separate electric power supplies were capable of delivering average power ~1kW each, at a repetition frequency of 18910 Hz. That moderate power input was appropriate just for the basic preliminary research we aimed to carry out. An electronic delay unit was used for managing the delay between the triggering pulses for the MO and PA.

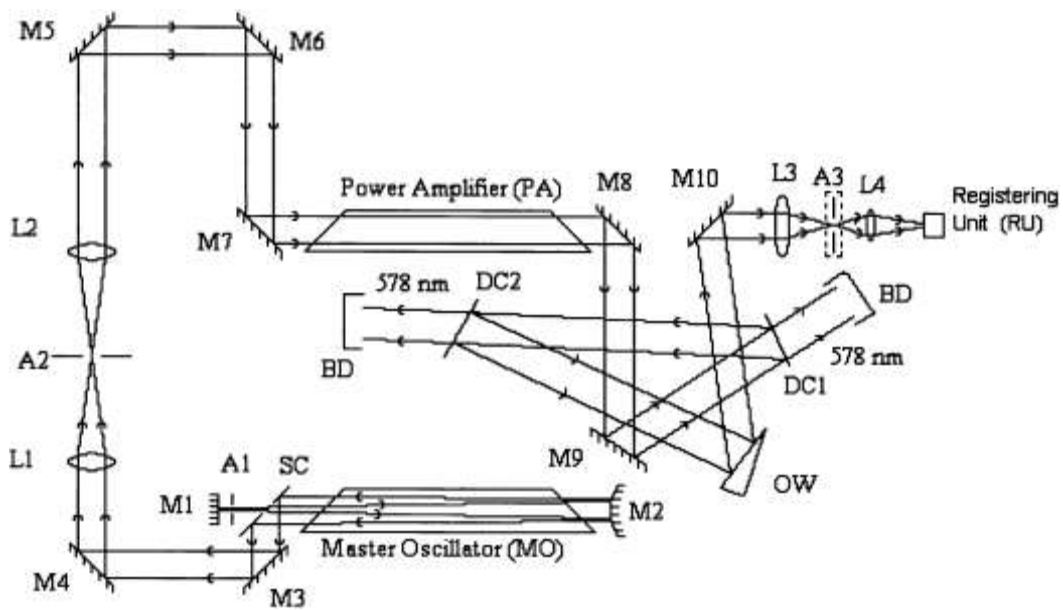

**Figure 4.2.1 Schematic of the optical system of MOPA CuBr laser (herein fitted with GDFR) and the**

measurement arrangement for line separation, attenuation and pulse recording of the green (λ510nm) laser radiation: M1-M10 (except of the convex M2) - plane high-reflection mirrors for both laser lines; SC - scraper mirror; DC1, DC2 - dichroic mirrors; L1, L2 - lenses of intermediate spatial filter; L3, L4 - lenses of far-field filtering optics; A1-A3 - circular submillimeter apertures; OW - optical wedge; BD - beam dumps; RU – registering unit

Three types of resonators for the MO were studied in experiments. The optical setup consists of, namely a stable plane-plane resonator (PPR; $R_1 = R_2 = \infty$, L = 1.3 m ), a confocal unstable resonator of positive branch (PBUR; $R_1 = $ 3 m, $R_2 = $ -40 cm, M = 7.5, L = 1.3 m) and a filtering resonator (GDFR; $R_1 = $ -28 cm, $R_2 = \infty$, diameter of diffracting aperture, A1 at plane mirror is 0.6 mm, L = 1.3 m ) and also the beam guiding, spatial filtering and measuring optics in between oscillator-amplifier/after the amplifier, PA (Figure 4.2.1). The MO beam filtered (by L1-A2-L2, spatial filter) from amplified spontaneous emission, ASE was sent through PA. The optical MOPA measurement setup was being changed and adapted to the specific tasks of each of a number of measurements. For MOPA output divergence measurements, the green radiation (λ=510nm) was selected by means of dichroic mirrors (DC1, DC2) and an optical wedge (OW), and then focused by a lens (L3) of focus length of 100 cm. A magnified image of this far-field spot, made by the relay lens L4, was placed onto the registering unit (RU). The RU consists of either a pin-holed photodiode placed on a precision linear carriage (for time-resolved divergence studies) or a set of calibrated pinhole and power meter for power-in-bucket measurement (time-averaged divergence). As per the demand of accuracy, the image magnification was about 20 for estimation of temporal divergence evolution (lens L4, F = 7.6 cm) and the magnification was about 60 (lens L4, F= 3.7 cm) for average divergence estimation. The MOPA beam measurement with/without ASE (amplified spontaneous emission) was carried out by removing/inserting a proper-sized pinhole, A3 within an L3--L4 spatial filter. For measurements of only MO characteristics, everything remains same except putting off the laser power amplifier (PA).

*Results and discussion*

As the most of the performance properties of an MOPA system are normally predetermined by the MO output properties, some measurements were done with the MO only. The temporal evolution of beam divergence during the laser pulse was measured by the technique described by Stoilov *et al.* (2000). It was done for PPR, UR, GDFR MO as well as GDFR MOPA with and without external ASE filtering. This method is based on radial scanning of magnified (x20) far-field distribution and recording pulses at chosen radial locations. Then the waveforms recorded were numerically processed to get a 3D pattern of laser beam profile evolution. Instantaneous beam divergence was estimated from these scans through measuring the focal spot size (at $1/e^2$ point) and dividing it by the distance from the focusing element. That was repeatedly done at different time slices of the 3D beam profiles. The instantaneous divergence is given in terms of diffraction-limit divergence, DL. In our experiments DL is calculated for a flat near-field profile - DL=2.44 x $\lambda$/ID , where ID is beam diameter (proofed sufficiently equal to the inner diameter of laser tube) and this fits well with the study. Divergence measurements were carried out for the more intense green ($\lambda$=510nm) laser line. The diffraction-limited divergence ($\lambda$510nm) is 89 μrad and 62.3 μrad for the MO of 14mm ID and MOPA of 20mm ID, respectively.

In Figure 4.2.2 we present the temporal evolution of beam divergence within a laser pulse from PPR, PBUR, and GDFR CuBr MO. (Hereafter, all timing is given in relation to the start of the discharge current pulse.)

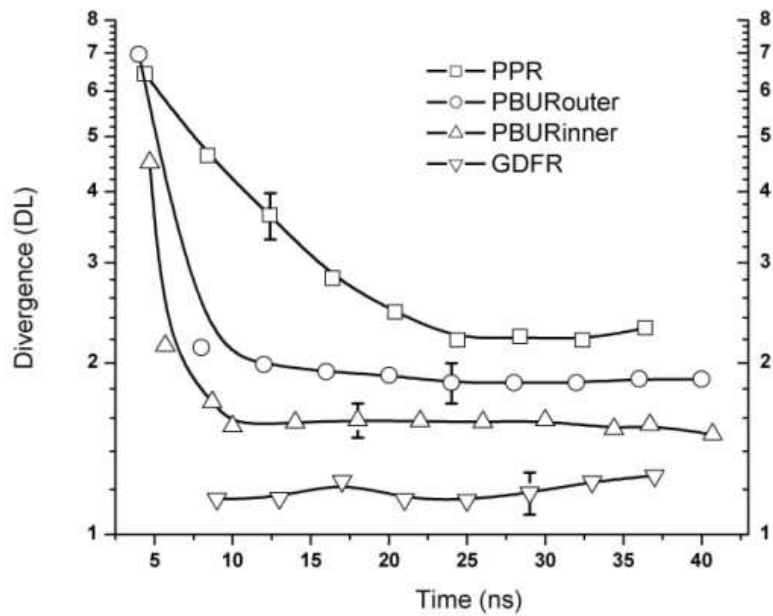

**Figure 4.2.2 Temporal throughout-pulse evolution of laser beam divergence of CuBr MO fitted with PPR, PBUR, and GDFR**

As can be seen, the PPR/PBUR divergence has high values at the beginning of the pulse.

Afterward follows a decrease down to steady values, which stay constant till the end of the pulse. The divergence drops much faster and to lower values with PBUR compared to PPR. For PBUR ~10ns are enough to get to a steady value of 1.6-1.8 DL, while PPR goes down to 2.2 DL for ~25ns. The faster fall of divergence in PBUR is consistent with the geometrical picture (neglecting diffraction) of mode build up in UR that divergence is reduced by a factor of resonator magnification, M (= 7.5) after every round trip is completed (Eggleston, 1988, Dixit *et al.,* 1994). However, as Dixit *et al.* (1994) stated, for PPR (M=1) divergence reduction in successive round trips, n is due to the increased effective resonator length, nL by a much smaller factor of n = 1, 2, 3,…. However diffraction is always present and is much more effective in later round trips when circulating radiation is more coherent. The diffraction slows down the rate of reduction in divergence as laser pulse progress, ultimately reaching close the diffraction limit (1DL). This slowing down is adequately reflected in later part of PPR/UR CuBr MO pulse. In contrast, GDFR starts lasing 3-4 ns later but the divergence exhibits no change throughout the pulse. It is as low as 1.2 DL. This is so, as right from the beginning only central uniphase lobe of far-field (*i.e.* definition of DL beam) is only allowed to build up as the GDFR mode.

In Figure 4.2.2 two evolution traces are given for PBUR (Note: this resonator was of off-axis type). As seen, the beam divergence of PBUR had two areas of slightly differing behavior: the outer part of the beam had a divergence of higher magnitude throughout the pulse with a final value of 1.9 DL while the inner part got 1.5 DL at the end. Similar patterns are observed with PPR but they are not shown in separate traces in Figure 4.2.2. As an illustration of that difference in the radial profile of PPR output beam, laser pulse waveforms at two different places (spots) were recorded and plotted in Figure 4.2.3.

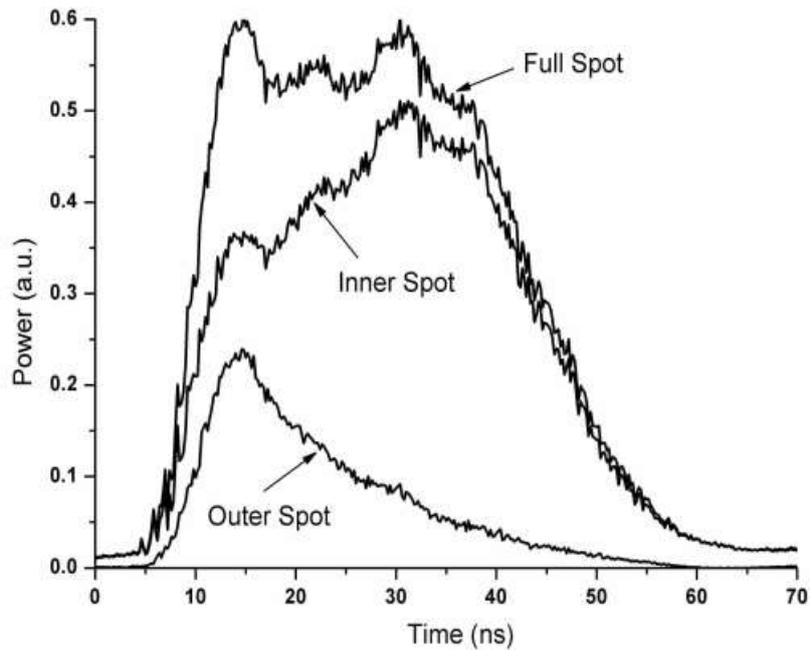

**Figure 4.2.3 Laser pulse waveforms of different parts of far-field PPR beam cross section**

The outer spot (annulus, in reality) peaks earlier than the inner spot of the PPR beam. It might be the main contributor to the high divergence during first 25ns. The power concentrated in these two areas was measured and the ratio of laser power amounted to 4 in favor of inner spot. So we might consider the outer part of the laser beam a culprit for the high divergence during first 25ns in both cases – PPR and PBUR CuBr master oscillator.

The next Figure 4.2.4 depicts the temporal divergence of the GDFR-fitted MOPA in two cases of with and without amplified spontaneous emission, ASE. The divergence of MOPA output containing ASE is 1.9 DL during the pulse. When ASE is spatially filtered, the throughout-pulse divergence goes down to the diffraction limit - it is just 1.06 DL.

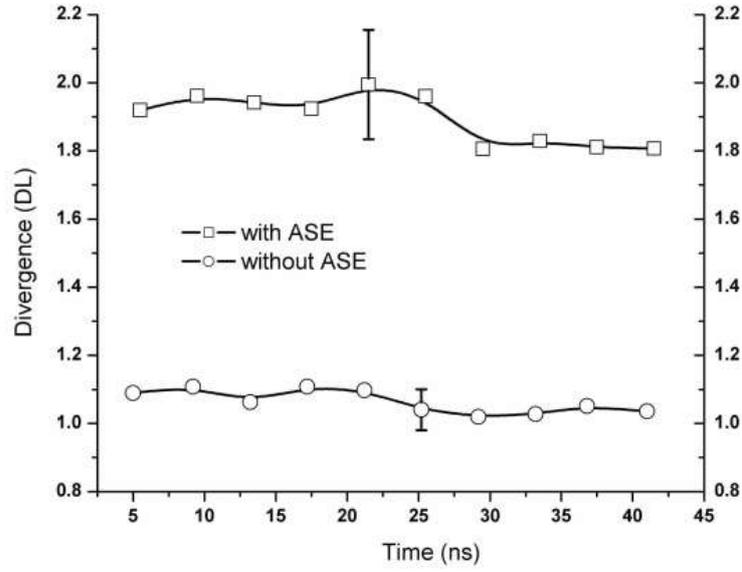

**Figure 4.2.4 Temporal divergence of the GDFR MOPA: with ASE and without ASE (eliminated by an output spatial filter)**

As we discussed in a paper (Stoilov *et al.,* 2000), the laser power density, $P_s$ obtained in the focal spot is an important laser property. If F is the focal length of output focusing optical system, it is equal to

$$P_s = \frac{4}{\pi F^2} P/\theta^2 \ .$$

Then we specified a quantity called "spatial intensity" as $P/\theta^2$, where P and θ are the power and the full divergence of the laser beam, respectively. Being an inherent laser emission property, it concerns the focusability of laser radiation and by definition can be recognized as brightness. In this study, we shall use the term "brightness" for it. Moreover, we shall show how brightness evolves during the laser pulse. So, the brightness, B is considered as time-dependent *i.e.* $B(t) = P(t)/\theta^2(t)$. Here, P (t) is the isotropic laser pulse power calibrated by means of the average output of λ510nm. This quantity spotlights the advantages of GDFR in comparison with other types of laser resonators usually utilized in CVLs. Also, we can specify the "average brightness", $B_a$ of the MOPA resonator configuration as the integral of B(t) over the pulse length, τ

$$B_a = \frac{1}{\tau} \int_0^\tau B(t)dt \ .$$

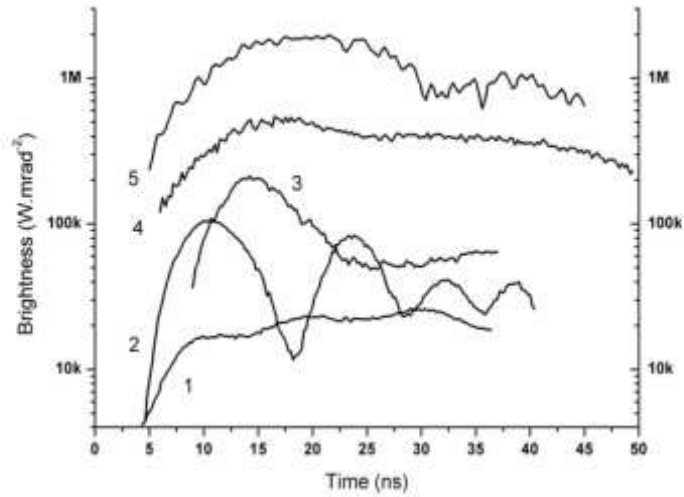

**Figure 4.2.5 Time dependence of brightness for different MO resonators without amplification and for GDFR MOPA output: with ASE and without ASE (eliminated by an output spatial filter): 1-PPR, 2-PBUR; 3-GDFR MO, GDFR MOPA with ASE; and 5—GDFR MOPA without ASE**

In Figure 4.2.5 we plotted the time dependence of B (t) for different MO resonators without amplification, 1-3, and for two cases of the GDRF MOPA output: one with ASE, 4, and another, 5, with ASE eliminated by the output spatial filter. Figure 4.2.6 illustrates the average brightness (Ba) and maximum instantaneous brightness (Bm) of different MO and MOPA resonator configurations in a graph and Table 4.2 gives their numerical values.

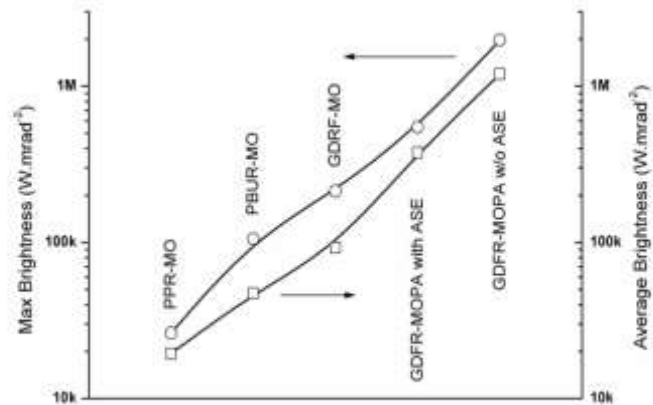

**Figure 4.2.6 Average brightness and maximum instantaneous brightness of different MO and MOPA resonator configurations**

**Table 4.2 Average and maximum instantaneous brightness of different MO and MOPA resonator configurations**

| Laser resonator configuration | Maximum brightness, kWmrad$^{-2}$ | Average brightness, kWmrad$^{-2}$ |
|---|---|---|
| MO-PPR | 26.3 | 19.5 |
| MO-PBUR | 105 | 47.1 |
| MO-GDRF | 213 | 92.4 |
| MOPA-GDFR with ASE | 546 | 375 |
| MOPA-GDFR without ASE | 1980 | 1200 |

The presented results should be regarded rather as qualitative, since as B(t), (and consequently, $B_a$) was not calculated strictly for lack of detailed information about the angular distribution of laser power, which was very difficult (sometimes impossible) to get in our experiments. Nevertheless, they are indicators of the practical capabilities of different MOPA configurations. Because the average laser output of λ510nm did not vary too much with the different cavity configurations (the highest power was with PPR - 0.77 W and the lowest power was with GDFR - 0.5 W), B (t)-dependence reflects the principal differences in the instantaneous divergence between cavity configurations. While for PPR, $B_a \approx$ 20kWmrad$^{-2}$ and the maximum instantaneous brightness, $B_m \leq$ 27kWmrad$^{-2}$, for PBUR we have $B_a \approx$ 47kWmrad$^{-2}$ and $B_m \approx$ 105kWmrad$^{-2}$, and for GDFR - $B_a \approx$ 92kWmrad$^{-2}$ and $B_m \approx$ 213kWmrad$^{-2}$. Thus, even for the non-amplified MO radiation, the brightness of GDFR is as much as twice higher than the brightness of PBUR. The brightness of GDFR-MOPA radiation is of an order of magnitude higher: with ASE it is $B_a \approx$ 375kWmrad$^{-2}$ and $B_m \approx$ 546kWmrad$^{-2}$, and without ASE - $B_a \approx$ 1.2MWmrad$^{-2}$ and $B_m \approx$ 2MWmrad$^{-2}$. This is an outcome not only of the higher levels of laser power (despite the MOPA output is of an order higher than that of MO, in fact, the MOPA power varied within 20% with different MO resonators) and reduced beam divergence but it comes from the permanently-very-low divergence of laser emission throughout the laser pulse. This is an intrinsic characteristic of GDRF MOPA laser radiation, which features it out of the other MOPA resonator configurations. When the MOPA output is spatially filtered (*i.e.* without ASE), the GDRF MOPA laser radiation is not only of the brightness of factor of 2-3 higher. Because the ASE background is removed, it no longer affects the adjacent areas of laser spot produced by the focused low-divergence laser beam.

In Figure 4.2.7 we give the relative power-in-the-bucket of GDRF MOPA *without* output spatial filtering of laser radiation *i.e.* ASE is present in the low-divergence beam as a high-divergence background.

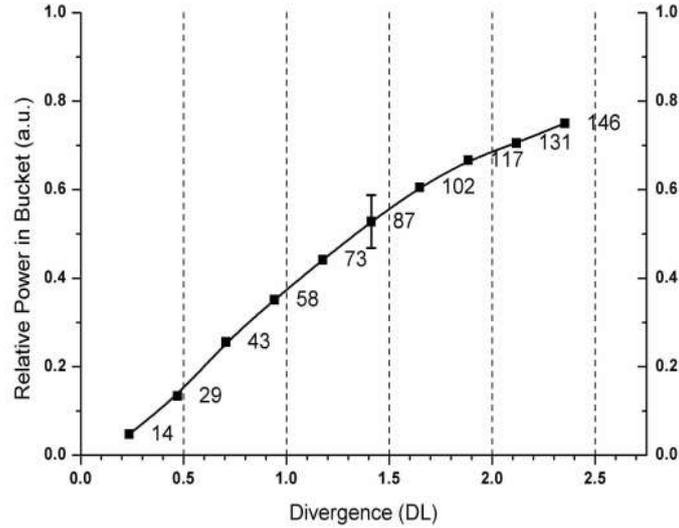

**Figure 4.2.7 Relative power-in-bucket of GDRF MOPA *without* output spatial filtering of laser radiation *i.e.* ASE is present as a high-divergence background. The divergence (in μrad) at the measurement points is shown nearby**

The relative power-in-the-bucket is the fraction of power within a given divergence angle. It is evident that the fraction of the diffraction-limited (62 μrad) beam is ~ 0.4, whereas of $\theta \leq$ 100 μrad is ~ 0.6. As mentioned in the beginning, the MOPA system has not been projected for high laser output. The average MOPA output power was only ~5W on λ510nm, which means that ~2W were within the diffraction-limited divergence. However, the low-divergence fractions of laser radiation are high enough to be a factor in the justified implementation of the GDFR MOPA.

*Conclusion*

We demonstrated a generalized diffraction filtered resonator as a cavity for the master oscillator in a CuBr MOPA laser system and analyzed its performance. A diffraction-limited high-brightness emission is obtained with the novel MO resonator, which manifested its exceptional properties as a light source for power amplification in an MOPA laser assembly.

## *4.3. Propagation factor M² of MOPA CuBr laser radiation*

Between the middle of the 1990s and first years of XXI century, a number of papers have been devoted to the analysis of copper vapor laser radiation. Laser radiation properties are qualified as different optical resonators have been applied in single stand-alone lasers or in MOPA (master oscillator, power amplifier) laser systems (Dixit *et al.,* 1994; Chang, 1994, 1995; Chang *et al.,* 1996; Salimbeni, 1996; Dixit, 2001 *etc*.). The free-space MOPA or their fiber-connected alternative (Coutts, 2002) give diametrically different laser output with respect to the beam divergence, as the latter claims uniform (flat-top) beam profiles (*cf.* Brown *et al.,* 1996; Guyadec *et al.,* 1996; Brown *et al.,* 2001). Assembled by Astadjov *et al.* (2005) an MOPA CuBr laser with GDFR (generalized diffraction filtering resonator) has already demonstrated unique properties.

The present study (Astadjov *et al.,* 2007a, 2007b) is an extended and deeper investigation of the relations of the near and far optical fields of our GDFR MOPA CuBr laser system, applying also the specifications of ISO11146 (ISO, 2005). The beam propagation factor, $M^2$ of the master-oscillator power-amplifier (MOPA) CuBr laser emission compliant with ISO 11146 is studied methodically. Statistical parameters of 2D intensity profile of the near and far fields of MOPA laser radiation are measured by a beam analyzing technique as functions of timing delay between master oscillator (MO) and power amplifier (PA). As a matter of statistics, a dozen of samples (for each point) were averaged and their statistical parameters were derived. For the first time, the influence of the gas buffer which causes the radiation profile to change from annular to top-hat and Gaussian-like, and the polarization of light on CuBr laser beam focusability (namely, $M^2$) were under a systematical investigation.

### *The method of ISO11146*

Here we shall outline the method of ISO11146 in brief. ISO11146 introduces a quantity, $M^2$ ('m-squared') which is referred as "times-diffraction-limited factor". It is defined as

$$M^2 = \frac{\pi}{4\lambda F} Dd$$

D is the diameter of the input beam and d is the diameter of the focus spot of a lens of focal distance, F. The point is that the diameters are specified as being equal to 4σ (4 x sigma). Sigma is the second momentum of distribution along the major/minor axis of beam intensity profile ellipse. $M^2$ is also called 'propagation factor' since it does not change with distance as the light beam propagates. According to Siegman (1997, 2004), $M^2$ is the only invariant regardless beam intensity distribution.

For the calculation of the second momentum of intensity distribution (aka sigma) we employed Spiricon LBA software (v.3.03) and a CCD camera (FINE model) which allows fast reliable in-line measurement of a number of statistical parameters of laser beam profile. In terms of Fourier optics, the optical field in front of the lens and the optical field in the lens focus are referred as near field and far field, respectively.

*Experimental setup*

Two MOPA CuBr laser systems with GDFR-fitted master oscillator were examined. The first system (MOPA1) was the same as we reported in Astadjov *et al.* (2005). It comprised a master oscillator (MO) and a power amplifier (PA) having respectively bore diameter of 14 and 20 mm and electrode separation of 60 and 55 cm. The second system (MOPA2) had an MO tube of a different design but electrode separation was same - 60 cm. The total tube length was 123 cm which ensured longer zones at tube ends to reduce contamination of the optical windows coming from the active discharge volume of the laser tube. The optical laser windows were set at Brewster angle which decreased the optical losses in the laser cavity and produced polarized laser output. This MO tube was of 20 mm diameter and had *no diaphragms* inside the tube. The laser beam diameter was ~14 mm just after the Brewster windows. MO and PA were optically folded by mirrors and a mirror telescope. The telescope collimated and got the MO output beam diameter matched with the PA active zone diameter. The optical connection path was long enough (~10m) to prevent the MO from the PA feedback radiation. A schematic of the optical arrangement of MOPA CuBr laser system is plotted in Figure 4.3.1.

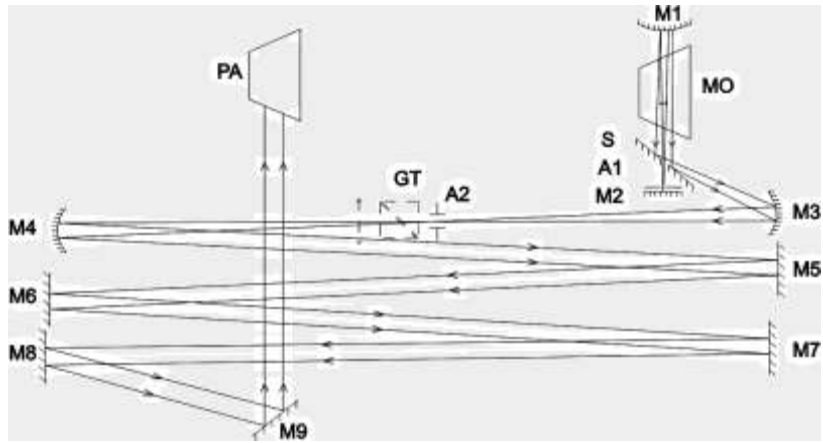

**Figure 4.3.1 MOPA laser optical setup for the feed-in of PA from MO. The high-reflection mirrors for both laser lines are M1—a convex mirror of radius of 28 cm, M3, and M4—concave mirrors of radii of 115 cm; M2 and M5-M9—planar mirrors; S—scraper mirror; A1 and A2—circular submillimeter apertures and GT—Glan-Taylor prism. The arrows after GT show the polarization plane**

The measurement setup is shown in Figure 4.3.2. It had several functions - attenuation of the MOPA emission, separation of the green (λ510nm) laser line and projecting it onto a screen with minimum distortions. The multiplied image of the optical field was then recorded with a CCD camera. The in-line analysis was done by a beam analyzing program which computed the sigma parameter for the calculation of $M^2$-factor. The program (Spiricon LBA software, v.3.03) ran in sample-averaging mode and provided a full statistics of each sampling series of minimum 10 samples.

The $M^2$-factor measurements in this study are time-averaged over the laser pulse (*cf.* time-dependent measurements during the laser pulse reported in Astadjov *et al.,* 2005). The dependence of $M^2$ on the time delay (abbr. delay) of electric excitation pulses of MOPA components (*i.e.* MO and PA) was studied. The timing between MO and PA is carried out by a digital triggering electronic unit. The electronic triggering delay of laser excitation pulses ranges from nanoseconds to microseconds with steps of 3ns (minimum). The high-voltage thyratron pulsers are separate for each tube. The excitation schemes are well-known and similar to other pulsed lasers (excimers, nitrogen laser, *etc.*). The excitation current pulses were measured by Rogowski-coil probes. For laser pulses, a fast photodiode was used. All pulses were recorded with a digital oscilloscope (Tektronix TDS 420A). The average laser power was measured by a Scientech power meter.

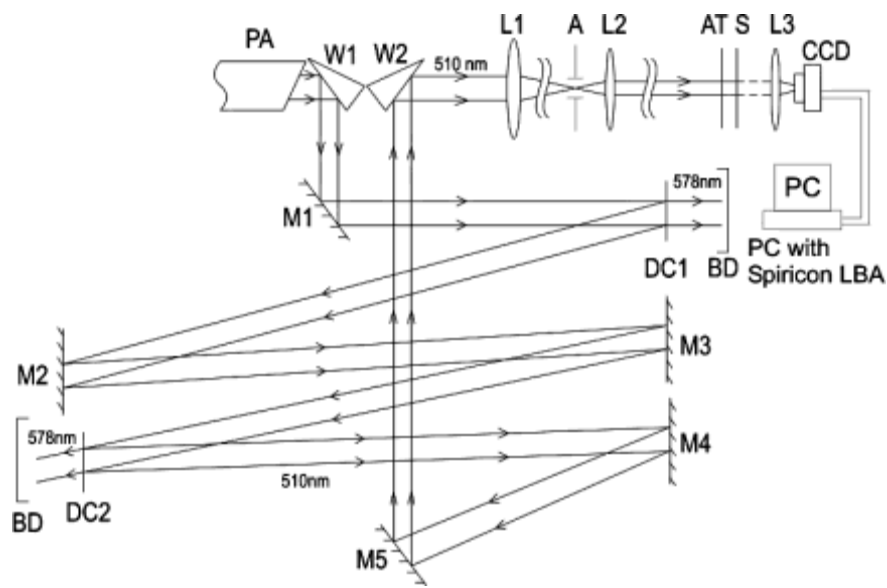

**Figure 4.3.2 Optical measurement setup of MOPA laser for line separation, attenuation and far-field profiling of the green (λ510nm) laser radiation: M1-M5—plane high-reflection mirrors for both laser lines; DC1, DC2—plane dichroic mirrors; L1-L3—lenses of far-field profiling/filtering optics; A—circular submillimeter aperture; W1, W2—optical wedges; BD—beam dumps; AT—neutral attenuator; S—semi-transparent screen and CCD—CCD camera**

In the 1980s, we established that hydrogen additives lead not only to a great increase in laser power but the laser beam profile pattern changes from annular to top-hat and Gaussian-like (Astadjov *et al.,* 1985, 1988a). Here, within MOPA1 experiments, we examined that influence of gas buffer composition on laser beam focusability and $M^2$. Light polarization effect on $M^2$ was investigated within MOPA2 experiments. MOPA2 tubes had a buffer gas containing a mixture of neon and hydrogen at pressures of 17 Torr and 0.3 Torr, respectively.

*Results*

The dependence of $M^2$-factor and laser power in MOPA1 on delay is given in Figure 4.3.3. The laser power dependence (section B) can be referred as the 'gain curve' of MOPA. As seen in section A, $M^2$ is not constant during the time when gain occurs. It decreases with the delay increase. In the case of 'nonhydrogen' buffer gas (when annular emission is prevailing) $M^2$ is much higher but goes down to ~5 at the end of gain time period. In the 'with-hydrogen' case (neon of 17 Torr and hydrogen 0.3 Torr) of the buffer, gas $M^2$ is ~7 at small delays and ~4 at large delays. The 'with-hydrogen' gain curve is shifted ~40 ns to larger delays where $M^2$ is low and laser power is highest. The 'nonhydrogen' case is unlike. The maximum of laser power appears at quite high $M^2$. At the same time, the shapes of gain curves are similar with some small level variations as follows. So at the basis (zero level) the duration of 'nonhydrogen' gain is higher - 170ns vs. 150ns. At 50% level the gain durations are equal (~70-80ns) and at 80% level, the 'with-hydrogen' gain is longer - 45ns vs. 35ns.

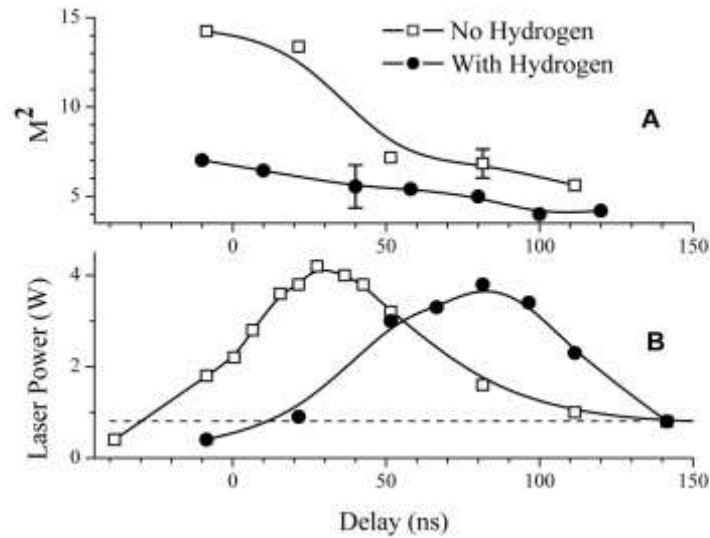

**Figure 4.3.3 $M^2$ (section A) and laser power (section B) versus delay of excitation pulses between MO and PA in gas-dependence experiment. The horizontal dash line is MO power**

The influence of light polarization on $M^2$-factor was studied in MOPA2 where the buffer gas was of 'with-hydrogen' type (*i.e.* neon of 17 Torr and hydrogen 0.3 Torr). Due to the Brewster angle windows of MO, there is anisotropy of light polarization in the MO radiation. Using a polarizer the vertical and horizontal polarization components were estimated to be in the proportion of 4 to 1. With a Glan-Taylor prism, we were able to select only the vertical or the horizontal polarization component. Either polarization component had an identical impact on $M^2$. Since the vertical component is more intensive we plotted the experimental results with it in Figure 4.3.4. From section A (if discarding the curves variations within the time limits of gain) one can infer that independently of degree of polarization, $M^2$ is ~6 at the beginning and drops to 3 at the end of gain curves. The difference occurs around the laser power maximum (*i.e.* during the highest gain of MOPA). Here, $M^2$ of the fully-linearly-polarized beam is 25% less than $M^2$ of the partially-polarized MO beam. We can propose that the horizontal component of partially-polarized seed beam is the culprit for higher $M^2$ (i.e. higher divergence).

Section B of Figure 4.3.4. is about laser power output (aka gain curves) of MOPA in light polarization experiment. The gain curve of partially polarized (Ver:Hor=4:1) seed beam is wider and higher in comparison to gain curve of fully-linearly-polarized (vertical) seed beam. Clearly, the changes in gain curves (section B) due to the polarization degree of seed beam are not significant and we have two options to treat them. One option is to perceive the differences as produced by the horizontal component of partially-polarized seed beam. The amplified horizontal component can contribute to the length and height of gain curve in a manner that we experimentally observed. So for the partially polarized beam FWHM is ~80ns and for the vertically-polarized seed – ~65ns. Laser power from the partially-polarized seed beam exceeds laser power from the vertically-polarized seed for all delays. This surplus power due to the amplified horizontal component is ~0.5W but at some time delays, namely 90 ns and 150 ns, this power increase is of 0.7-0.8 W which as relative extra power percentage amounts to 30-50%.

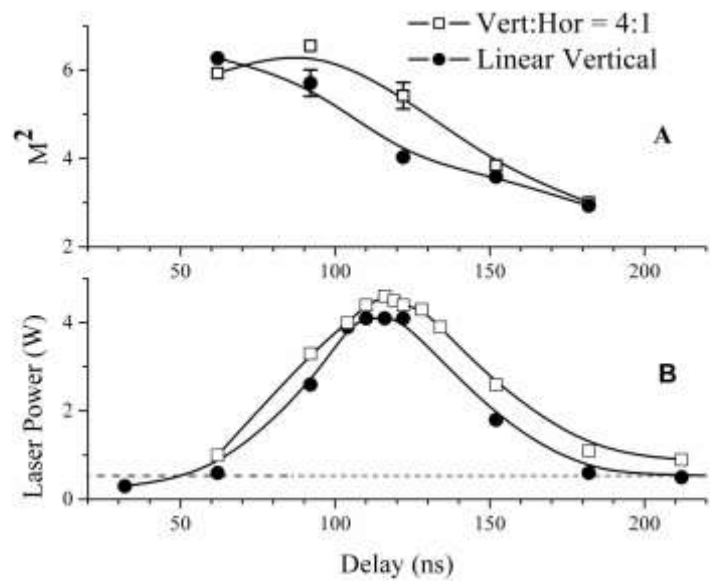

**Figure 4.3.4 $M^2$ (section A) and laser power (section B) versus delay of excitation pulses between MO and PA in light polarization experiment. The horizontal dash line is MO power**

However, the second explanation can be simpler. The Larger envelope of the partially-polarized gain curve could be the result of its higher input power compared to vertically-polarized seed beam. We should note that although our efforts, fluctuations of the laser power of ~ ±20% are unavoidable in real experimentation.

The brightness of laser emission is an important characteristic for most of the laser applications. Astadjov *et al.* (2005) already studied the time dependence of brightness during the laser pulse of MOPA fitted with different resonators. Here we define the brightness, $B_M$ with reference to $M^2$: $B_M = P / (M^2)^2$, where P is the average laser power. This brightness can be one more way for comparison of MOPA1 and MOPA2. As seen in Figure 4.3.5 in the 'with-hydrogen' case (neon of 17 Torr and hydrogen 0.3 Torr) of buffer gas, $B_M$ is 3-5 times higher than the 'nonhydrogen' $B_M$ (*i.e.* of annular emission). If the light is polarized the brightness tends to increase further. The brightness of linearly-polarized beam is at least 40% higher than that of partial or non-polarized beams. So, despite the higher laser output (≥10%), the worse divergence (i.*e.* higher $M^2$) of the partially polarized beam (see Figure 4.3.4), linearly-polarized beam wins in brightness. However, the $B_M$-curves of polarized light tend to be narrower (*i.e.* the operational delay range is shorter).

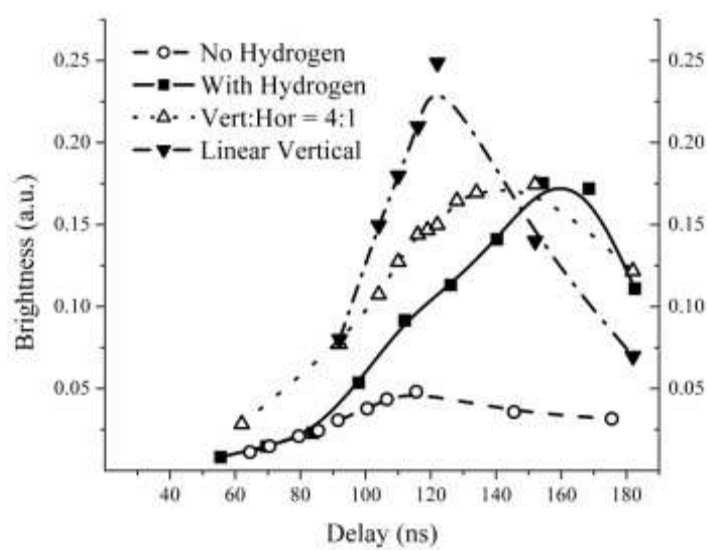

**Figure 4.3.5 Brightness of MOPA laser versus delay of excitation pulses between MO and PA in gas dependence and light polarization experiments**

Now, we are going to concern some aspects of gain characteristics of our MOPA CuBr laser system which are not the main target of the recent study but could not be neglected. They are important because they can affect findings and determine their interpretation. Here, in brief, we present the dependences of MOPA laser power output (abbr. output power) on MOPA laser power input (abbr. input power), and their ratio (*i.e.* output power/input power which defines MOPA amplification) on input power. As an output power, we use the highest average power obtained at the optimum delay and as an input power is the average MO power measured before MOPA entrance. In Figure 4.3.6 we plotted output power and amplification versus input power for unseparated (green λ510nm and yellow λ578nm) laser lines and for green λ510nm line only.

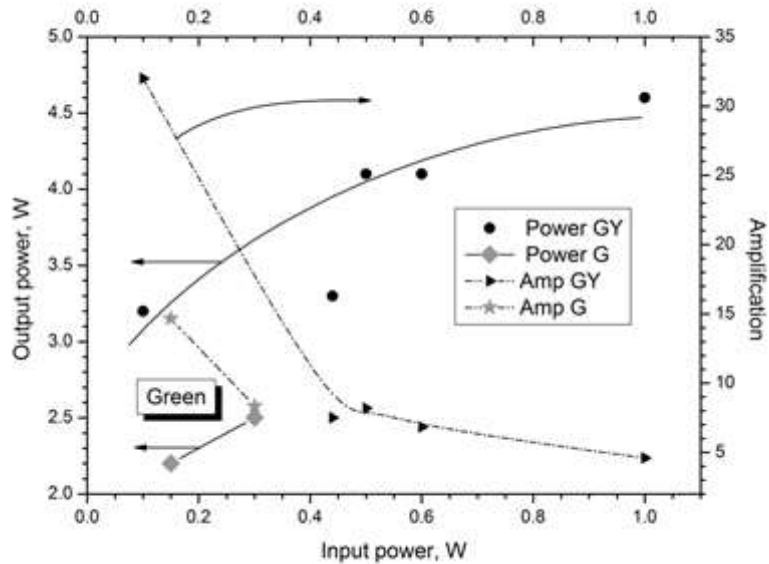

**Figure 4.3.6 MOPA laser output power and amplification versus input power for unseparated (green λ510nm and yellow λ578nm) laser lines (GY) and for green λ510nm line only (G)**

As seen for unseparated lines, at the low input (~ 0.1-0.4W), amplification is quite high – 30-10. However, for input power ≥ 0.5 W amplification is 7-8 and finally tends to 4-5 which is a sign of saturation of MOPA amplification. The green line is shown only at low input powers because of lack of other certain experimental data. Nevertheless, the green line must follow same tendency and exhibits saturation of amplification too. Since experiments are at MOPA laser power of levels ≥4W (at maximum!), we conclude that all (if not - majority of them) are conducted at conditions of amplification saturation (or very close).

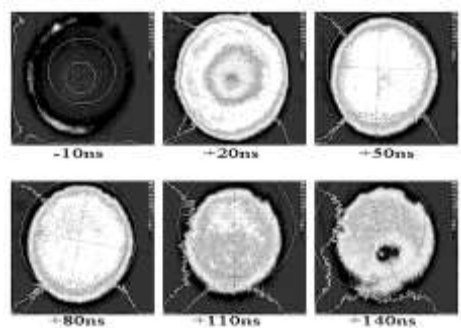

**Figure 4.3.7 Six near-field patterns of MOPA laser radiation at different delays of excitation pulses between MO and PA in the experiment with hydrogen additives**

*Discussion and conclusions*

The $M^2$-factor measurements revealed that $M^2$ depends on the delay *i.e.* the moments of time the MO laser pulse passes through the PA gaining volume. This necessitated more attention to be paid to the near-field properties. In the nonhydrogen case of buffer gas, the net laser output is normally annular. With hydrogen additives, despite the fact that the net laser output is not annular, during the late stage of the gas discharge in PA (which is the early part of the gain curve *i.e.* small delays) gain may cancel and then, light absorption takes place in the beam central area. Figure 4.3.7 shows this behavior of the near-field patterns at different delays with MOPA1 in the 'with-hydrogen' case. The fall in gain in the center of PA can be the result of many factors: faster depletion of copper ground state in the center area, bromine (or hydrogen bromide) radial inhomogeneity, *etc*. The last implication comes from casual observation of discharge instabilities at small delays (which corresponds to late stages of the pulsed gas discharge in PA). The effect of polarization of MO laser input to PA is subtle and yet, unclear. The explanation of polarization influence needs additional experiments and is out of the scope of this paper.

Comparing with MOPA1, the performance of MOPA2 is better. The reason is probably the partial polarization due to the MO Brewster windows. The polarization effect is furthermore manifested with fully polarized beams as a Glan-Taylor prism is employed. Concerning the $M^2$ values experimentally obtained, we should note that the minimum value of $M^2$-factor is 1 and it can be achieved with perfect Gaussian beam (as $TEM_{00}$) while the theoretical value of $M^2$ for a top-hat beam profile is 1.9 (Siegman, 1997, 2004). Our lowest value is $M^2=3$, which is as much as ~1.6 times the minimum value for a top-hat beam profile. Recently, $M^2=4.9$ was reported in a HyBrID copper laser by Giao *et al.* (2006) which is much less than the best propagation factor reported here.

Measurements similar to ours, in elemental copper vapor lasers by Prakash *et al.* (2003) have shown the constancy of beam divergence over a range of ± 24ns. Under our conditions, for the same range of ±30ns (that corresponds to the FWHM of PA gain curve), the variation of $M^2$ is ±20%. If presuming that the normal experimental error/variation for such kind of measurements is of the same order, we can conclude that if not identical, yet MOPA CuBr laser with GDFR has similar behavior like MOPA based on elemental copper vapor lasers.

As a final conclusion, we can say that MOPA CuBr laser systems (having an optimized neon-hydrogen gaseous buffer) with GDFR-fitted master-oscillator in which linearly-polarized light is used to feed the power amplifier can produce beams of $M^2$ as low as of 3-4.

## *4.4. Optical coherence of CuBr laser radiation. Novel reversal shear interferometer*

The CuBr laser is a well-established and proven member of the copper vapor lasers (CVL) family. Over 30 years[2] of existence have demonstrated its excellent properties. The CuBr laser features are well-known: simple construction, short warm-up time, long-time sealed-off operation, high power, record efficiency *etc*.

Coherence is one of radiation properties of great importance for holography, harmonics generation *etc*. Coherence has been studied for years. Early experiments (Zemskov *et al.,* 1978; Bakiev *et al.,* 1991; Chang, 1994; Omatsu & Kuroda, 1996 *etc.*) with elemental CVL showed the non-constancy and/or low beam coherence of usual resonators (stable or unstable). CuBr laser systems are of much lower operating temperature (~1000 degree less) compared to elemental CVLs and coherence of copper halide lasers should be better (Kreye & Roesler, 1983). Geng (1996) studied coherent properties of CuBr lasers aimed to high-speed holograph. *Dixit et al.* (1993) introduced a new type of filtering resonator, the generalized diffraction filtered resonator and Prakash *et al.* (2003) reported that it is free from volume utilization limitation of the self-filtering unstable resonator and offers diffraction-limited beam with steady coherence throughout the laser pulse.

Here we present results from our research on CuBr laser beam coherence (Astadjov *et al.,* 2006a). We compared three types of laser cavity: stable plane-plane resonator (PPR), the confocal unstable resonator of the positive branch (so called PBUR) and generalized diffraction filtered resonator (GDFR). With the MOPA system, only a GDFR was utilized. We greatly improved CuBr laser beam coherence by a special design of the MO laser resonator, namely the GDFR. With it, a diffraction-limited beam divergence can be easily obtained throughout the laser pulse. Since coherence (strictly, *spatial*) is in inverse relation with the beam divergence, decreasing the latter we increase the former (Svelto, 1984). Thus, it is clear that any improvement (*i.e.* decrease) of beam divergence will lead to an improvement (*i.e.* increase) of beam coherence. Because of the pulsed character of copper vapor laser radiation, the constancy of divergence at diffraction-low values *as long as possible* during each self-terminating laser pulse is of principal importance. Moreover, the longer duration of CuBr pulsed emission implies improvement of temporal coherence too.

---

[2] The work was published in 2006.

The temporal evolution of beam divergence during the laser pulse was measured by the technique described in (Stoilov *et al.,* 2000). The far-field intensity distribution was obtained at the focal plane of a focusing lens (or mirror). By another lens, a magnified focal plane image was relayed. This magnified image was radially scanned by a pin-holed photodiode. Then the waveforms recorded were numerically processed to get a 3D pattern of laser beam profile evolution. Instantaneous beam divergence was estimated from these scans through measuring the focal spot size (at $1/e^2$ point) and dividing it by the distance from the focusing element. That was repeatedly done at different time slices of the 3D beam profile.

The instantaneous divergence is given in terms of diffraction-limit divergence, DL. It is calculated for a flat near-field profile (DL=2.44 x $\lambda$/ID), which fits well for our experiments. Divergence measurements were carried out for the more intense green ($\lambda$510nm) laser line rather than the yellow ($\lambda$578nm) one because both laser lines have similar behavior (Dixit *et al.,* 2003). The diffraction-limited divergence is DL=89 μrad and DL=62.3 μrad for the MO of 14mm ID and MOPA of 20mm ID, respectively, at $\lambda$=510nm. A spatial filter was put between MO and PA to prevent from optical feedback and minimize the MO amplified spontaneous emission (ASE) input into PA. An aperture was also placed after the PA as a spatial filter in the MOPA operation experiment to compare the final PA output laser emission with ASE and without ASE.

*Experimental setup*

The basic CuBr MOPA optical setup is given in Figure 4.4.1. It consists of CuBr master oscillator (MO), a beam folding (M4-M7) and ASE filtering optics (L1-A2-L2), and a CuBr power amplifier (PA). Between mirrors M8 and M10, the two laser lines were separated and attenuated by means of two filtering dichroic mirrors (DC1 and DC2) and an optical wedge (OW). The green line was focused by a lens (L3) of focus length of 100 cm. In most of the cases, a magnified and relayed image was produced by a lens (L4) of focus length of 7.6 cm. That part of the setup needed to get adjusted over and over again, so the magnification was not constant but varied ~20. For ASE elimination in the MOPA output, an aperture (A3) is inserted forming this way an output spatial filter (namely, the combination L3-A3-L4)**.** For bad luck, an additional detrimental factor deteriorating experimental data was the inappropriate laboratory environment having a pretty high level of mechanical vibrations. In spite of that, in some experiments (MOPA with GDFR) the diffraction limit of beam divergence was actually reached.

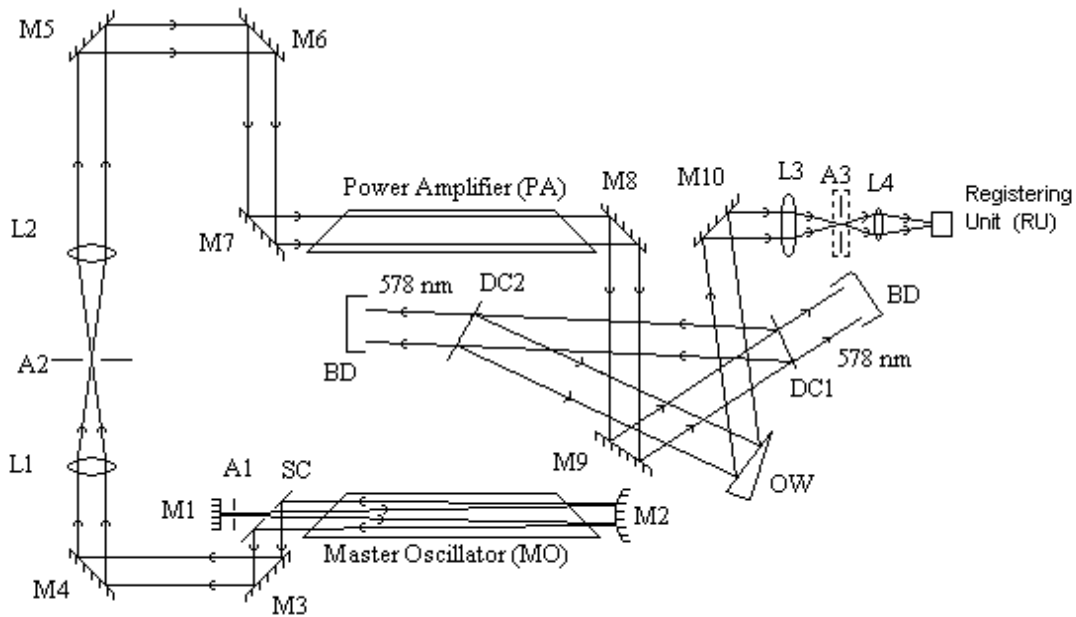

**Figure 4.4.1 Schematic of the general optical system of MOPA (herein fitted with GDFR) and the measurement optics for line separation, attenuation and far-field profiling of the green (λ510nm) laser radiation: M1-M10 (except of the convex M2) - plane high-reflection mirrors for both laser lines; SC - scraper mirror; DC1, DC2 - dichroic mirrors; L1, L2 - lenses of intermediate spatial filter and L3, L4 - lenses of far-field profiling/filtering optics; A1-A3 - circular submillimeter apertures; OW - optical wedge; BD - beam dumps**

As mentioned, we tested three types of resonators for the MO. They were a stable plane-plane resonator (PPR), a confocal unstable resonator of the positive branch (PBUR) and a generalized diffraction filtered resonator (GDFR). The PPR comprised a plane plate as an output coupler and a plane high-reflection mirror. The PBUR had two high-reflection mirrors: a convex button mirror of the curvature radius of 40 cm and a concave mirror of 3m-radius of curvature. The GDFR consisted of a convex mirror of the curvature radius of 28 cm and a small-sized plane mirror with an aperture of a diameter of 0.6 mm in front of it. The optical cavity length was 130 cm for all the resonators.

## *Results and discussion*

Figure 4.4.2 depicts the time evolution of laser beam divergence with the tested resonators. As can be seen, the PPR/PBUR divergence has high values at the beginning of the pulse. Afterward follows a decrease down to steady values, which stay constant till the end of the pulse. The divergence drops much faster and to lower values with PBUR compared to PPR. For PBUR ~10ns are enough to get to a steady value of 1.6-1.8 DL, while PPR goes down to 2.2 DL for ~25ns. In contrast, GDFR starts lasing later but the divergence exhibits no change throughout the pulse. It is as low as 1.2 DL. The divergence of MOPA output containing ASE is 1.9 DL during the pulse. When ASE is spatially filtered, the throughout-pulse divergence goes down to the diffraction limit - it is just 1.06 DL.

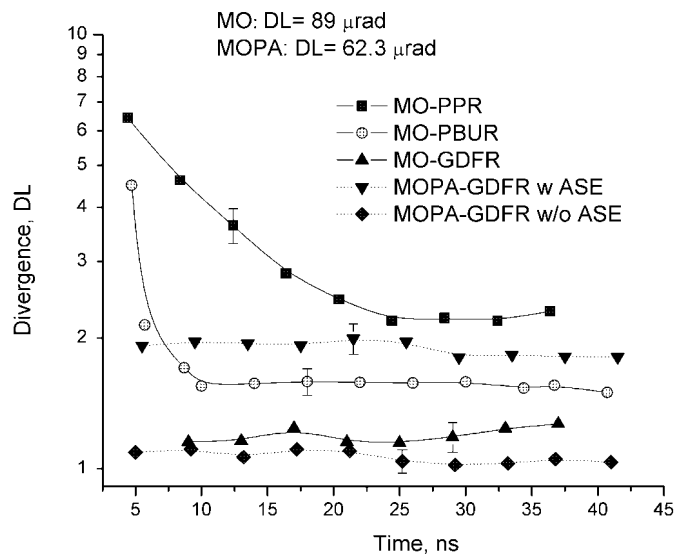

**Figure 4.4.2 Temporal throughout-pulse evolution of laser beam divergence of the MO fitted with PPR, PBUR and GDFR and the GDFR-fitted MOPA (two cases: one - with ASE and other - ASE is eliminated by an output spatial filter)**

As a reasonable indicator of the spatial coherence degree, we can use the normalized inverted beam divergence. Based on the data from Figure 4.4.2 we estimated and plotted in Figure 4.4.3 the spatial coherence (its degree) as an inverse function of beam divergence during the laser pulse. The results give an opportunity to compare the tendency of changes of spatial coherence with the tested resonators. As seen, the coherence degree increases from PPR through PBUR to GDFR. Moreover, with GDFR it is time-independent. With MOPA system, the coherence degree goes up further when output is spatially filtered.

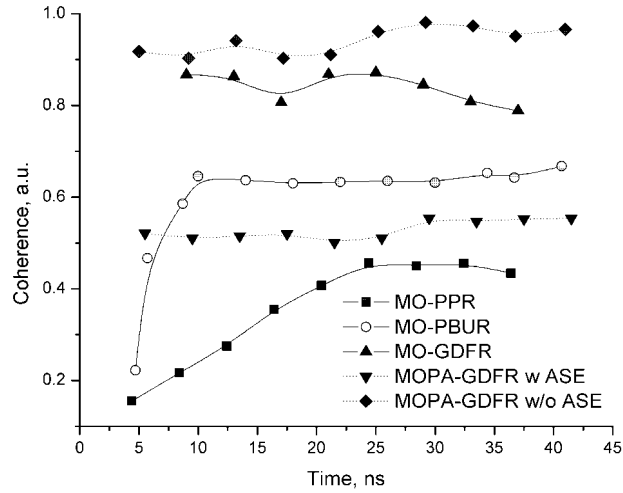

**Figure 4.4.3 Time-dependence of the estimated spatial coherence (its degree) during the laser pulse of different MO and MOPA resonator configurations**

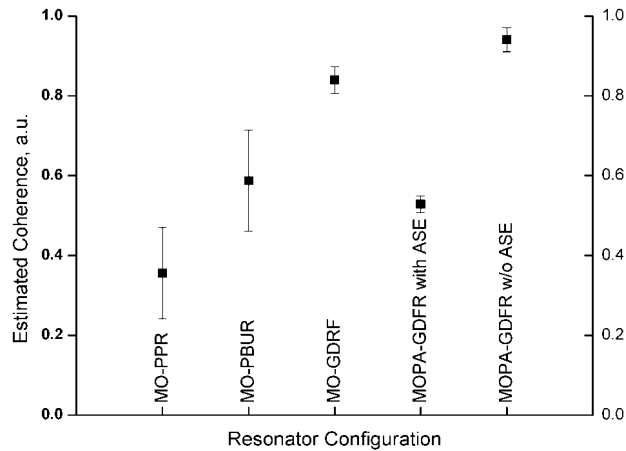

**Figure 4.4.4 Time-averaged estimated coherence of different MO and MOPA resonator configurations**

Figure 4.4.4 illustrates the great increase of quality of CuBr laser radiation with respect to spatial coherence as GDFR is employed. The coherence is much higher (~0.8) and its variation is lower with GDFR while with an unstable resonator as PBUR it is ~ 0.6 and variations are 3-4 times higher (see error bar lengths). With the MOPA system and appropriate spatial filtering, a very high coherence (~0.9) is obtained. The coherence of non-filtered MOPA-GDFR emission is poorer (~0.5) than those of PBUR and GDFR of MO. All these findings point to the importance of spatial filtering of MOPA output radiation.

## *Novel reversal shear interferometer*

Coherency estimations should be verified and we did it by direct interferometric measurements. For that purpose, we set up a reversal shear interferometer which was of new design and proved to work fine. A schematic of the arrangement for interferometric coherence measurements including this reversal shear interferometer is shown in Figure 4.4.5. The green ($\lambda$=510nm) line is separated and attenuated by means of two filtering dichroic mirrors (DC1 and DC2) and an optical wedge (OW1); the ASE is eliminated by a spatial filter formed by two lenses (L1 and L2) of focus lengths of 25 cm and a circular aperture (A) of 1 mm.

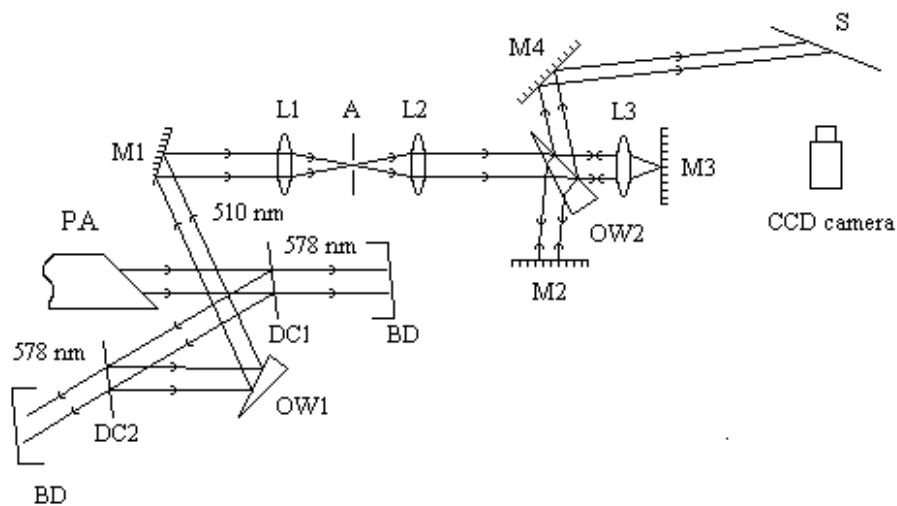

**Figure 4.4.5 Experimental setup for measurement of spatial coherence: DC1, DC2 - dichroic mirrors; OW1, OW2 - optical wedges; L1-L3 – lenses; M1-M4 - plane high-reflection mirrors; A - an aperture and S - a screen**

Details of the novel design of reversal shear interferometer (a variety of the Michelson interferometer), are shown in Figure 4.4.6 and explained below. The interferometer consists of two plane mirrors M1 and M2 (100% reflectivity), a beam-splitting optical wedge OW (7.5° wedge angle) and a lens L (focal length of 7.6cm) in one of interferometer arms (Dixit, 2005).

The interferometer works as follows. Beam 0 split into two parts on the rear surface of the wedge. The reflected beam 1 goes refracting to mirror M1, reflects back on it and travels the same path after refraction to the rear surface of the wedge. The refracted beam 2 goes to the lens L, reflects on M2, passes through L again and thus reverses its wavefront. After reflection on the wedge, beam 2 combines with refracted beam 1 on the rear surface of the wedge to form beam 3. The fringe pattern is relayed on a screen by the mirror M3 and recorded by a CCD camera. By this way, spatial coherence is measured. If the lens L is absent it is possible to measure temporal coherence of the incident beam 0 as well. As an illustration, the pattern for the MOPA-GDFR emission is given in Figure 4.4.7.

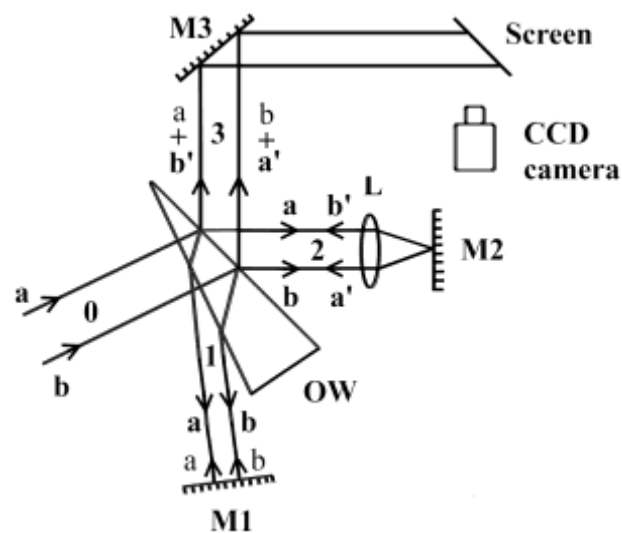

**Figure 4.4.6 Design of the reversal shear interferometer employed in the coherence measurement: M1 – M3 - plane mirrors, OW - an optical wedge and L - a lens**

The pattern recorded by the CCD camera was analyzed by computer programs and the degree of spatial coherence was calculated as the visibility of fringes (Svelto, 1984) *i.e.* $(I_{max}-I_{min})/(I_{max}+I_{min})$, where $I_{max}$ and $I_{min}$ being fringe intensity maxima and minima, respectively. If the intensity of the two interfering beams in the Michelson interferometer is equal, visibility is equal to the degree of coherence. It should be noted that the degree of spatial coherence across the laser beam almost does not vary in contrast to other reports (Omatsu *et al.,* 1991, Prakash *et al.,* 2006). The results plotted in Figure 4.4.8 represent the actually-measured degree of coherence of different MO and MOPA resonator configurations.

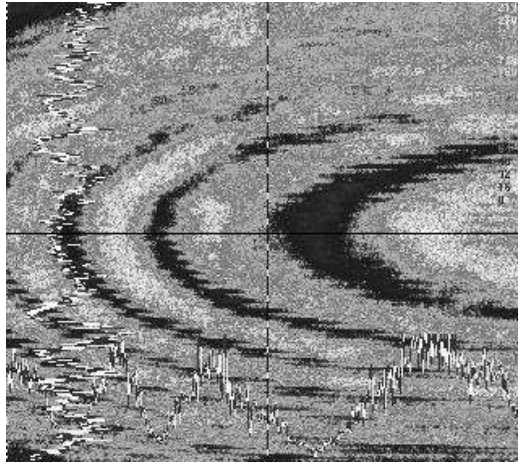

**Figure 4.4.7 A fringe pattern of the MOPA-GDFR emission. The vertical and horizontal intensity profiles of fringes crossed by the straight-line cursors are seen within the interferogram (left and bottom)**

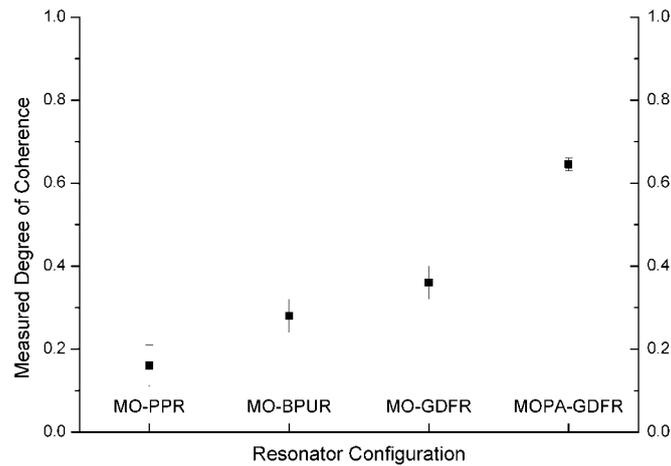

**Figure 4.4.8 Degree of spatial coherence of different MO and MOPA resonator configurations measured by the novel reversal shear interferometer**

The comparison starts with MO: the PPR coherence is very poor (0.16), the PBUR coherence gets better (0.28) and that of GDFR is pretty good - 0.36. As MOPA is run the measured degree of coherence reaches 0.65. The estimated and the measured coherence are obviously qualitatively similar. The discrepancy in the magnitude can be attributed to the roughness of the approach applied to the *estimation* of the degree of spatial coherence on the basis of beam divergence and the mechanical vibrations accompanying the real interferometric measurements. And at the end, we could say that the *estimation* values of spatial coherence degree shown in Figure 4.4.4 can be treated a sort of upper limits of spatial coherence degree in our experimental status.

*Conclusions*

Implementation of a generalized diffraction filtered resonator to CuBr laser allows us to achieve a great rise of the spatial coherence quality of laser radiation. Exact coherence values are experimentally measured by a newly designed version of reversal shear interferometer (a Michelson interferometer variety). The interferometric data confirm the trend predicted using estimations upon beam divergence changes. The novel reversal shear interferometer employed is of simple construction comprising just four optical components, namely a spherical lens, an optical wedge and two plane mirrors of high reflectivity. It is a fast and rigid instrument for coherence analysis of optical beams.

## 4.5. Green, yellow lines and laser pulse shaping by MOPA CuBr laser system

### Green and yellow lines of MOPA CuBr laser

Copper bromide (vapor) laser is a variety of copper vapor laser which uses CuBr as a donor of copper atoms. It is well known for quite a long time of 40 years. The demand for high-performance copper lasers fosters research of so-called MOPA (master-oscillator power-amplifier) systems. The investigations are mostly focused on beam divergence, spatial coherence, focal spot size, brightness *etc*. An interesting point for CuBr laser investigation is the behavior and control of the green ($\lambda$511nm) and the yellow ($\lambda$578nm) laser lines (Stoychev *et al.,* 2009). This is an important feature in many applications of CuBr laser. The MOPA CuBr laser system is a way of management of magnitude and proportions of laser lines output in same time.

### Setup of CuBr MOPA system

The basic CuBr MOPA optical setup is given in Figure 4.5.1. It consists of a CuBr master oscillator (MO, which is fitted with a generalized diffraction filtering resonator, GDFR), a beam folding (M3-M7), ASE filtering optics (L1-A2-L2) and a CuBr power amplifier (PA). Between mirrors M8 and M10, the two laser lines were separated and attenuated by means of two filtering dichroic mirrors (DC1 and DC2) and an optical wedge (OW). For ASE elimination in the MOPA output, an aperture (A3) is inserted forming this way an output spatial filter with a lens (L3) of focus length of 100 cm and another lens (L4) of focus length of 7.6 cm. The registering unit (RU) for laser pulse waveforms was a fast photodiode. Average laser power was measured by a Scientech power meter.

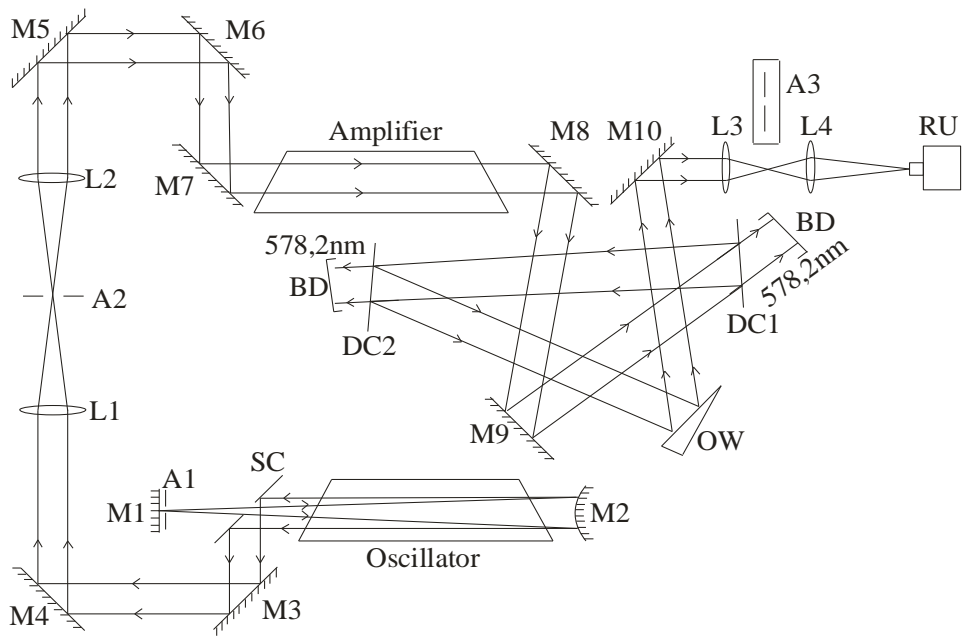

**Figure 4.5.1 Schematic of the optical system of MOPA CuBr laser system (here, oscillator is with generalized diffraction filtering resonator) and the measurement arrangement for line separation, attenuation and pulse recording of the green (λ510nm) laser radiation: M1-M10 (except of the convex M2) - plane high-reflection mirrors for both laser lines; SC - scraper mirror; DC1, DC2 - dichroic mirrors; L1, L2 - lenses of intermediate spatial filter; L3, L4 - lenses of far-field filtering optics; A1-A3 - circular submillimeter apertures; OW - optical wedge; BD - beam dumps; RU – registering unit**

Laser lines behavior from CuBr MO is studied first. MOPA CuBr laser lines behavior is in the section after that.

### *CuBr MO laser lines behavior*

In Figure 4.5.2 we give the waveforms of green (a) and yellow (b) laser line pulses for different types of resonator configurations: PPR − plane-plane resonator; PBUR − positive branch unstable resonator; GDFR − generalized diffraction filtering resonator. In all cases, pulses expose spiky structure due to successive cavity round trips of light. As overall radiation loss increases in direction PPR → PBUR → GDFR so interim peaks develop. This is more valid for the green line which has higher gain coefficient than the yellow line.

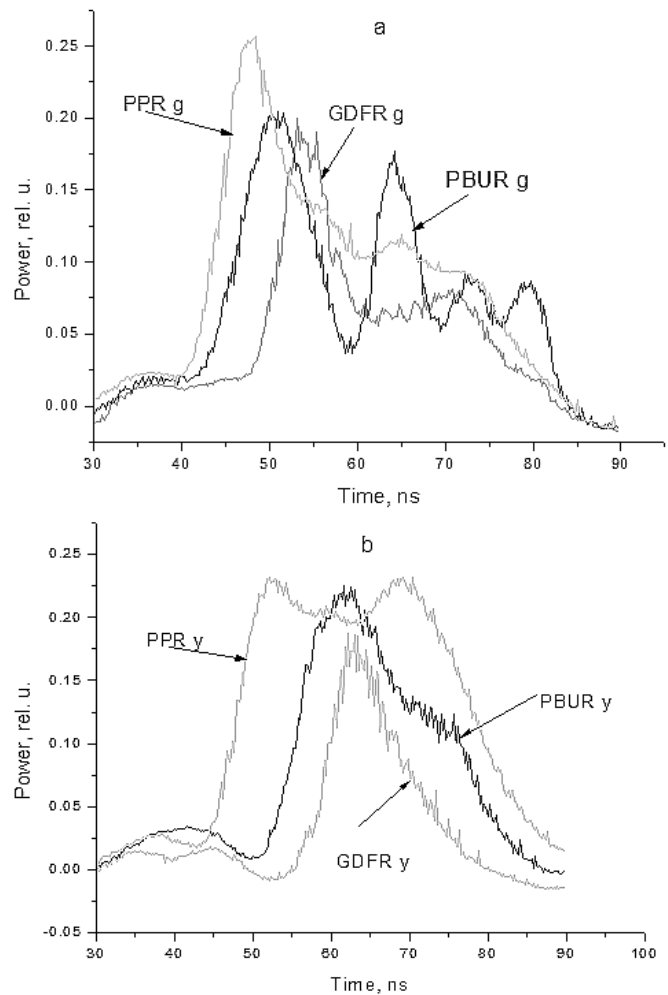

**Figure 4.5.2 Waveforms of green (a) and yellow (b) laser line pulses for different types of resonator configurations: PPR – plane-plane resonator; PBUR – positive branch unstable resonator; GDFR – generalized diffraction filtering resonator**

The similar explanation goes for the reducing of total power and its components (green and yellow) in the same succession PPR →PBUR →GDFR which is given in Figure 4.5.3.

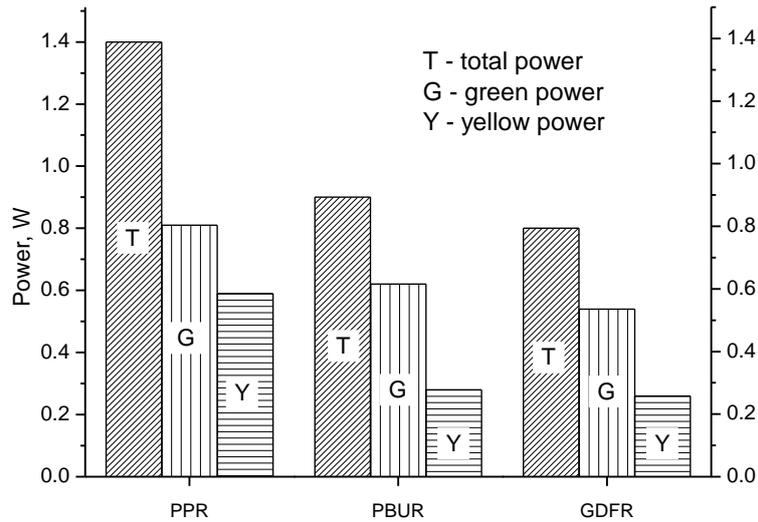

**Figure 4.5.3 CuBr oscillator laser power with different types of optical resonators for green λ510nm (G), yellow λ578nm (Y) and totally for both lines (T)**

The total laser power of CuBr oscillator decreases in the sequence PPR →PBUR →GDFR. Such decrement (with slight differences) is observed for both laser lines: the green line power is 0.80 W for PPR, 0.62 W for PBUR and 0.54 W for GDFR; the yellow line power is 0.59 W for PPR but it halves for GDFR and PBUR - 0.26 W and 0.28 W, respectively. This comes out from the differences of laser radiation formation in optical resonators employed in the CuBr oscillator arrangement. In PBUR and GDFR, the radiation quits the active medium quickly and the higher gain of the green line determines its higher output power. The PPR geometry allows more round trips to be done by the radiation which leads to saturation of the green line while the yellow line pays off for its lower gain. This is evident from the proportions of green and yellow components plotted in Figure 4.5.4. Since radiation conditions do not differ too much for PBUR and GDFR, their green/yellow laser power proportions are practically same - G/Y power ratio ≈ 2/1. For PPR, the G/Y power ratio is ≈ 3/2. This value is a hint of saturation conditions for both lines in PPR.

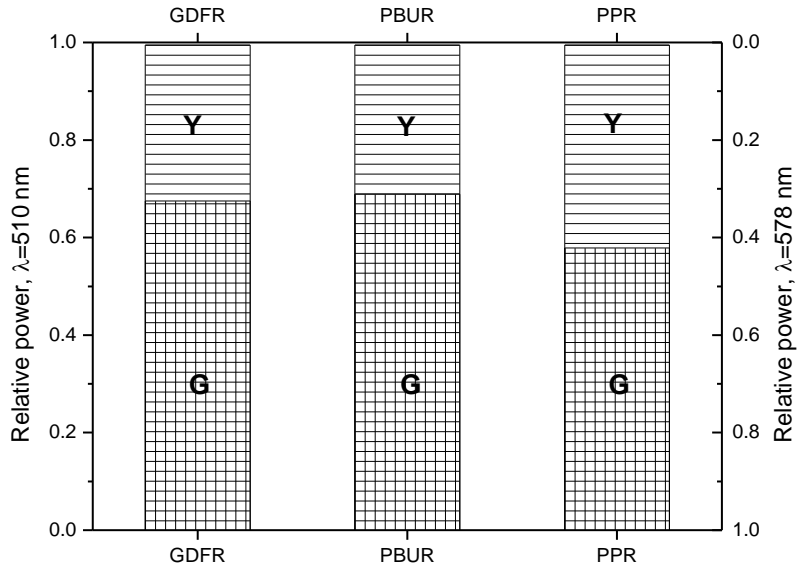

**Figure 4.5.4 Relative laser power (proportion) of the green line λ510 nm (G) and the yellow line λ578 nm (Y) of CuBr master oscillator with different resonators**

## *GDFR MOPA CuBr laser lines behavior*

In next paragraphs, we describe the behavior of laser lines from GDFR MOPA CuBr laser system as functions of triggering time delay in the case of hydrogen addition to the dominant buffer gas neon and in the case of no hydrogen added to it. In this section, all optima are defined in reference to maximum laser power from MOPA system.

In Figure 4.5.5 we show the waveforms of GDFR MOPA CuBr laser pulses - green (λ510nm) and yellow (λ578nm) lines without hydrogen additives at different delay times.

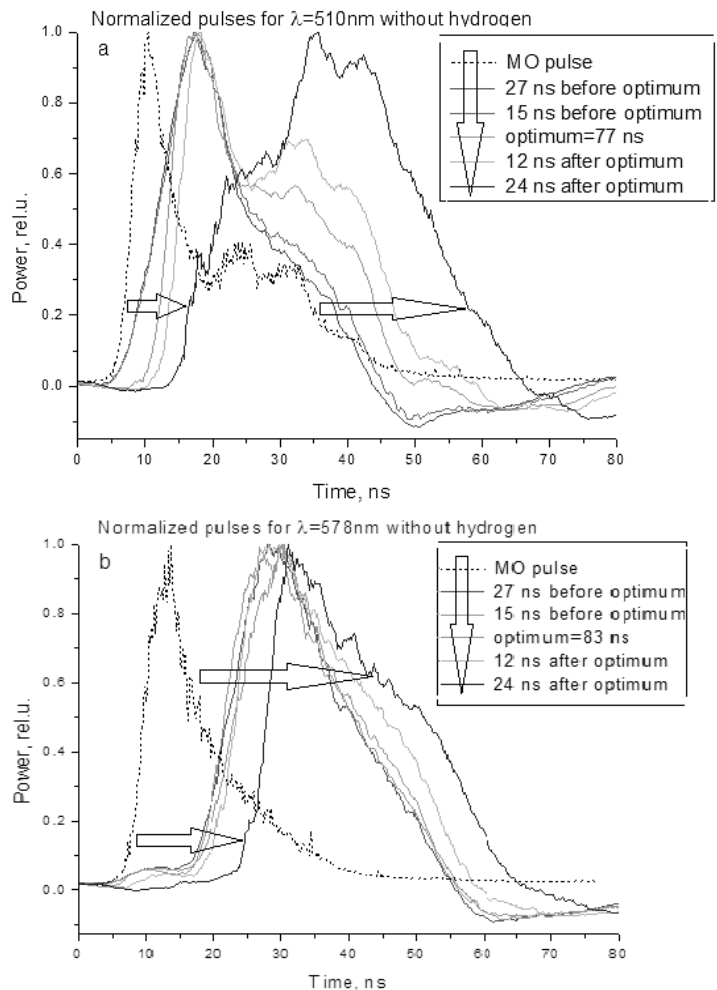

**Figure 4.5.5 Waveforms of GDFR MOPA CuBr laser pulses - green (λ510nm) and yellow (λ578nm) lines *without* hydrogen additives at different delay times. Optimum is defined in reference to maximum laser power. Arrows show direction of shift of ascending (front) and descending (rear) slopes**

As can be seen, the waveforms of MO and MOPA pulses are quite different. This is caused by the difference in conditions the amplification of MO pulse occurs in the PA. So the green (λ510nm) line from MO is ~30 ns long (from here on, at basement) and does not change much excepting for the longest time delay - of 24 ns after optimum. Here it is ~50 ns long and the shape is reversed – the secondary spikes (from the middle to end of the pulse) are more prominent. Last means the gain in that time (second half of pulse) is higher than before and this part of the signal goes through higher amplification. The other pulses also get more amplified in their second half but not that much.

The yellow (λ578nm) line does not change so mush in shape. From ~35 ns to ~40ns is the time of its pulse duration. And here the second half of MO pulse gets more amplified and the rear of pulse rises obviously.

In Figures 4.5.6 and 4.5.7, we show GDFR MOPA CuBr laser lines pulses in hydrogen case. In that case, we have hydrogen added to the dominant buffer gas neon.

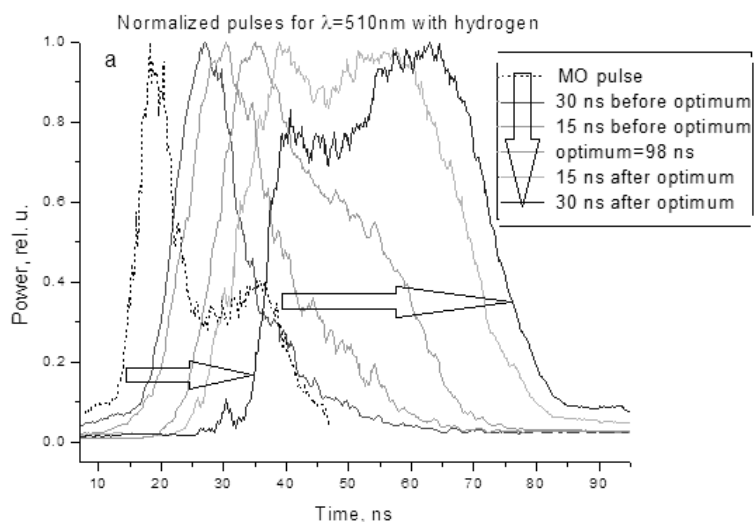

**Figure 4.5.6 Waveforms of GDFR MOPA CuBr laser pulses of green (λ510nm) line *with hydrogen* additives at different delay times. Optimum is defined in reference to maximum laser power. Arrows show the direction of shift of ascending (front) and descending (rear) slopes.**

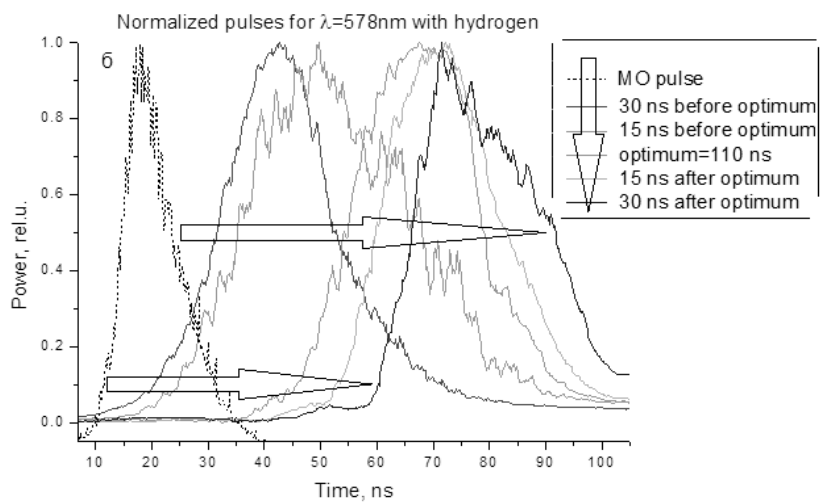

**Figure 4.5.7 Waveforms of GDFR MOPA CuBr laser pulses of yellow (λ578nm) *with hydrogen* additives at different delay times. Optimum is defined in reference to maximum laser power. Arrows show direction of shift of ascending (front) and descending (rear) slopes**

The situation with hydrogen does not differ too much from the nonhydrogen case. MO green (λ510nm) pulse of ~30 ns lengthens to ~50 ns after PA. Its waveform evolves with enhanced second part but without great suppression of the first part of it (*cf.* the nonhydrogen case). The green pulse looks more compact. The yellow (λ578nm) line appears like in the nonhydrogen case.

In Figure 4.5.8 we present the average output power (green, yellow and total) of GDFR MOPA CuBr laser system as a function of the triggering time delay in the nonhydrogen case and hydrogen case.

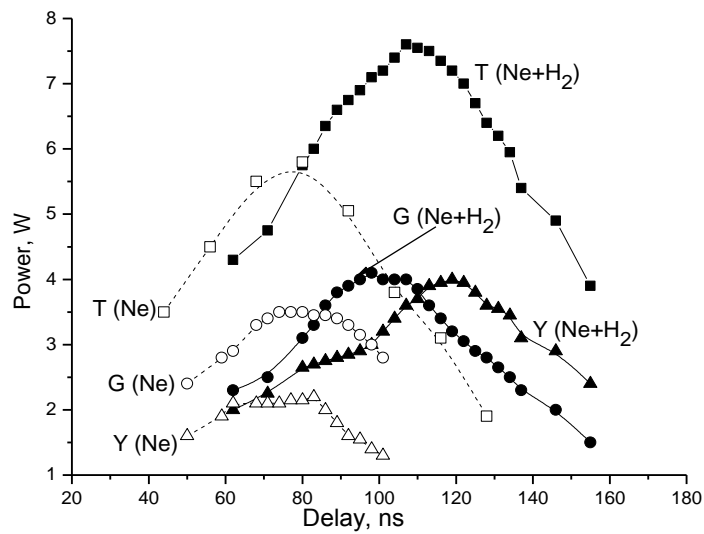

**Figure 4.5.8 Laser line power of GDFR MOPA CuBr laser system as a function of triggering delay in the nonhydrogen case and hydrogen case: λ510nm (G), λ578nm (Y) and total (T). Solid symbols are for hydrogen case**

The laser power dependence of GDFR MOPA CuBr laser system on the delay is very sensitive to the presence of hydrogen. In nonhydrogen case, the maxima of output power for the green and yellow laser lines are almost at same delays. These maxima are quite wide - ~15ns for the green line and ~20ns for the yellow line.

In hydrogen case, the maxima of the output of green and yellow lines occur later and the shifts of optimal delay (optimal delay corresponds to the delay when the laser power is at maximum) are different. The addition of hydrogen increases all of the optimal delays. The optimal delay shift for the green laser line is ~20-25ns and for the yellow laser line, it is as much as ~30-40ns. The maximum of total output power moves ~30ns further. The magnitude of PA signal amplification is ~ 10 and hydrogen does not affect it in a remarkable way (however the problem was not studied in details).

In some applications, the proportions of green and yellow can be of interest. Next Figure 4.5.9 depicts the change in color fractions in light emission of GDFR MOPA CuBr laser. It deals with the relative green/yellow power of GDFR MOPA CuBr laser as a function of triggering time delay in the nonhydrogen case and hydrogen case.

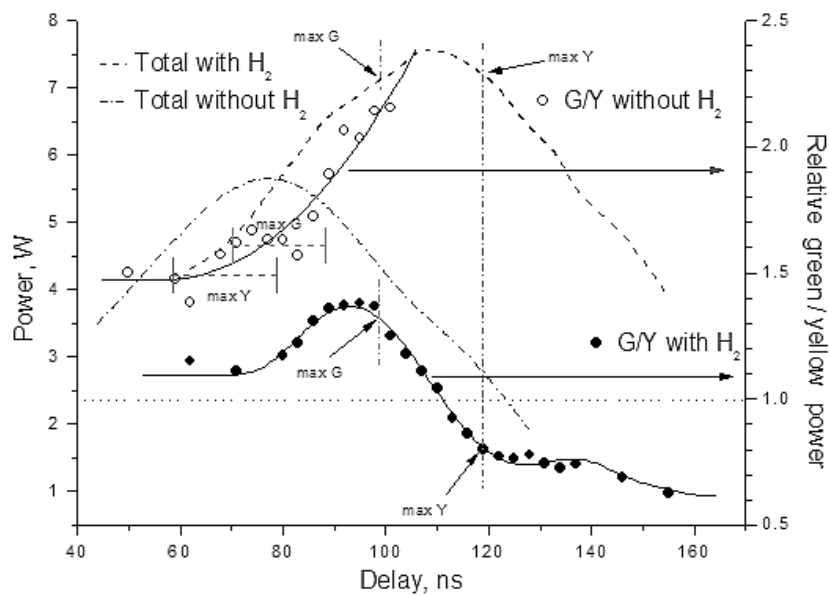

**Figure 4.5.9 Total and relative green/yellow power of GDFR MOPA CuBr laser as a function of triggering time delay in the nonhydrogen case and hydrogen case. Solid symbols are for hydrogen case**

As seen the green/yellow (G/Y) power ratio varies quite a lot. In nonhydrogen case the green line power is higher than the yellow line power: starting (for low delays < 60ns) from G/Y~1.5 it rises up to G/Y=2.2 for longer delays of 100ns. Hydrogen case is different. Between 60 and 110ns green line is more intensive with a peak of G/Y~1.4 at 90-95ns. For delays > 110ns the yellow line becomes stronger than the green line. Further for delays >120ns the yellow power is twice the green power - G/Y~0.7.

This way, by simply varying the delay time we can easily control the green/yellow (G/Y) power ratio of MOPA CuBr laser which can be important for many applications.

## *CuBr laser pulse shaping by MOPA system*

Laser beam shaping techniques have long been a center of scientific and practical interest. That comes out of the increasing demands to the laser radiation quality. Frequently, laser beam quality is crucial for applications as fine material processing (automotive and aircraft industry, biological tissues, therapy), science (laser fusion research) and many other.

Studies on laser temporal beam/pulse shaping for quality micromachining have been carried in many laboratories for a long time and a great number of papers have been published on the subject. So, the effect of laser temporal pulse shaping on weldability of crack sensitive alloys by laser spot welds was shown (Michaud *et al.,* 1995, Collins *et al.,* 1999 *etc*.) They concluded that the occurrence and severity of weld defects such as spattering, cratering, occluded vapor porosity and solidification cracking demand a systematic research to provide a better understanding of the effects of temporal pulse shaping on the resultant weld pool dimensions (depth and width). They found that the degree of nonuniformity imprinted on the surface of a target depends in part on the time dependence of the laser pulse. Therefore, pulse shaping techniques are being sought that can be used to enhance energy coupling efficiencies and prevent these commonly observed laser spot weld defects (Metzler *et al.,* 1999). In brittle dielectric materials, other authors reported that manipulated laser pulses and temporally tailored pulses are promising new tools for quality micromachining, making possible adapting the energy delivery rate to the material properties for optimal processing quality (Stoian *et al.,* 2003). The sequential energy delivery with judiciously chosen pulses may induce softening of the material and change the energy coupling. This can result in lower stress, cleaner structures, and allow for a material-dependent optimization process.

Temporal pulse shaping is performed by phase and/or amplitude modulation. General approaches are two: multiple pulses (from double pulse to pulse train) with adjustable distance; or tailoring special pulse shapes according to experimental requirements.

Employment of MOPA (master-oscillator power-amplifier) system is a simple and straight way to control laser pulse shape on the two lines, λ510 nm and λ578 nm of CuBr vapor laser. Here, we present a survey (Astadjov *et al.,* 2006b) on the feasibility of pulse shaping by the management of the triggering delay time between MO (master oscillator) and PA (power amplifier). This approach is good enough for changing (*to some extent*) the instant power distribution within the laser pulse. Pulse shaping allows the investigation of the processes taking place in the region between the ablation and surface and having a definite effect on the quality of performed operations. As Collins et al. (1999) reported the width of this region or smoothing distance, is proportional to the laser intensity. Larger smoothing distances result in decreased imprint of laser perturbations on the target during early times. The smoothing distances and imprint of three basic pulse shapes have also compared: the standard continuous pulse consisting of a foot pulse followed by a rise to a flat-top pulse, and the same pulse, preceded by a single prepulse (toe), with and without a thermal relaxation period between it and the foot pulse (Figure 4.5.10).

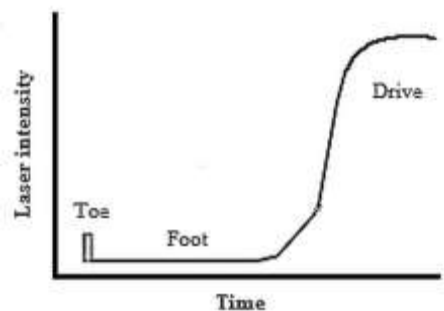

**Figure 4.5.10 Intensity spike (toe) at the beginning of foot pulse (Collins et al., 1999)**

Recently, the most attracting is the ultrafast (femtosecond) laser micromachining and the main research is concentrated there. Nevertheless, lasers producing longer pulses (from picoseconds to milliseconds) are being explored likewise because many problems of laser-material interaction are still unclear. The present report is not a comprehensive study but a practical demonstration of MOPA system as a way of pulse shaping approach to CuBr lasers.

## *Results and conclusion*

The MOPA setup is as the described by Astadjov et al. (2005). The electronic timing unit has a time resolution of 3 ns which is satisfactory for the present research. The two laser lines are separately measured: directly, the average power by a laser power meter; and the waveform – after an optical integrator (a semi-transparent paper making the radiation almost isotropic) placed in front of a fast photodiode.

In Figure 4.5.11 we plot waveforms of the green and the yellow lines of MOPA CuBr laser emission at four delay times that are representative for the variations of the shape of laser pulses.

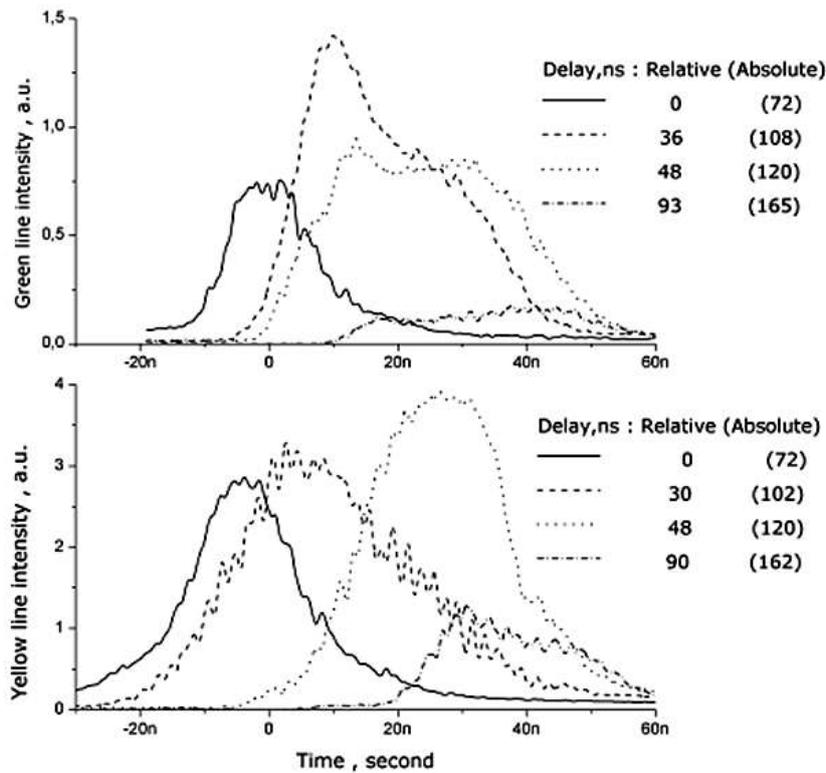

**Figure 4.5.11 Waveforms of green and yellow lines of MOPA CuBr laser emission at four typical delay times**

These delay times specify the extent of overlapping of optical gain of the two parts of MOPA laser system, namely the oscillator and the amplifier. The gain overlapping is low at delaying < 0 ns as well as at delaying > 90 ns while it has maximum at medium delays (30-50 ns). As can be seen, waveforms vary from Gaussian-like or triangular shape to more complex shape – trapezoid with a spiky structure that is more or less pronounced. Pulses have also different skewness, S that describes their different rise- and fall-time characteristics (Tyth *et al.,* 2003). In all cases $S \geq 0$: that corresponds to a symmetric pulse shape ($S = 0$) or represents a fast increase and a long tail ($S > 0$).

A variety of waveforms is pretty good. At short delays, one or two spikes are prominent; at longer delays, prepulse structure appears resembling foot, step or toe.

The pulses given are samples that should be treated merely as an illustration for the diversity of laser shapes produced by the MOPA CuBr laser system. However, an elaborate study may uncover more details of this laser pulse shaping technique in future. In a similar way, for fiber lasers (pulse length of typically less than 0.1 ms) has been shown that ramping up the laser pulse power during the laser cutting process may reduce the taper of the sidewalls of metal cuts (Kleine *et al.,* 2004). So, the laser cut will be of good surface quality with a minimum amount of slag and burr to reduce post-processing. Similarly, the heat affected zone and molten material recast will be small. This agrees with observations from other papers (*i.e.* Low *et al.,* 1999) that report a lower taper with a linear increasing sequential pulse delivery pattern.

Here we show a possible laser pulse shaping procedure applied to the CuBr laser operated as an MOPA system. With two laser lines and with pulses that can be manipulated not only in time but in space (e.g. retarded) we may have an unlimited number of laser waveforms possible. The optional single or dual-line action further extends the flexibility of the MOPA CuBr laser technique for laser pulse shaping.

Our experiments concerning CuBr laser temporal pulse shaping by MOPA system demonstrated feasibility and relevance of this method.

# 5. *MOPA CuBr laser beam transformations*

Many applications of lasers are closely related and determined by the properly chosen beams. Usually, that concerns the distribution (a.k.a. profile) of different light parameters - intensity, phase, coherence *etc.* Laser-produced radiation inherently shows up a great diversity of profiles. These raw profile beams are not always good enough for straightway use and therefore they need to be tailored. Last decades feature an enormous interest in this new area of research - laser beam shaping. The abundance of techniques and models occur that can produce profiles required for the great variety of laser applications. Dickey & Holswade (2000) summarized in a comprehensive reference book the most fundamental theories and techniques and provided all basic information to research, develop, and design beam shaping systems.

Copper vapor lasers are the most high-power and high-efficiency lasers which produce visible light in a straightforward way without any light conversion. Their beam intensity radial profile (hereafter we only treat and mean *intensity radial* profiles omitting these words in italics in most of the cases) varies from annular or top-hat to Gaussian-like.

In galore of applications, the laser radiation operates after being concentrated by a focusing instrumentation. In the process of focusing a near-field beam profile is transformed into a far-field beam of intensity distribution often quite different from the initial intensity distribution of that near field. Relations between near-field and far-field beam profiles (distributions) are of expanding interest due to the vast employment of optical (laser) beams in science and technology. This is an apparent reason to undertake the following study with attention to CuBr laser specificity.

## 5.1 *Focusing of MOPA CuBr laser beam*

CuBr laser intensity profiles vary from annular to top-hat or Gaussian-like. MOPA CuBr laser profile variations are reported in experiments on timing and buffer gas composition by Astadjov *et al.* (2007). While annular profiles are typical for a nonhydrogen buffer or for MOPA small trigger delays, top-hat profiles are predominant for a buffer containing hydrogen or for MOPA long trigger delays (Astadjov *et al.,* 1988a; Astadjov, D.N. & S.V. Nakhe, 2010). This is illustrated in next Figures 5.1.1-3 with profiles taken experimentally through a CCD camera controlled by laser beam analyzing Spiricon® software.

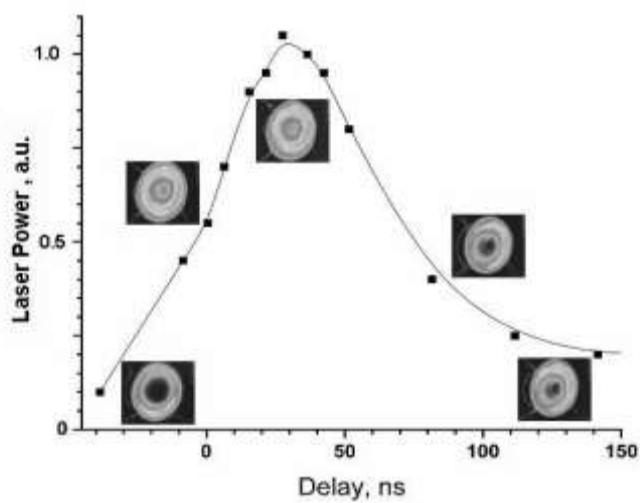

**Figure 5.1.1 Near-field profiles of MOPA CuBr laser beam as functions of the MOPA trigger delay. Buffer gas is neon of 17 Torr**

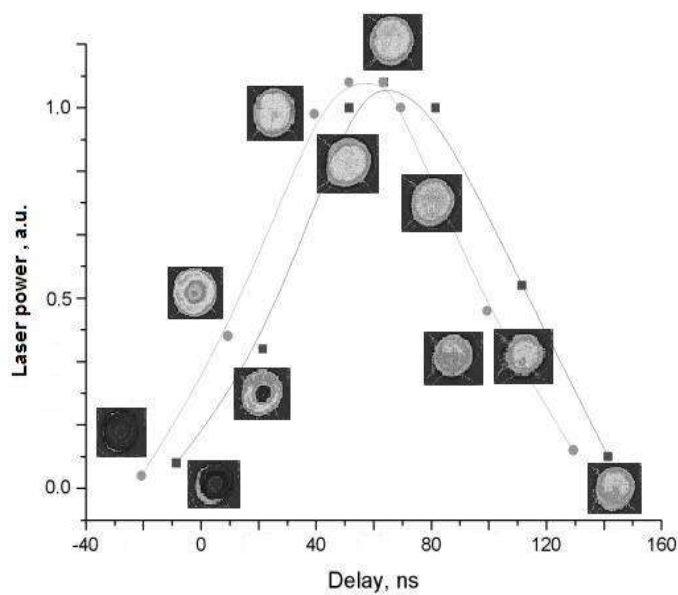

**Figure 5.1.2 Near-field profiles of MOPA CuBr laser beam as functions of the MOPA trigger delay. The buffer gas is a mixture of neon (17 Torr) and hydrogen (0.3 Torr). In the two random cases pictured here, laser beam profile mutability is exposed**

In nonhydrogen buffer case, the MOPA CuBr laser emission is annular for all trigger delays. In a buffer gas mixture of neon (17 Torr) and hydrogen (0.3 Torr) MOPA CuBr laser beam is annular for small trigger delays only.

In Figure 5.1.3 some more details are shown in pictures of experiments with far-field beam profiles produced from near-field MOPA CuBr laser beam profiles.

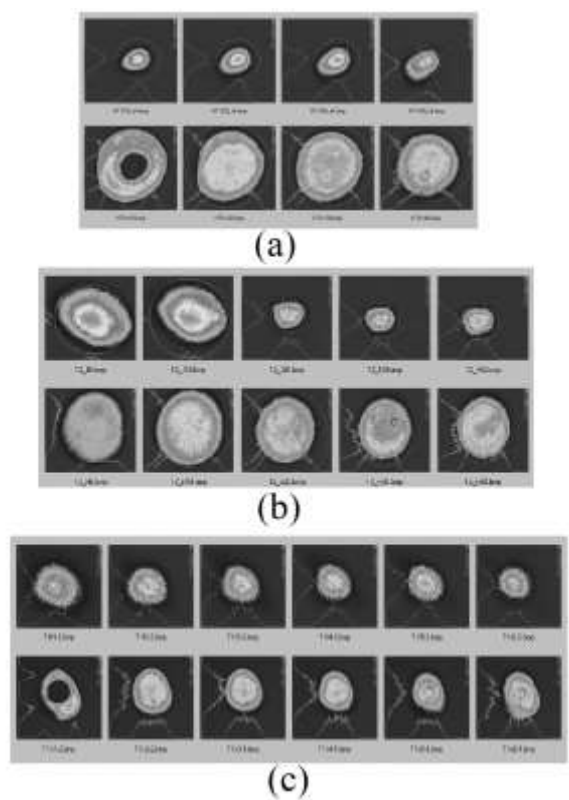

**Figure 5.1.3 Three picture sets of near-field profiles (the lower traces) and their far-field profiles (the upper traces) taken experimentally of MOPA CuBr laser beam using laser beam analyzing Spiricon® software**

From Figure 5.1.3 we can infer that independently of near-field beam profile patterns (annular or top-hat) the far-field beam profiles exhibit a pattern with intensity peak in the central area (Gaussian-like pattern).

## 5.2 Fourier transform of annular beams

As well established, optical lenses perform Fourier transform of a near-field beam profile into its far-field beam profile which is located at the lenses' focal plane. Two-dimensional (2D) far-field beam profiles can be calculated by 2D Fourier transform of near-field beam profiles. The profile of focal spot depicts the far-field beam profile of the beam near-field profile located just before the lens (within the accuracy of a phase coefficient) (Goodman, 1996). In essence, the Fourier transform is a very powerful tool for signal processing (including optical signals). The most popular and simple version of Fourier transform is the fast Fourier transform (FFT). As a derivative of Fourier transforms the FFT is applied to the complex optical field. The optically detectable signal is the intensity of the optical field. To find it, we have to calculate the product of the optical field complex amplitude and its conjugated optical field complex amplitude (Longhurst, 1957; Lee, 1981). We completed all calculations via MATLAB® incorporated functions (Astadjov, 2009, 2010).

### Annular beam parameters and parameters of 2D FFT output beam

### Near-field parameters of annular beams

As a first step, parameterization of the annular source beam is necessary in order to perform simulations. The simplest type of annulus is the normalized flat two-level circular annulus (*i.e.* with a flat top and a flat bottom): the inner concentric circular area is of intensity $I_1$, which varies from 0 to 1; the annulus itself is of intensity $I_2 \equiv 1$. If $I_1 = I_2$, the profile is top-hat; if $I_1 = 0$, the profile is pure annular (aka dark annulus, hollow beam *etc.*). The pattern of such annular beam profile can be described by two parameters: (a) annularity $k$ – defined as $k = d_1/d_2$, where $d_1$ and $d_2$ are the inside and the outside diameters of annulus, and (b) $I_{dip}$, which is the intensity dip of the central area (the flat bottom): $I_{dip} = I_2 - I_1$. Figure 5.2.1 illustrates the parameters according to these definitions.

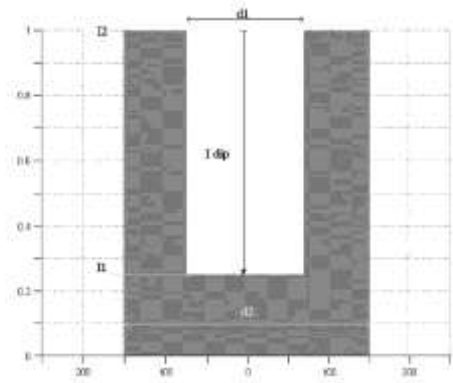

**Figure 5.2.1 Intensity axial section of normalized flat two-level circular annulus; d1 and d2 are the inside and the outside diameters of annulus; I1 is intensity of the circular core area (I1 varies from 0 to 1); I2 is annulus intensity: I2≡1**

## *Far-field parameters of 2D FFT output beam*

After applying the two-dimensional fast Fourier transform (2D FFT) onto near-field beam profile we get the type of a calculated pattern of far-field beam profiles shown in Figure 5.2.2. The profile can be described as having a prominent central peak and side peaks concentrically surrounding it (better seen in the figure inset). The appearance of side peaks comes from the discrete form of calculations done by 2D FFT. Actually, they should smear into uninterrupted concentric rings.

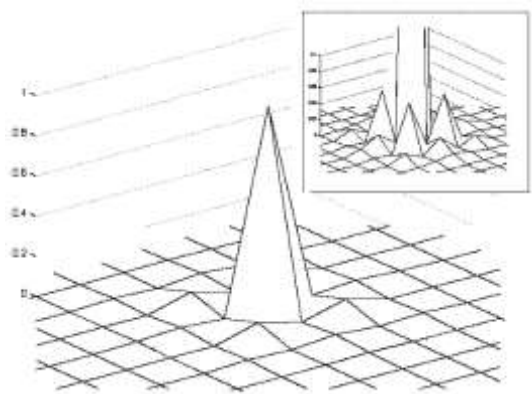

**Figure 5.2.2 Pattern of far-field beam profiles of a two-level circular annulus; inset - the central area magnified (x10)**

In Figure 5.2.3 we give the patterns of far field calculated from pure annuli (Idip=1 *i.e.* zero-intensity / dark core) of different values of annularity k. (Z-axis is 10% of central peak intensity for a better view of side peaks structure.) The simulation shows that when annularity k goes up (*i.e.* black core widens) so does the side peaks intensity. In the particular case of k=0 (no dark core at all *i.e.* top-hat), the near-field product of FFT is a far-field profile having an Airy pattern.

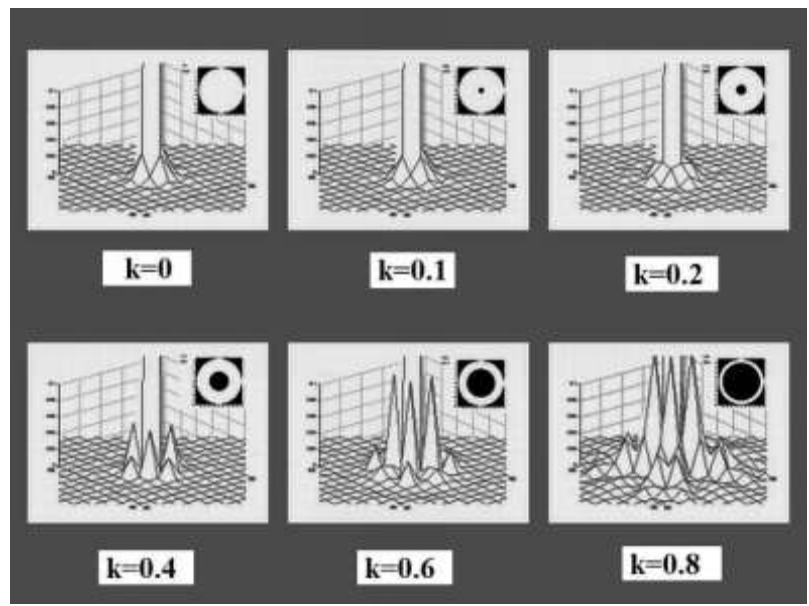

**Figure 5.2.3 Patterns of far fields calculated from dark annuli (insets - Idip=1) of different values of annularity k. Z-axis is 10% of far-field central peak intensity**

Here we are going to specify far-field profile parameters related to laser beam utilization in applications which are not susceptible to phase characteristics of light. We pay attention to the power (energy) component of light. Nevertheless, the coherence of light is a requirement for all our simulations. As a major parameter, we introduce the fraction of the central peak energy to the whole energy of the beam, PF0. This parameter gives a notion about energy spread within the far-field spot. The higher PF0 the lower energy spread is. In practice, that means less affected surrounding (the central peak) area by light radiation. The dependence of PF0 on Idip (the intensity dip of the central area: Idip=I2-I1) is plotted in Figure 5.2.4.

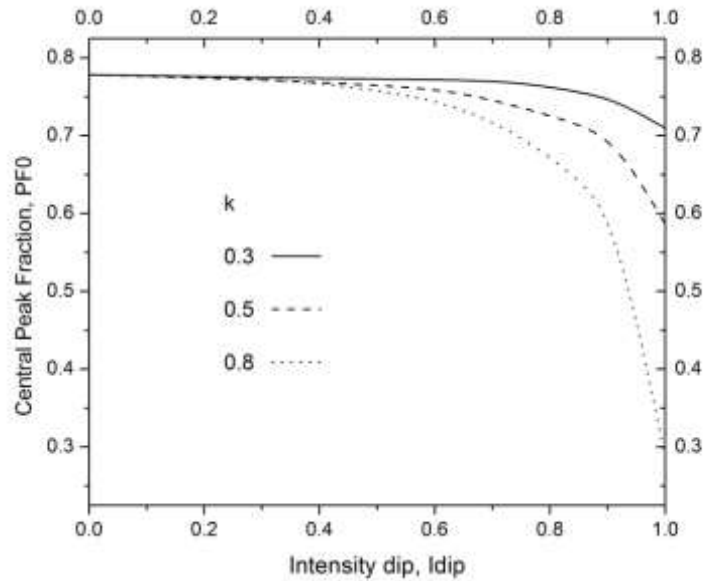

**Figure 5.2.4 Dependence of the far-field central peak energy fraction (PF0) on Idip (the intensity dip of the central area: Idip=I2-I1) for three values of annularity: k = 0.3, 0.5 and 0.8**

As seen in Figure 5.2.4, the increase of Idip leads to the decrease the central peak energy fraction. For higher k the decrease starts at lower Idip. So for k=0.3 and Idip<0.75, the change of PF0 is less than 1.5%: while the maximum of PF0 is 0.778, for Idip=0.75 the fraction of the central peak energy PF0 is 0.767. However for k=0.8 and the same Idip of 0.75, PF0 is 0.7 - the drop is 10%.

The central peak energy fraction dominates the far-field energy distribution nearly over the whole range. The exception is for narrow (k>0.65) annular beams of core intensity near zero (Idip>0.9) which occurs seldom in practice. Taking into account that the central peak energy is concentrated on a smaller spot area, the net impact of side energy spread diminishes furthermore.

For the sake of usability, a two-variable formula of PF0 dependence on Idip and k *i.e.* PF0=f(Idip,k) of a fairly good approximation is calculated using web-based program created

by Phillips, J. R. (2010). The formula is[3]:

$$z = a(\sinh(x)y^{1.5}) + b(\tan(x)y^{1.5}) + c \qquad \text{Eq.(1)}$$

where  a =  2.0310451454356748E+00, b = -1.9232938348115700E+00  and
       c =  7.8402321357391169E-01; Consider that $z \equiv PF0$, $x \equiv Idip$ and $y \equiv k$.

Also, in Figure 5.2.5 we have a 3D graph view of this PF0 dependence on Idip and k according to Eq.(1).

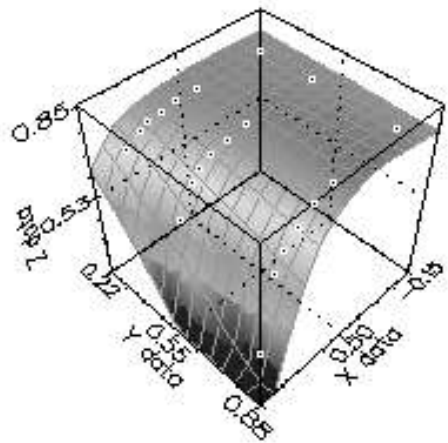

**Figure 5.2.5 Dependence of the central peak energy fraction (PF0) on Idip (the intensity dip of the central area: Idip=1-I1) and annularity k.**  *Note:*  $x \equiv Idip$, $y \equiv k$ and $z \equiv PF0$

---

3   User-selectable polyfunctional approximation based on lowest sum of squared (SSQ) absolute error; herewith sum of squared absolute error is 0.0116978139713.

## *5.3. Verification of focusability of coherent annular laser beams*

Next step was a study conducted to validate issues of our coherent dark (aka hollow, zero-intensity core) annular beam simulation by means of comparing them with experimental results from focusing the light passed through annular apertures (Astadjov & Prakash, 2013a). Therefore, the major target was to measure the dependence of the central peak power to the whole power of beam versus the annularity (*i.e.* 'inside diameter /outside diameter' ratio) of annular laser beams.

The experimental set-up is shown in Figure 5.3.1. An annular aperture is illuminated by a full-coherency laser beam (28 mm diameter) of a copper vapor laser (λ510nm) with generalized diffraction filtered resonator (Prakash *et al.,* 1998).

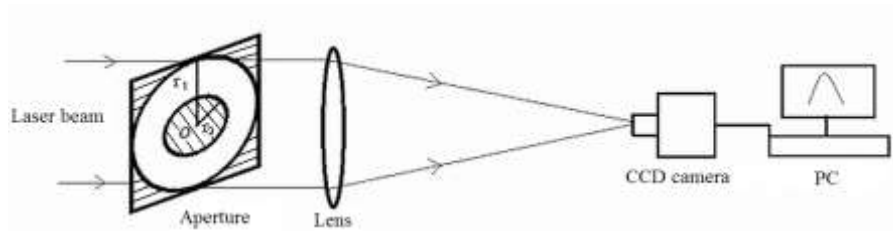

**Figure 5.3.1 Experimental set-up for recording far-field diffraction patterns produced by dark annular laser beams**

Four annular slits with an open circular aperture of 12.5 mm radius ($r_1$) and having an opaque central disc of radius ($r_2$) of 2.5 mm, 5 mm, 7.5 mm and 10 mm respectively were placed in sequence in the beam. Far-field diffraction patterns were recorded at the focal plane of a lens of focal length of 500 cm. These patterns images were captured by a CCD camera connected to a PC with a frame grabber card and then analyzed by image processing software.

## *Data analysis and results*

After images of focused laser light intensity patterns had been captured, they went through a series of mathematical operations to calculate the fraction (denoted as PF0) of the central peak power to the whole power of beam focused. For that purpose, we did the integration of the three-dimensional image of intensity patterns of each slit. The first integral was over the central peak area and the second integral - over the whole beam area of the output image; then the first integral was divided by the second integral. That is the PF0 of the focused output of the beam coming from a slit of annularity k ($k=r_2/r_1$). Results are given in Table 5.3.

**Table 5.3 Results from the experiment with annular slits of different annularity (k = 0, 0.2, 0.4, 0.6 and 0.8). Here are also shown the aperture views, the output images at the focus (aka of their far-field intensity), image intensity profiles (horizontal) and the power fraction PF0 calculated from focal images. PF0 is the ratio of the central peak power to the whole beam power**

| Annularity k | 0 | 0.2 | 0.4 | 0.6 | 0.8 |
|---|---|---|---|---|---|
| Annular aperture | 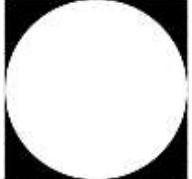 | 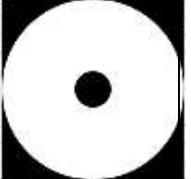 | 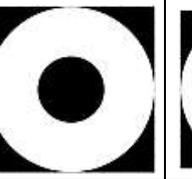 | 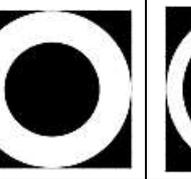 | 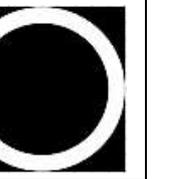 |
| Output image at focal plane | 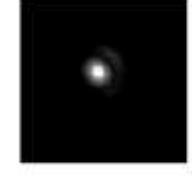 | 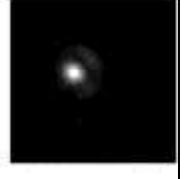 | 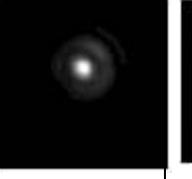 | 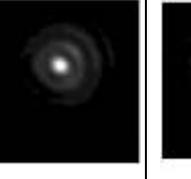 | 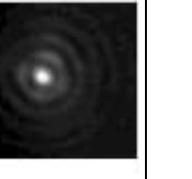 |
| Image intensity profile (*horizontal*) | 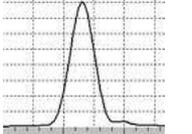 | 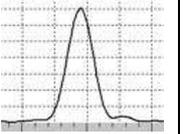 | 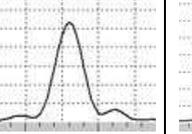 | 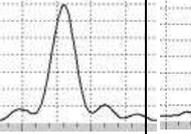 | 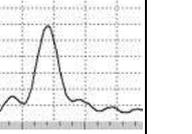 |
| Power fraction PF0 | 0.91 | 0.79 | 0.64 | 0.42 | 0.18 |

The table shows that the intensity of rings of focal images goes up with the increase of annularity and the PF0 goes down. So the relative power concentrated in the center diminishes and the spread of energy increases in accordance with simulation outcome. Both the experimental and simulation results are plotted in Figure 5.3.2.

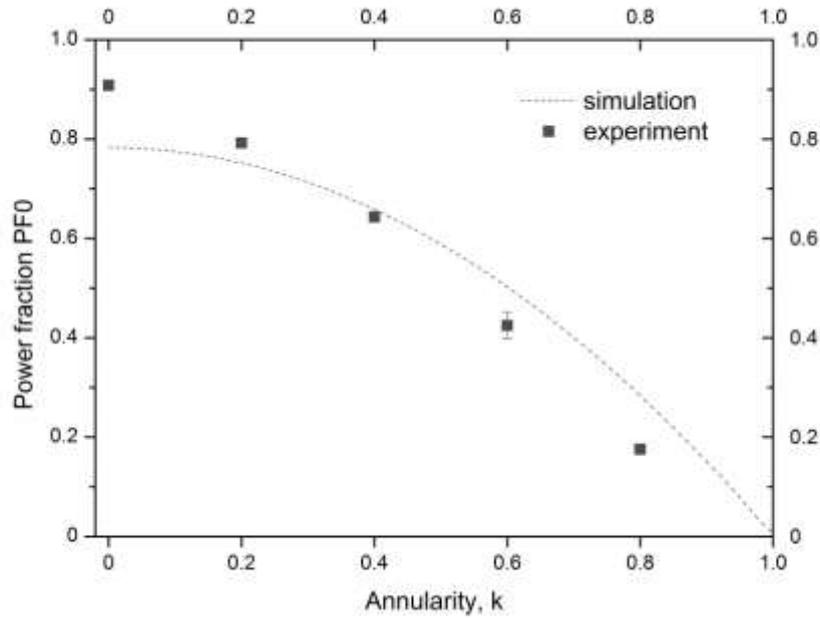

**Figure 5.3.2 Central peak power fraction PF0 versus the annularity k for full coherence dark annular laser beam. Experimental points are for five different values of k = 0, 0.2, 0.4, 0.6 and 0.8; FFT simulation is plotted by the dash line**

As seen in Figure 5.3.2 the experimental points follow the simulation curve. However at low annularity (k=0 and 0.2) they are above the curve, then the middle point (k=0.4) is very near it, and at the high annularity (k=0.6 and 0.8) the points are below the simulation curve. For a better estimation of deviation between both set of data (experimental and simulated) we calculated the difference between the experimental and simulated data and divided it by the simulated value. This PF0 difference is plotted in Figure 5.3.3. Four of the five experimental points deviate from simulated curve less than 16% but the point of highest annularity (k=0.8) is 38% deviation.

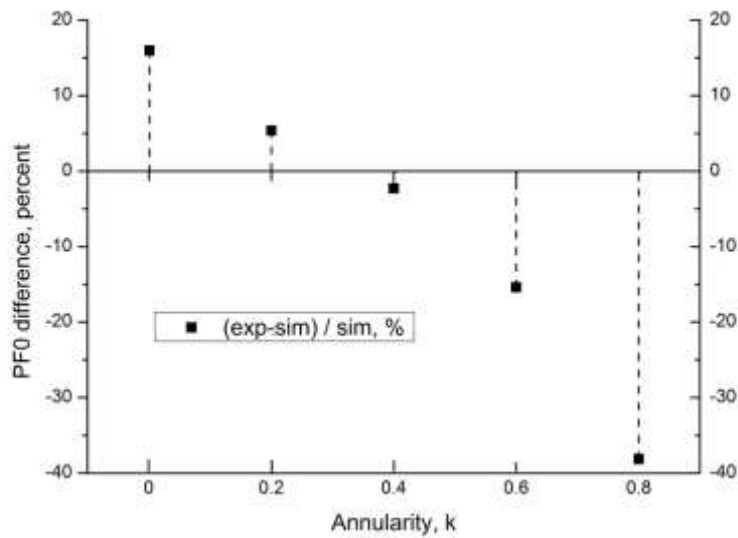

**Figure 5.3.3 PF0 difference versus beam annularity of full coherence dark annular laser beam for the five values of k = 0, 0.2, 0.4, 0.6 and 0.8. PF0 difference is estimated as a deviation between both set of data – experimental (exp) and simulated (sim)**

## *Discussion and conclusions*

We found that from the five experimental points four of them are within 16% error as to the simulated dependence. We conceive it as tolerable in such a measurement. An exception is the point of highest annularity (k=0.8). Its deviation is 38% which is much higher. With regard to this, it should be good to note that our simulation was based on MATLAB®-incorporated functions which use a matrix (discrete) form of calculation. In the output pattern, instead of concentric rings we got side peaks concentrically surrounding the center (*cf.* Figures 5.2.2-3). The appearance of side peaks is an artifact which comes from the calculation method we employed. When annularity gets higher (thinner annulus) the pixeled structure is more apparent and the divergence from the ring shape (uninterrupted and smooth) becomes very distinct. The experiment contributed to total error increase too. Because this error is due to the visual asymmetry of experimental output data that is seen in the images and intensity profiles in Table 5.3.1.

Considering that we can conclude that the experiment verifies the issues of a two-dimensional Fast Fourier Transform simulation of coherent dark (hollow, zero-intensity core) annular flat beams done before. The discrepancy is tolerable and we may expect to reduce it by improvement of simulation and experiment.

## 5.4 Simulation of hollow annular beams as dependents of the annularity parameter

During the FFT simulation of hollow annular beams of Idip=1 (aka dark *etc.*) we found interesting correlations between certain parameters of the far-field intensity profile of these beams (Astadjov, D.N., 2009, 2010).

Here we shall define some parameters which concern energy relations in the central area of the far-field intensity profile of these beams. So PF0 (the power fraction of "zero" center) is the magnitude of the highest intensity pixel divided by the sum of the values of all pixels of the far-field intensity profile. We regard it as the power fraction of the very peak of the far-field intensity profile with respect to whole beam power. PF1 (the power fraction of "zero+1" center) is defined same way but here numerator yet includes the sum of the intensities of nearest 8 pixels around the peak intensity pixel (the "zero" center) plus the "zero" center itself (*i.e.* the numerator comes out from a 3x3 matrix). Also, we can make use of the quantity pf=PF0/PF1, which actually is the magnitude of the highest intensity pixel over the sum of the intensities of nearest 8 pixels around it plus the peak pixel itself. All three quantities are plotted in Figure 5.4.1 as functions of beam annularity.

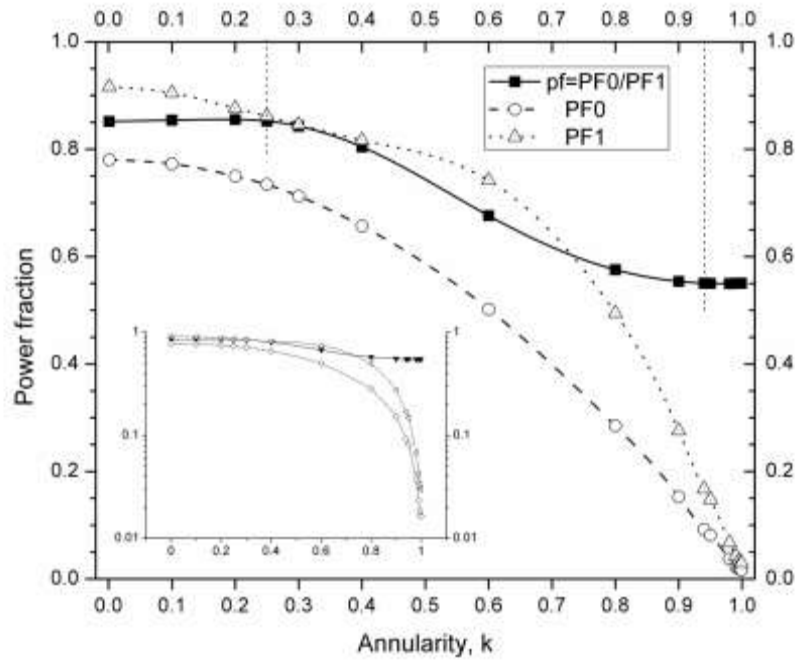

**Figure 5.4.1 Power fraction parameters PF0, PF1 and pf as functions of annularity of dark annular beams**

As seen, PF0 and PF1 go down with the increase of annularity k; then they plunge for k>0.9 when k approaches unit (see the inset). Meanwhile, the behavior of pf can be depicted as having two stable constant values which pf tends to. So in the left part of the line, pf=0.85 when k<0.25; and in the right end, pf=0.550 for k>0.94. These two points divide the pf-curve into three sections.

The behavior of far-field 3D pattern is quite distinctive in the different sections. As a whole, the far-field 3D pattern changes in form and amplitude. This is shown in Figure 5.4.2, where for explicitness we fix the z-axis and feeble patterns (as of k≥0.8) are magnified.

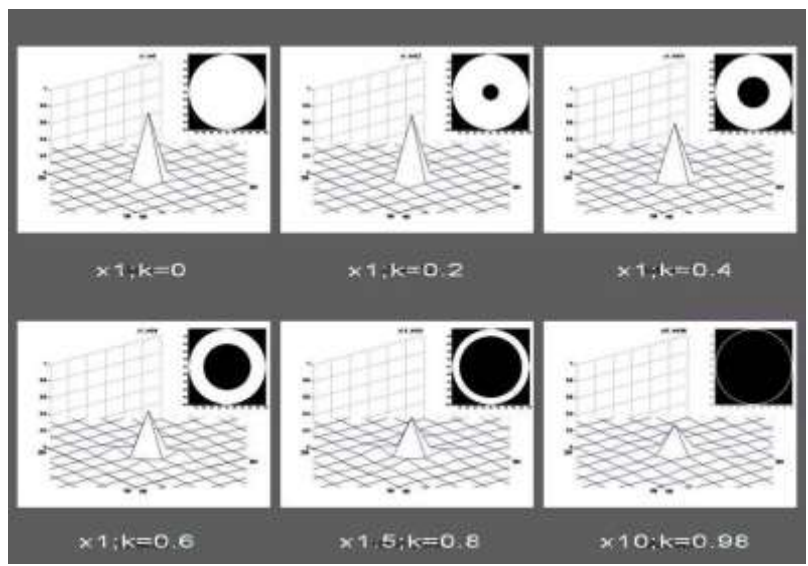

**Figure 5.4.2 Far-field patterns of dark annular beams of different annularity; feeble patterns of k≥0.8 are magnified**

Far-field 3D patterns in four points of the left section (k≤0.25) and four more of the right section (k≥0.94) of the pf-line are given in Figure 5.4.3 and Figure 5.4.4, respectively. Despite the changes in near fields (the inlets) we do not observe considerable variations of far-field structures. The only exception is the magnitude of the "zero" peak amplitudes of the right section of the pf-line (Figure 5.4.4). So, for k=0.999 the amplitude (PF0~8.10$^{-3}$) is an order of magnitude less than for k=0.94 (PF0≈0.09).

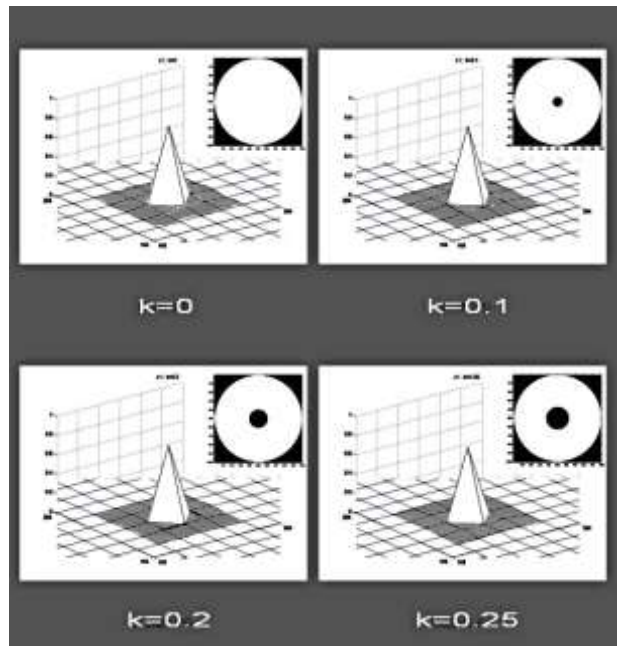

**Figure 5.4.3 Far-field patterns of dark annular beams of annularity k≤0.25**

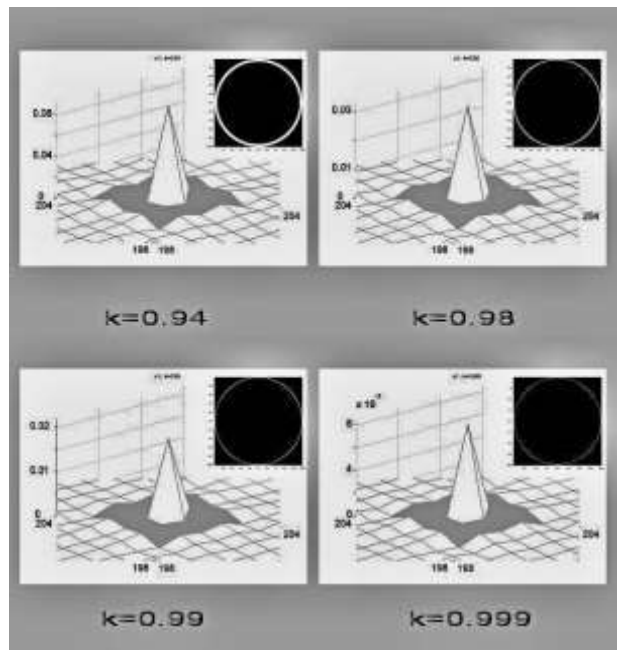

**Figure 5.4.4 Far-field patterns of dark annular beams of annularity k≥0.94**

*Discussion and conclusions*

The simulation findings can be commented from several points of view. If we are interested in the technology-relevant power laser applications the simulation of beam annularity effect can give us a notion of the spread structure of focused light in the focal plane. As simulation showed serious malformation (say, the central peak fraction in Figure 5.4.1 becomes less than 0.5) of the far-field pattern is probable when the near-field annularity is high (k>0.65) and the annulus is almost dark (Idip>0.9 *i.e.* the central sag intensity is less than 10% of the ring intensity). This situation does not occur frequently. However, even in that case, since the central peak energy is concentrated into a smaller spot area, the net impact of side energy spread could not be so detrimental.

A quite unexpected result was the simulation finding that up to an annularity of 0.25 the far-field intensity pattern is actually not disturbed by the black hole in the near-field center (*cf.* Figure 5.4.3). A parameter, pf was introduced as characteristic of different kinds of behavior of far-field intensity structure (Figure 5.4.1). The range of pf is $0.55 \leq pf \leq 0.85$. So if pf=0.85 we have a steady-state far-field pattern produced from near field with a "small" (k<0.25) black hole in the center. The lower limit (0.55) of pf also corresponds to a steady-state kind of far-field pattern (*cf.* Figure 5.4.4). This time near field is actually a thin annulus of k>0.94. The shape of the far-field structure is invariable but its amplitude diminishes with the increase of k. Having in mind that very thin annular beams are required for generation of Bessel beams (Durnin *et al.,* 1987), it is quite possible that annularity k>0.94 is a reflection of a kind of threshold conditions for Bessel beam production.

# *Afterword*

In this book, I made an effort to expose the variety of research carried out in the last 20 years in the Laboratory of metal vapor lasers of Institute of solid state physics, Bulgarian academy of sciences which yielded to the realization of a unique scientific product – the copper bromide vapor laser. On the ground of worldwide perception and appreciation of this originally-Bulgarian invention, I dare to entitle the book as high-end state-of-art copper bromide vapor lasers.

Here I just outline the major fields we dealt with success which is recognized by the research community in the world.

We proved the leading position of CuBr laser in terms of specific average power per unit volume (1.4 W/cm$^3$ from a small-bore laser), the average power of 120W with 3% efficiency (a large-bore laser) and a lifetime of thousand hours of completely sealed-off operation (30W/40W lasers).

We entered deeply the physical phenomena occurring in the body of CuBr laser with the investigation of energy dissipation and dynamics of the electric field in the gaseous active zone and electrical interface – the electrodes. We pointed out that hydrogen effect (addition of a few tenths of Torr of hydrogen to main buffer gas neon) is beneficial to the distribution of energy and electric field in CuBr laser in the favor of an upgraded laser performance in power, efficiency and lifetime. Electric field measurements are conducted by a *nonperturbation* technique employed in our laboratory, for the first time (to our best knowledge).

We operated the copper bromide vapor laser in the step-like mode – one master oscillator seeded a power amplifier (forming MOPA system) and further enhanced laser characteristics as low beam divergence, high brightness, high optical coherence *etc*. With an MOPA using generalized diffraction filtering resonator as a cavity for the master oscillator, we produced diffraction-limited high-power pulses with *constant* divergence and spatial coherence. The throughout-pulse divergence was just 6 % above the diffraction limit. This is the perfect pulsed laser light source to be utilized.

For the experimental measurement of spatial coherence, we assembled a *novel* reversal shear interferometer (a variety of the Michelson interferometer) of very simple construction comprising just four optical components (namely, a spherical lens, an optical wedge and two plane mirrors of high reflectivity). It is a fast and rigid instrument for coherence analysis of optical beams. We applied this interferometer to measure the spatial coherence of other lasers. So the spatial coherence of a low-cost 532nm green laser appeared to be pretty high (Astadjov & Prakash, 2013b).

Theoretically and experimentally we examined the behavior of laser beams out of the laser. We studied the impact of intensity beam profile on beam propagation and focusability. We found that the intensity beam profile has a low influence on the applications we normally employ the copper bromide vapor lasers.

The summarized issues approve the classification - high-end state-of-art copper bromide vapor lasers.

## *Acknowledgements*

At the end of this book, I would like to thank the people who all these years have been supporting me in doing research and thus, my career. Here are my co-workers: prof. Nikola Sabotinov (a Real Member of BAS - academician, President of BAS 2008-2012, and lab head for most of the time), prof. Nikolai Vuchkov, assoc. prof. Margarita Grozeva, and assoc. prof. Petar Telbizov, and all other guys from the Lab (Department) of metal vapor lasers.

I owe special thanks to acad. Sabotinov who pioneered and steered R&D&T of metal vapor laser in Bulgaria, and invited me to work with the lab for all these years of creativity and success.

Indispensable partners are also Ivan Angelov (senior expert and head of Department of glass blowing) and Lina Todorova (senior expert and head of Department of optical devices).

I highly appreciate the attitude of prof. Alexander Petrov (a Real Member of BAS – academician and Director of ISSP 1999-2015) to my work.

I want to acknowledge the vital financial support for our work the following institutions/programs: NATO Science for Peace Program, EU Copernicus Program, National Science Fund, Indo-Bulgarian Program of Cooperation in Science & Technology, and Bulgarian Academy of Sciences.

Last but not least, I am thankful to my wife, Elena who lovingly cares for our family and makes possible my endeavor in the professional life.

# *Bibliography*

Akirtava, O.S., V.L. Dzhikiya & Y.M. Oleinik (1975) *Kvant.elektronika* **2** 181-1832.

Astadjov, D. (1989) Hydrogen effect on the laser oscillation and the amplification properties of copper bromide vapor *Ph.D. Thesis*, Sofia, Bulgaria.

Astadjov D., (1996) Influence of hydrogen on the kinetics of copper bromide lasers, in *Pulsed Metal Vapour Lasers* (C.E. Little & N.V. Sabotinov Eds.) Kluwer Academic Publishers, Dordrecht, 169-174.

Astadjov, D.N. (2009) Fourier transform of annular beams *Arxiv* http://arxiv.org/ftp/arxiv/papers/0904/0904.1911.pdf

Astadjov, D.N. (2010) Energy focusability of annular beams *AIP Conf. Proc.* **1203**, 472-476.

Astadjov, D. & N. Sabotinov (1996) Electrode power losses in the pulsed discharge of CuBr laser *CLEO Europe'96, IEEE Tech.Dig.* Piscataway NJ, CThI41.

Astadjov, D.N. & N.V. Sabotinov (1997) Energy dissipation in the electrodes of CuBr laser-like discharges *J. Phys. D: Appl. Phys.* **30** 1507–1514.

Astadjov, D.N. & N.V. Sabotinov (1999) Dynamics of electric field in CuBr laser-like discharges *Meas. Sci. Technol.* **10** N67–N70.

Astadjov, D.N. & S.V. Nakhe (2010) CuBr laser beam transformations *J. Phys.: Conf. Series* **253** 012076.

Astadjov, D.N. & Om Prakash (2013a) Experimental verification of focusability of coherent annular laser beams *Proc. SPIE* **8770** 87701O.

Astadjov, D.N. & Om Prakash (2013b) Spatial coherence of low-cost 532nm green lasers *Proc. SPIE* **8770** 87701P.

Astadjov, D.N., L.I.Stoychev, S.K.Dixit, S.V.Nakhe & N.V.Sabotinov (2005) High-brightness CuBr MOPA laser with diffraction-limited throughout-pulse emission *IEEE J.Quantum Electronics* **41**, 1097-1101.

Astadjov, D.N., L.I.Stoychev & N.V.Sabotinov (2006a) Improvement of CuBr laser coherence properties *Proc. SPIE* **6252**, 1/ 625229-5.

Astadjov, D.N., L.I. Stoychev & N.V. Sabotinov (2006b) CuBr laser pulse shaping by MOPA system *Proc. 4th Intern. Symp."LTL 2005* Plovdiv Bulgaria, 111-114.

Astadjov, D., L. Stoychev & N. Sabotinov (2007a) $M^2$-factor for MOPA CuBr laser system *Proc. SPIE* **6604**, 66040Z.

Astadjov, D., L. Stoychev & N. Sabotinov (2007b) $M^2$ of MOPA CuBr Laser Radiation *Opt Quant Electron* **39**, 39603–39610.

Astadjov, D., N. Sabotinov & N. Vuchkov (1985) Effect of hydrogen on CuBr laser power and efficiency *Opt. Commun.* **56**, 279-282.

Astadjov, D. N., N K Vuchkov, A A Isaev, G G Petrash, I V Ponomarev & N V Sabotinov (1986) Efficient CuBr laser with hydrogen additives *Kratkie Soobshcheniya po Fizike RIAN* (N G Basov Ed.) 11 RIAN Moscow, 58–60.

Astadjov, D.N., N.K. Vuchkov & N.V. Sabotinov Parametric study of the CuBr laser with hydrogen additives (1988a) *IEEE J. Quantum Electronics* **24**, 1927-1935.

Astadjov, D.N., N.K. Vuchkov, K.I. Zemskov, A.A. Isaev, M.A. Kazaryan, G.G. Petrash & N.V. Sabotinov (1988b) Active optical systems with a copper bromide vapor amplifier *Kvantovaya Elektronika* **15**, 716-719.

Astadjov, D.N., K.D. Dimitrov, C.E. Little, N.V. Sabotinov & N.K. Vuchkov (1994) A CuBr laser with 1.4 W/cm$^3$ average output power *IEEE J. Quantum Electronics* **30**, 1358-1360.

Astadjov, D. N., K. D. Dimitrov, D. R. Jones, V. Kirkov, L. Little, C. E. Little, N. V. Sabotinov & N. K. Vuchkov (1997a) Influence on operating characteristics of scaling sealed-off CuBr lasers in active length *Opt. Commun.* **135**, 289-294.

Astadjov, D.N., K.D. Dimitrov, D.R. Jones, V.K. Kirkov, C.E. Little, N.V. Sabotinov & N.K. Vuchkov (1997b) Copper bromide laser of 120-W average output *IEEE J. Quantum Electronics* **33**, 705-709.

Astadjov, D.N., L.I. Stoychev, S.K. Dixit, S.V. Nakhe & N.V. Sabotinov (2005) High-brightness CuBr MOPA laser with diffraction-limited throughout-pulse emission *IEEE J. Quantum Electronics* **41**, 1097-1101.

Astadjov, D.N., L.I. Stoychev & N.V. Sabotinov (2006) Improvement of CuBr laser coherence properties *Proc. SPIE* **6252** 625229.

Astadjov, D., L. Stoychev & N. Sabotinov (2007) $M^2$ of MOPA CuBr laser radiation *Opt Quant Electron* **39**, 603–610.

Bakiev, A.M., S.Kh. Valiev & N.V. Kryazhev (1991) Investigation of coherent properties of the radiation of active optical systems with a copper vapor amplifier, *Kvantovaya Elektronika* **18**, 1256.

Bhatnagar, R., S.K. Dixit, B. Singh & S.V. Nakhe (1989) Performance of copper vapor laser with self- filtered unstable resonator *Opt. Commun.* **74**, 93-95.

Boettocher W, H Lueck, S Niesner & A Schwabedissen (1992) Small volume coaxial discharge as precision testbed for 0D-models of XeCl lasers *Appl. Phys. B* 54, 295–302.

Brown, D. J. W. & D. W. Coutts (1996) Beam quality issues in copper vapour lasers, in *Pulsed Metal Vapour Laser* (C. E. Little and N. V. Sabotinov Eds.) Kluwer Academic Publishers Dordrecht, 241-254.

Brown, D. J. W., M. J. Withford & J. A. Piper (2001) High-power, high-brightness master-oscillator power-amplifier copper laser system based on kinetically enhanced active elements *IEEE J. Quantum Electron.* **37**, 518-524.

Chang, J.J. (1994) Time-resolved beam quality characterization of copper vapor laser with unstable resonators *Appl. Opt.* **33**, 2255- 2265.

Chang, J. J. (1995) Copper laser oscillator with adjoint-coupled self-filtering injection *Opt. Lett.* **20**, 575-577.

Chang, J.J., C.D. Boley, M.W. Martinez, W.A. Molander & B.E. Warner (1994) in *Gas, Metal Vapor and Free Electron Lasers, and Applications* (V.N Smiley & F.K. Tittel Eds.) *SPIE* **2118**, 2-8.

Chang, J.J., B.E. Warner, C.D. Boley & E.P. Dragon (1996) High-power copper vapour lasers and applications, in *Pulsed Metal Vapour Lasers* (C.E. Little & N.V. Sabotinov Eds.) Kluwer Academic Publishers, Dordrecht, 101–112.

Collins, T. J. B. & S. Skupsky (1999) The effects of pulse shaping on imprint *Report at 41st Annual Meeting of the American Physical Society (Division of Plasma Physics)* Seattle WA.

Coutts, D.W. (1995) Time-resolved beam divergence from a copper vapor laser with unstable resonator *IEEE J. Quantum Electronics* **31**, 330-342.

Coutts, D. W. (2002) *IEEE J. Quantum Electronics* **38**, 1217

Daalder J E (1975) Erosion and the origin of charged and neutral species in vacuum arcs *J. Phys. D: Appl. Phys.* 8, 1647–1659.

Daalder J E (1977) Energy dissipation in the cathode of a vacuum arc *J. Phys. D: Appl. Phys.* 10, 2225–2234.

Dickey, F. M. & S. C. Holswade (Eds.) (2000) Laser beam shaping: theory and techniques *Opt. Eng. Series* **70** Marcel Dekker Inc.

Dixit, S.K. Ed. (2001) *Filtering resonators* Nova Science Publishers, New York.

Dixit, S.K. (2005) Private communication.

Dixit, S.K., J.K. Mittal, B. Singh, P. Saxena & R. Bhatnagar (1993) A generalized diffraction filtered resonator with a copper vapor laser *Opt. Commun*. **98,** 91-94

Dixit, S.K., B. Singh, J.K. Mittal, R. Choubey & R. Bhatnagar (1994) Analysis of the temporal and spatial characteristics of the output from short inversion time self-terminating lasers with various resonators *Opt. Eng.* **33**, 1908-1920.

Dixit, S.K., Om Prakash, S. Talwar & R. Bhatnagar (2001) Generalized diffraction filtered resonator (GDFR) copper vapor laser and its applications: a study, in *Filtering Resonators*, S.K. Dixit, Ed. Nova Science Publishers, New York, 253-301.

Dixit, S.K., Om Prakash, D.N. Astadjov & R. Bhatnagar (2003) Unpublished observations.

Durnin, J., J. J. Miceli & J. H. Eberly (1987) Diffraction-free beams *Phys. Rev. Lett.* **58**, 1499-1501.

Eggleston, J.M. (1988) Theory of output beam divergence in pulsed unstable resonators *IEEE J. Quantum Electronics* **24**, 1302-1311.

Elaev, V.F., G.D. Lyakh & V.P. Pelenkov (1989) CuBr laser with average lasing power exceeding 100 W *Atm. Optics* **2,** 1045-1047.

Geng, J. Spectral control in copper bromide lasers and its application in high-speed holography (1996) in *Pulsed Metal Vapour Lasers* (C.E. Little & N.V. Sabotinov Eds.) Kluwer Academic Publishers, Dordrecht, 269-274.

Giao, M. A. P., W. Miyakawa, N. A. S. Rodrigues, D. M. Zezell, R. Riva, M. G. Destro, J. T. Watanuki & C. Schwab (2006) High beam quality in a HyBrID copper laser operating with an unstable resonator made of a concave mirror and a plano-convex BK7 lens *Optics & Laser Technology* **38,** 523-527.

Goodman, J. W. (1996) *Introduction to Fourier optics* McGraw-Hill, 4-31; 101-107.

Guyadec, E. Le., P. Countance, G. Bertrand & G. Peltier (1999) *IEEE J. Quantum Electronics.* **35**, 1616.

Hargrove, R.S., R. Grove & T. Kan (1979) Copper vapor laser unstable resonator oscillator-amplifier characteristics *IEEE J. Quantum Elect*ronics **QE15**, 1228-1233.

Ilyushko, V.G., E.K. Karabut, V.F. Kravchenko & V.S. Mikhalevskii (1985) *Sov. J. Quantum Electronics* **15**, 1442.

Isaev, A. A. & G. Yu. Lemmerman (1987) Power supply for pulsed metal vapours lasers *Trudi FIAN* (N G Basov Ed.) 181 Nauka Moscow, 164–179.

ISO 11146 (2005) Lasers and laser-related equipment - test methods for laser beam parameters - beam widths, divergence angle and beam propagation factor *International Organization for Standardization*.

Jones, D. R., A. Maitland & C. E. Little (1994) A high-efficiency 200 W average power copper HyBrID laser *IEEE J. Quantum Electronics* **30**, 2385-2390.

Kleine, K. F. & K. G. Watkins (2004) Pulse shaping for micro cutting applications of metals with fiber lasers *Proc. SPIE* **5339** San Jose USA

Knyazev I. I. (1971) Investigation of physical processes in pulsed gas-discharge lasers on molecules of hydrogen and deuterium, and the first positive band system of nitrogen molecule *Trudi FIAN* (N G Basov Ed.) 56 Nauka Moscow, 119–190.

Kravchenko, V.F., E.K. Karabut, A.A. Gudkov & V.E. Bogoslavskii (1982) *Sov. J. Quantum Electronics* **12,** 143.

Kreye, W.C. & F.L.Roesler (1983) High-resolution line-shape analyses of the pulsed cuprous-chloride oscillator and amplifier *Appl.Opt.* **22**, 927

Kukharev V. N. (1984) Space-time field characteristics in longitudinal continuous-pulsed discharge, typical for pumping of lasers on self-terminating transitions *J. Tech. Phys.* 54, 1910–1914.

Lee, S. H. (1981) *Optical information processing - fundamentals topics in applied physics* (S.H. Lee, Ed.) **48** Springer Berlin, 2-41.

Little, C. (1999) *Metal vapour lasers: physics, engineering, and applications*, John Wiley & Sons Ltd.

Little, C.E. & N.V. Sabotinov Eds. (1996) *Pulsed metal vapour lasers*, Kluwer Academic Publishers, Dordrecht, Netherlands.

Liu, C.S., D.W. Feldman, J.L. Pack & L.A. Weaver (1977) *IEEE J. Quantum Electronics* **QE13**, 744.

Livingstone, E. S., D. R. Jones, A. Maitland & C. E. Little (1992) Characteristics of a copper bromide laser with flowing Ne–HBr buffer gas *Opt. Quantum Electron.* **24**, 73-82.

Loeb L.B. (1959) Significance of formative time lags in gaseous breakdown *Phys. Rev.* **113**, 7–12

Longhurst, R. S. (1957) *Geometrical and physical optics* Longmans, Green & Co, 194.

Low, K. Y., L. Li, A. G. Corfe & P. J. Byrd (1999) Taper control during laser percussion drilling of NIMONIC alloy using sequential pulse delivery pattern control (SPDPC) *Proc. 18 Int. Cong. Appl. of Laser & Electro-Optics (ICALEO '99)* **87C**, 11.

Metzler, N. A., L. Velikovich & J.H.Gardner (1999) Reduction of early-time perturbation growth in ablatively driven laser targets using tailored density profiles, *Physics of Plasmas* **6**, 3283

Michaud, E.J., H.W. Kerr & D.C. Weckman (1995) Temporal pulse shaping and solidification cracking in laser welded Al-Cu alloys *Proc. 4th Int'l. Conf. Trends in Welding Research* (H.B. Smartt, J.A. Johnson, S.A. David, Eds.), ASM Int'l. Gatlinburg, TN, 153

Omatsu, T. & K.Kuroda (1996) Beam quality issues for the second harmonic generation of copper vapor radiation, in *Pulsed Metal Vapour Lasers* (C.E. Little & N.V. Sabotinov Eds.) Kluwer Academic Publishers, Dordrecht, 263-268.

Omatsu, T., K. Kuroda, T. Shimura, K. Chihara, M. Itoh & I. Ogura (1991) Measurement of spatial coherence of copper vapour laser beam using a reversal shear interferometer *Optical and Quantum Electronics* **23**, S477

Omatsu, T., K. Kuroda & T. Takase (1992) Time-resolved measurement of spatial coherence of a copper vapor laser beam using a reversal shear interferometer *Opt. Commun*. **87**, 278-286.

Omatsu, T., T. Takase & K. Kuroda (1993) Intra-pulse decrease of $M^2$ of a copper vapor laser beam *Opt. Commun*. **101**, 199-204.

Phillips, J R (2010) Online curve fitting and surface fitting website zunzun.com *Unavailable now*

Pini, R., R. Salimbeni, G. Toci & M. Vannini (1992) Efficient second harmonic generation in β-barium borate by diffraction limited copper vapor laser *Applied Optics* **31**, 2747-2751.

Prakash, O., P K Shukla, S K Dixit, S Chatterjee, H S Vora & R Bhatnagar (1998) Spatial coherence of generalised diffraction filtered resonator copper vapour laser *Applied Optics* **37**, 7752-7757.

Prakash, O., S.K. Dixit & R. Bhatnagar (2002) On the role of coherence width and its evolution in a short pulse fundamental beam in second harmonic generation from beta barium borate *IEEE J. Quantum. Electronics* **38**, 603-613.

Prakash, O., G.N. Tiwari, S.K. Dixit & R. Bhatnagar (2003) Single pulse time-resolved comparative study on the performance of a master-oscillator power amplifier copper vapor laser system with generalized diffraction and unstable resonator as master oscillator *Applied Optics* **42**, 3538-3545.

Prakash, O., R. Mahakud, H.S. Vora & S.K. Dixit (2006) Cylindrical-lens-based wavefront-reversing shear interferometer for the spatial coherence measurement of UV radiations *Opt. Eng.* **45**, 055601

Prakash, O., D.N. Astadjov, P. Kumar, R. Mahakud, J. Kumar, S.V. Nakhe & S.K. Dixit (2013) Effect of spatial coherence on the focusability of annular laser beams *Optics Communications* **290** 1–7.

Raizer Yu. P. (1987) *Fizika Gazovogo Razryada (Physics of Gas Discharge)* Nauka, Moscow.

Sabotinov, N. (1991) Copper bromide vapor laser *D.Sc. Thesis*, Sofia.

Sabotinov, N.V. (1996) Copper bromide lasers, in *Pulsed Metal Vapour Lasers* (C.E. Little & N.V. Sabotinov Eds.) Kluwer Academic Publishers, Dordrecht, 113-124.

Sabotinov, N.V. & C.E. Little (1993) Unpublished observations.

Sabotinov, N.V., P.K. Telbizov & S.D. Kalchev (1975) *Bulgarian patent* N28674.

Sabotinov, N. V., N. K. Vuchkov & D. N. Astadjov (1986) Progress in CuBr lasers *Int. Conf. CLEO'86* San Francisco, 160.

Sabotinov, N. V., N. K. Vuchkov & D. N. Astadjov (1990) Copper bromide lasers: discharge

tubes and lifetime problems *Proc. SPIE* **1225**, 289-298.

Sabotinov, N.V., F. Akerboom, D.R. Jones, A. Maitland & C.E. Little (1995) *IEEE J. Quantum Electronics* **QE31**, 747.

Sabotinov, N.V., K.D.Dimitrov, N.K.Vuchkov, D.N.Astadjov, V.K.Kirkov, C.E.Little & D.R.Jones (1996) A copper bromide laser of 120W output power *CLEO Europe'96, Tech.Dig.IEEE* Piscataway NJ, CWE3.

Salimbeni, R. (1996) Beam quality issues in CVL applications, in *Pulsed Metal Vapour Laser* (C. E. Little and N. V. Sabotinov Eds.) Kluwer Academic Publishers Dordrecht, 229-240.

Satoh, K, R Morrow & R J Carman (1997) Modeling of the plasma kinetics in the electrode sheath regions of a pulsed copper vapour laser *XXIII ICPIG* **2** (M. C. Bordage & A. Gleizes Eds) Universit´e Paul Sabatier Toulouse, 258.

Shuhtin A.M., V.G. Mishakov, G.A. Fedorov & A.A. Ganeev (1975) *Optika i spektroskopiya* **39** 785-786.

Siegman, A.E. (1997) Tutorial presentation *OSA Annual Meeting* Long Beach, California.

Siegman, A.E. (2004) How to (maybe) measure laser beam quality *Stanford University*, http://www.stanford.edu/siegman/beamqualityseminar.pdf.

Stoian, R., M.Boyle, A.Thoss, A.Rosenfeld, G.Korn & I.V.Hertel (2003) Ultrafast laser material processing using dynamic temporal pulse shaping *Focused on Laser Precision Microfabrication* (LPM 2002), *RIKEN Review* No.50

Stoilov, V.M., D.N. Astadjov, N.K. Vuchkov & N.V. Sabotinov (2000) High-spatial intensity 10W-CuBr laser with hydrogen additives *Optical and Quantum Electronics* **32**, 1209-1217.

Stoychev, L.I., D.N. Astadjov & N.V.Sabotinov (2009) Green and yellow laser lines output of CuBr laser *Proc. DAE-BRNS NLS-08 High-power high-energy Lasers* Delhi India P1-022.

Svelto, O. (1984) *Principles of Lasers* Mir Moskva 288

Tyth, Cs., J. Faure, J. van Tilborg, C. G. R. Geddes, C. B. Schroeder, E. Esarey & W. P. Leemans (2003) Tuning of laser pulse shapes in grating-based compressors for optimal electron acceleration in plasmas *Optics Letters* **28**, 1823

Vorobev, V. B., S. V. Kalinin, I. I. Klimovski, I. Kostadinov, V. A. Krestov, V. N. Kubasov & O. Marasov (1991) A copper vapor laser with average specific emission power above 1 W/cm$^3$ *Kvant. Elektronika* **18**, 1178-1180.

Vuchkov, N. (1984) Physical factors limiting the lifetime of copper bromide vapor laser, *Ph.D. Thesis*, Moskva.

Vuchkov, N. K., D. N. Astadjov & N. V. Sabotinov (1991) A new circuit for CuBr laser excitation *Optical and Quantum Electron.* **23**, S549–S553.

Vuchkov, N. K., D. N. Astadjov, & N. V. Sabotinov (1994) Influence of the excitation circuits on the CuBr laser performance *IEEE J. Quantum Electronics* **30**, 750-758.

Westberg, R G (1959) Nature and role of ionizing potential space waves in glow-to-arc transitions *Phys. Rev.* 114, 1–17.

Zemskov, K.I., M.A. Kazaryan, V.G. Mokerov, G.G. Petrash & A.G. Petrova (1978) Coherent properties of a copper vapor laser and dynamic holograms in vanadium dioxide films *Kvantovaya Elektronika* **5**, 245